AI Ethics and Governance in Practice Programme

# AI Fairness in Practice

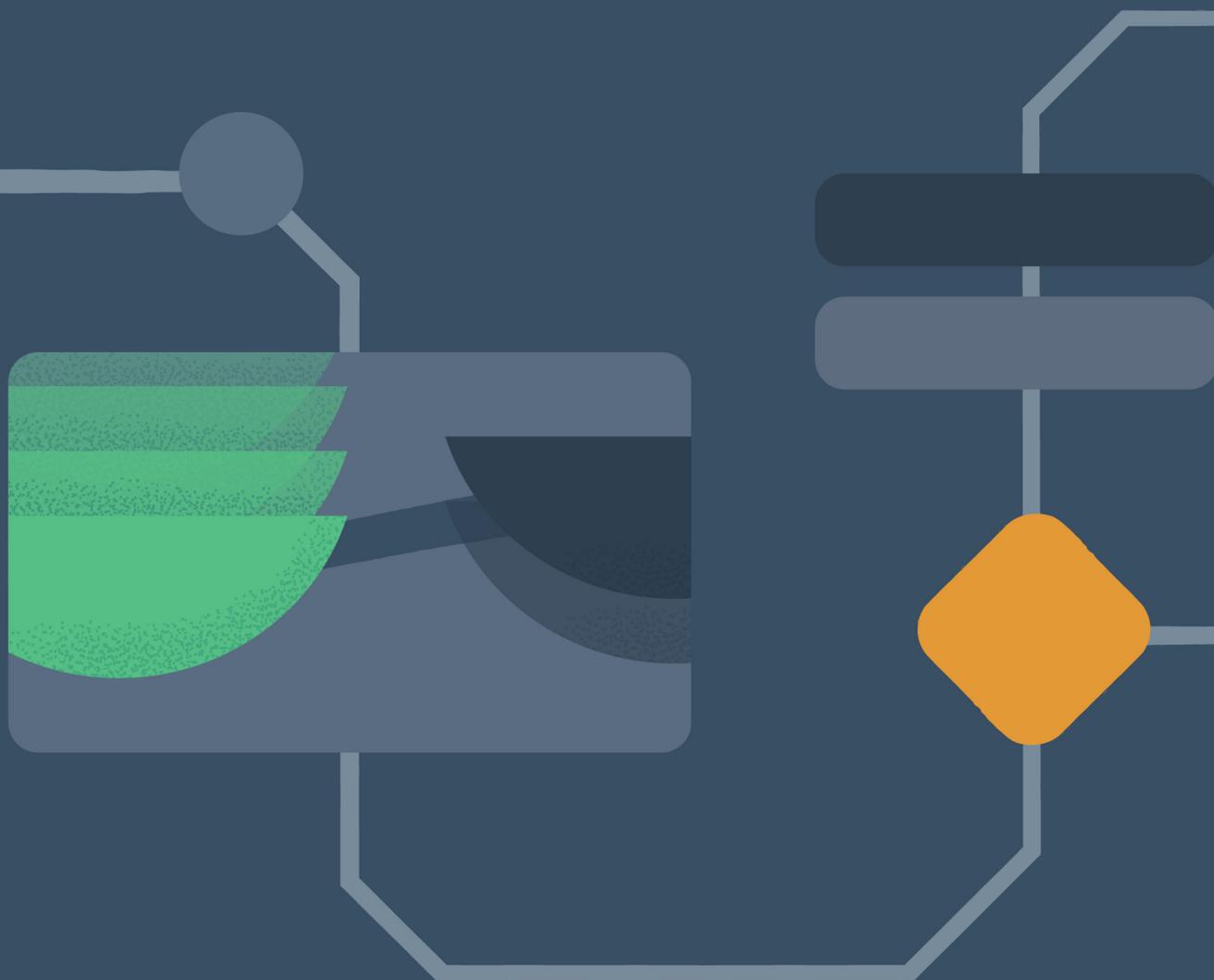

**For Facilitators**
This workbook is annotated to support facilitators in delivering the accompanying activities.

The Alan Turing Institute

# Acknowledgements

This workbook was written by David Leslie, Cami Rincón, Morgan Briggs, Antonella Perini, Smera Jayadeva, Ann Borda, SJ Bennett, Christopher Burr, Mhairi Aitken, Michael Katell, Claudia Fischer, Janis Wong, and Ismael Kherroubi Garcia.

The creation of this workbook would not have been possible without the support and efforts of various partners and collaborators. As ever, all members of our brilliant team of researchers in the Ethics Theme of the Public Policy Programme at The Alan Turing Institute have been crucial and inimitable supports of this project from its inception several years ago, as have our Public Policy Programme Co-Directors, Helen Margetts and Cosmina Dorobantu. We are deeply thankful to Conor Rigby, who led the design of this workbook and provided extraordinary feedback across its iterations. We also want to acknowledge Johnny Lighthands, who created various illustrations for this document, and Alex Krook and John Gilbert, whose input and insights helped get the workbook over the finish line. Special thanks must be given to the Information Commissioner's Office for helping us test the activities and reviewing the content included in this workbook, and Sophia Ignatidou (Information Commissioner's Office) and Claire Lesko (Equality and Human Rights Commission) for providing content inputs. Lastly, we would like to thank Sabeehah Mahomed (The Alan Turing Institute) and Janis Wong (The Alan Turing Institute) for their meticulous peer review and timely feedback, which greatly enriched this document.

This work was supported by Wave 1 of The UKRI Strategic Priorities Fund under the EPSRC Grant EP/W006022/1, particularly the Public Policy Programme theme within that grant & The Alan Turing Institute; Towards Turing 2.0 under the EPSRC Grant EP/W037211/1 & The Alan Turing Institute; and the Ecosystem Leadership Award under the EPSRC Grant EP/X03870X/1 & The Alan Turing Institute.

Cite this work as: Leslie, D., Rincón, C., Briggs, M., Perini, A., Jayadeva, S., Borda, A., Bennett, SJ. Burr, C., Aitken, M., Katell, M., Fischer, C., Wong, J., and Kherroubi Garcia, I. (2023). *AI Fairness in Practice.* The Alan Turing Institute.

# Contents

# About the AI Ethics and Governance in Practice Workbook Series

## Who We Are

The Public Policy Programme at The Alan Turing Institute was set up in May 2018 with the aim of developing research, tools, and techniques that help governments innovate with data-intensive technologies and improve the quality of people's lives. We work alongside policymakers to explore how data science and artificial intelligence can inform public policy and improve the provision of public services. We believe that governments can reap the benefits of these technologies only if they make considerations of ethics and safety a first priority.

## Origins of the Workbook Series

In 2019, The Alan Turing Institute's Public Policy Programme, in collaboration with the UK's Office for Artificial Intelligence and the Government Digital Service, published the [UK Government's official Public Sector Guidance on AI Ethics and Safety](). This document provides end-to-end guidance on how to apply principles of AI ethics and safety to the design, development, and implementation of algorithmic systems in the public sector. It provides a governance framework designed to assist AI project teams in ensuring that the AI technologies they build, procure, or use are ethical, safe, and responsible.

In 2021, the UK's National AI Strategy recommended as a 'key action' the update and expansion of this original guidance. From 2021 to 2023, with the support of funding from the Office for AI and the Engineering and Physical Sciences Research Council as well as with the assistance of several public sector bodies, we undertook this updating and expansion. The result is the AI Ethics and Governance in Practice Programme, a bespoke series of eight workbooks and a forthcoming digital platform designed to equip the public sector with tools, training, and support for adopting what we call a Process-Based Governance (PBG) Framework to carry out projects in line with state-of-the-art practices in responsible and trustworthy AI innovation.

# About the Workbooks

The AI Ethics and Governance in Practice Programme curriculum is composed of a series of eight workbooks. Each of the workbooks in the series covers how to implement a key component of the PBG Framework. These include Sustainability, Technical Safety, Accountability, Fairness, Explainability, and Data Stewardship. Each of the workbooks also focuses on a specific domain, so that case studies can be used to promote ethical reflection and animate the Key Concepts.

**Programme Curriculum: AI Ethics and Governance in Practice Workbook Series**

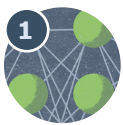

**1  AI Ethics and Governance in Practice: An Introduction**
*Multiple Domains*

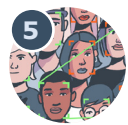

**5  Responsible Data Stewardship in Practice**
*AI in Policing and Criminal Justice*

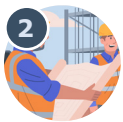

**2  AI Sustainability in Practice Part One**
*AI in Urban Planning*

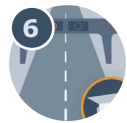

**6  AI Safety in Practice**
*AI in Transport*

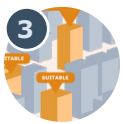

**3  AI Sustainability in Practice Part Two**
*AI in Urban Planning*

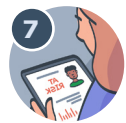

**7  AI Transparency and Explainability in Practice**
*AI in Social Care*

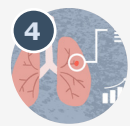

**4  AI Fairness in Practice**
*AI in Healthcare*

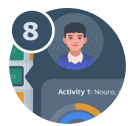

**8  AI Accountability in Practice**
*AI in Education*

Taken together, the workbooks are intended to provide public sector bodies with the skills required for putting AI ethics and governance principles into practice through the full implementation of the guidance. To this end, they contain activities with instructions for either facilitating or participating in capacity-building workshops.

Please note, these workbooks are living documents that will evolve and improve with input from users, affected stakeholders, and interested parties. We need your participation. Please share feedback with us at policy@turing.ac.uk.

**Programme Roadmap**

The graphic below visualises this workbook in context alongside key frameworks, values and principles discussed within this programme. For more information on how these elements build upon one another, refer to AI Ethics and Governance in Practice: An Introduction.

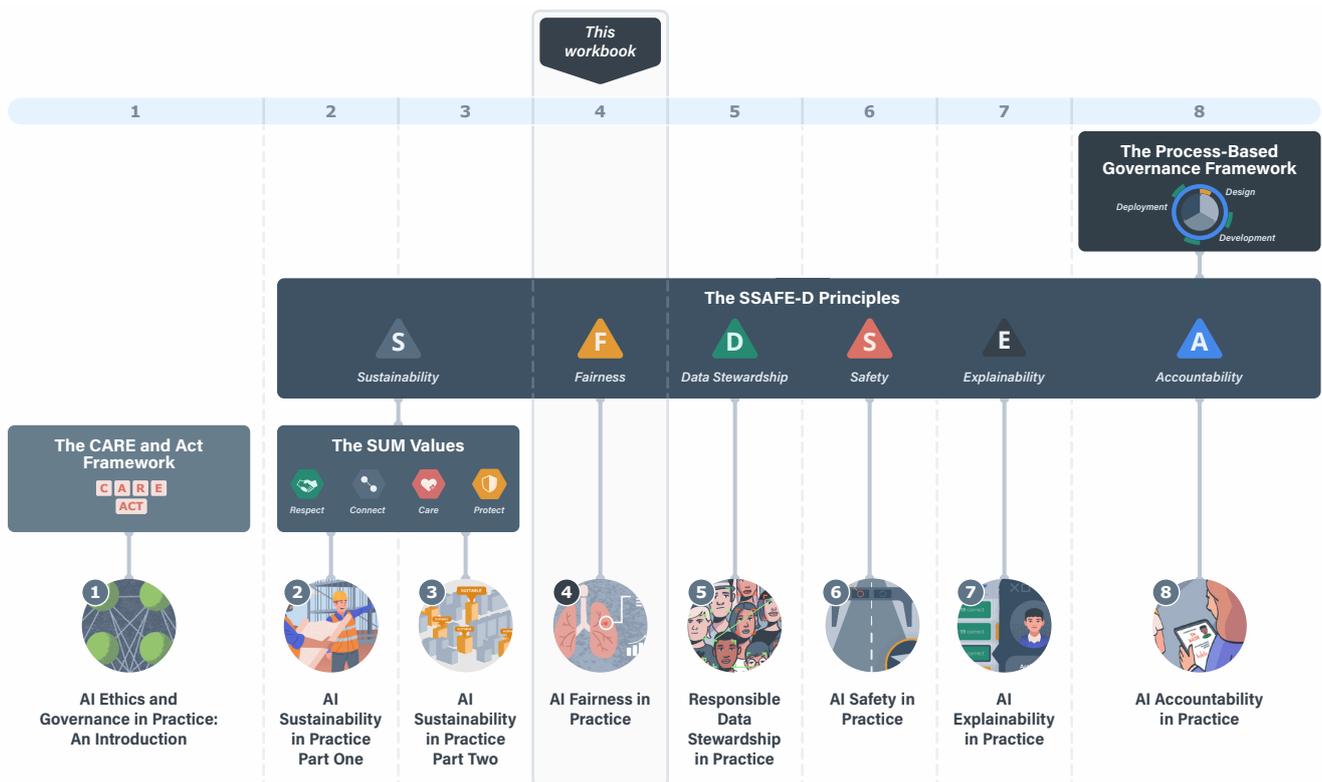

# Intended Audience

The workbooks are primarily aimed at civil servants engaging in the AI Ethics and Governance in Practice Programme as either AI Ethics Champions delivering the curriculum within their organisations by facilitating peer-learning workshops, or participants completing the programme by attending workshops. Anyone interested in learning about AI ethics, however, can make use of the programme curriculum, the workbooks, and resources provided. These have been designed to serve as stand-alone, open access resources. Find out more at turing.ac.uk/ai-ethics-governance.

There are two versions of each workbook:

- **Annotated workbooks** (such as this document) are intended for facilitators. These contain guidance and resources for preparing and facilitating training workshops.

- **Non-annotated workbooks** are intended for workshop participants to engage with in preparation for, and during, workshops.

# Introduction to This Workbook

The purpose of this workbook is to introduce participants to the principle of Fairness. Reaching consensus on a commonly accepted definition for fairness has long been a central challenge in AI ethics and governance. There is a broad spectrum of views across society on what the concept of fairness means and how it should best be put to practice. In this workbook, we tackle this challenge by exploring how a context-based and society-centred approach to understanding AI Fairness can help project teams better identify, mitigate, and manage the many ways that unfair bias and discrimination can crop up across the AI project workflow. This workbook is divided into two sections, Key Concepts and Activities:

**Key Concepts Section**

This section provides content for workshop participants and facilitators to engage with prior to attending each workshop. It covers the following topics and provides case studies aimed to support a practical understanding of fairness and bias within AI systems:

**Part One: Introduction to Fairness**

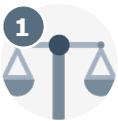 Introduction to Fairness

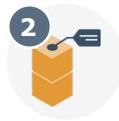 Data Fairness

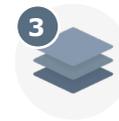 Application Fairness

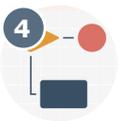 Model Design and Development Fairness

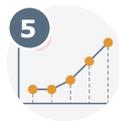 Metric-Based Fairness

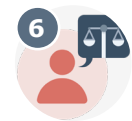 System Implementation Fairness

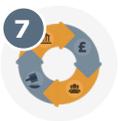 Ecosystem Fairness

**Part Two: Putting Fairness into Practice**

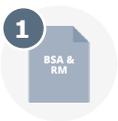 Bias Self-Assessment and Bias Risk Management

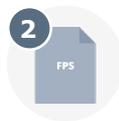 Fairness Position Statement

## Activities Section

This section contains instructions for group-based activities (each corresponding to a section in the Key Concepts). These activities are intended to increase understanding of Key Concepts by using them.

*Case studies within the AI Ethics and Governance in Practice workbook series are grounded in public sector use cases, but do not reference specific AI projects.*

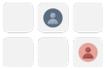
### Considering Application and Data Bias

Practise analysing real-world patterns of bias and discrimination that may produce data biases.

### Design Bias Reports

Practise assessing how bias may play out throughout the different stages of designing AI systems.

### Defining Metric-Based Fairness

Practise selecting definitions of fairness that fit specific use cases for which outcomes are being considered.

### Redressing System Implementation Bias

Practise identifying and redressing different forms of Implementation Bias.

> **Note for Facilitators**
>
> Additionally, you will find facilitator instructions (and where appropriate, considerations) required for facilitating activities and delivering capacity-building workshops.

AI Fairness in Practice

# Key Concepts

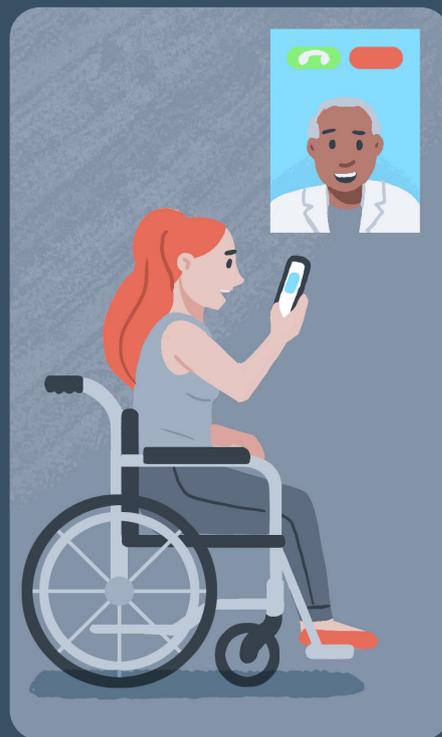

# Part One: Introduction to Fairness

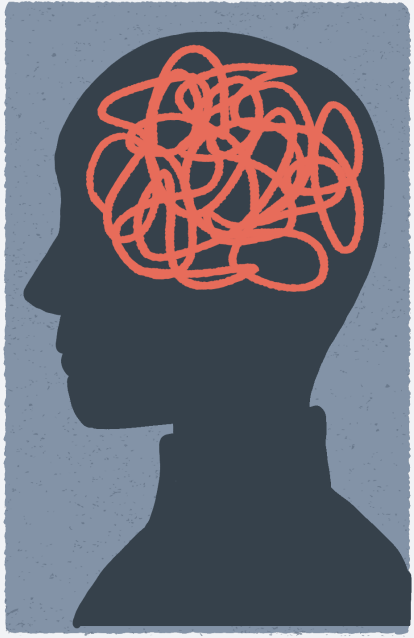 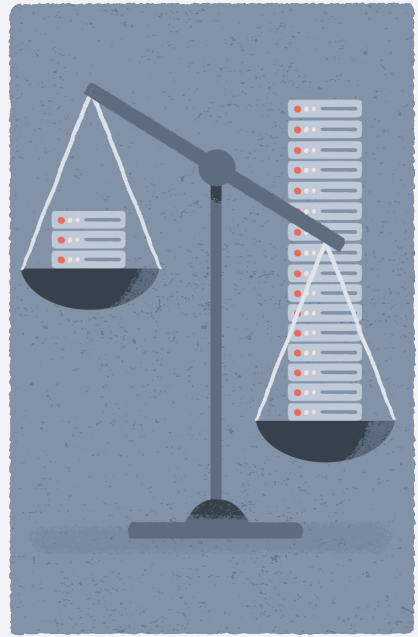

In this Key Concepts section, we will explore how concepts of fairness are used in the field of AI ethics and governance. This is a preliminary step towards understanding the ways in which existing biases manifest in the design, development, and deployment of AI systems. Relevant notions of fairness operate as ethical and legal criteria based upon which instances of unfair bias and discrimination can be identified across the AI project workflow. They also operate as yardsticks against which biases can be measured and then mitigated.

When thinking about AI Fairness, it is important to keep in mind that these technologies, no matter how neutral they may seem, are designed and produced by human beings. Human beings are bound by the limitations of their own given contexts. They are also bound by biases that can arise both in their cognitive processes and in the social environments that influence their actions and interactions. Pre-existing or historical discrimination and social injustice— as well as the prejudices and biased attitudes that are shaped by them— can be drawn into the AI innovation lifecycle and create unfair biases at any point in the project workflow.[1] This is the case from the earliest stages of Agenda Setting, Problem Selection, Project Planning, Problem Formulation, and Data Extraction to later phases of Model Development and System Deployment. Additionally, the datasets used to train, test, and validate AI/ML models can encode social and historical forms of inequity and discrimination. As a result, these datasets can embed biases in an algorithmic model's variables, inferences, and architecture.[2]

This wide range of entry points for bias and discrimination has complicated the notion of fairness, since fairness-centred approaches to AI innovaton such as 'discrimination-aware data mining' and 'fair machine learning' were developed more than a decade ago.[3] [4] [5] [6] [7] [8] The ability to reach consensus on a commonly accepted definition for AI/ML fairness and on how to put such a definition into practice has been hampered by the countless technical and sociotechnical contexts in which fairness issues arise.[9] Consensus has also been challenged by the broad spectrum of views in society on what the concept of fairness means and how it should best be put to practice.[10] [11] [12] For this reason, in this practical guidance, we take a context-based and society-centred approach to understanding AI/ML fairness, **anchored in two pillars**.

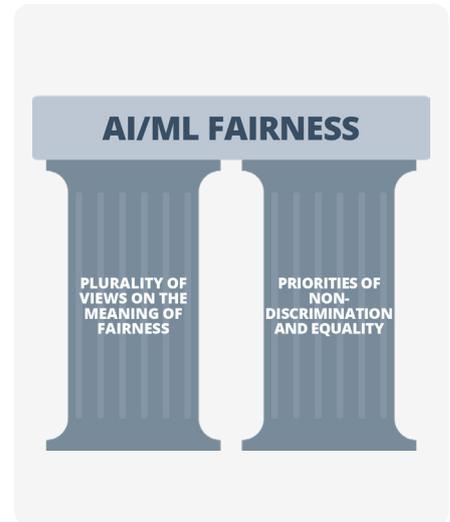

### 1. Plurality of Views on the Meaning of Fairness

To understand how concepts of fairness are defined and applied in AI innovation contexts, we must begin by acknowledging that there are myriad interpretations of the meaning of fairness within and across cultures, societies, and legal systems.[13] [14]

Therefore, fairness must be approached in a way that responds to:

  a. the many ways in which it can be interpreted; and

  b. the many contexts in which it can be applied.[15] [16] The same applies for adjacent notions like equity, impartiality, justice, equality, and non-discrimination.

**Navigating the Many Meanings of Fairness**

The meaning of the term 'fairness', in its contemporary English language usage, has dozens of interpretations. They include a range of related but distinctive ideas such as equity, consistency, non-discrimination, impartiality, justice, equality, honesty, and reasonableness.[17] Likewise, the translation of the word 'fairness' into other languages has proven to be notoriously difficult. Some researchers claim that it cannot be consistently understood across different linguistic groups.[18] [19]

## 2. Priorities of Non-Discrimination and Equality

Despite this diversity in the understanding and application of the concept of fairness, there has been considerable agreement around how the priorities of non-discrimination and equality constitute the core of the concept of fairness. While some claim that fairness is ultimately a subjective value that varies according to individual preferences and cultural outlooks,[20] general ethical and legal concepts of fairness are predicated on core beliefs in:

a. the equal moral status of all human beings; and

b. the right of all human beings to equal respect, concern, protection, and regard before the law.

On this view, it is because each person possesses an intrinsic and equal moral worth that everyone deserves equal respect and concern.[21] Respect and concern are grounded in the common dignity and humanity of every person.[22] [23] [24] **Fairness, therefore, has to do with the moral duty to treat others as moral equals and to secure the membership of all in a 'moral community' where every person can regard themself as having equal value.**[25] **Wrongful discrimination occurs when decisions, actions, institutional dynamics, or social structures do not respect the equal moral standing of individual persons.**[26] [27] [28]

The widespread agreement that equality and non-discrimination are central to fairness concerns has led to their acceptance as normative anchors of international human rights law and anti-discrimination and equality statutes.[29] [30]

In human rights law, interlocking principles of equality and non-discrimination are taken to be essential preconditions for the realisation of all human rights. They are implied in the guarantee of the equal enjoyment and protection of fundamental rights and freedoms to all human beings.[31] As such, principles of equality and non-discrimination are treated as peremptory or foundational norms that permeate all human rights provisions and from which no derogation is permitted.[32] [33] [34]

In anti-discrimination and equality statutes in the UK and beyond, equality and non-discrimination form principal aims and essential underpinnings of fairness concerns. This is manifested in the form of equal protection from discriminatory harassment, and from direct and indirect discrimination.

Let's examine each of these discriminatory harms in detail:

## 1. Discriminatory Harassment

Discriminatory harassment can be defined as unwanted or abusive behaviour linked to a protected characteristic that violates someone's dignity, degrades their identity, or creates an offensive environment for them.

**Example of Discriminatory Harassment**
An employer makes a racist remark about a protected group in the presence of an employee from that racial background. This would be considered harassment based on the protected characteristic of race.

**What are the Protected Characteristics?**

In the 2010 UK Equality Act, protected classes include age, gender reassignment, being married or in a civil partnership, being pregnant or on maternity leave, disability, race including colour, nationality, ethnic or national origin, religion or belief, sex, and sexual orientation.[35] The European Convention on Human Rights, which forms the basis of the UK's 1998 Human Rights Act, includes as protected characteristics 'sex, race, colour, language, religion, political or other opinion, national or social origin, association with a national minority, property, birth or other status.'[36]

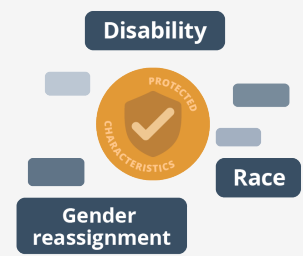

## 2. Direct Discrimination

In direct discrimination, individuals are treated adversely based on their membership in some protected class. This type of discrimination is also known as 'disparate treatment'. It involves instances where otherwise similarly positioned individuals receive different and more-or-less favourable treatment on the basis of differences between their respective protected characteristics.[37]

**Example of Direct Discrimination**
An otherwise well-qualified job applicant is intentionally denied an opportunity for employment because of their age, disability, or sexual orientation.[38]

## 3. Indirect Discrimination

In indirect discrimination, existing provisions, criteria, policies, arrangements, or practices—which could appear on their face to be neutral—disparately harm or unfairly disadvantage members of some protected class in comparison with others who are not members of that group. This type of discrimination is also known as 'disparate impact'. What matters here is not directly unfavourable treatment in individual cases. Rather, what matters is the broader disproportionate adverse effects of provisions, criteria, policies, arrangements, or practices that may subtly or implicitly disfavour members of some protected group while appearing to treat everyone equally.[39] Indirect discrimination can involve the impacts of tacitly unjust or unfair social structures, underlying inequalities, or systemic patterns of implicit Historical Bias that manifest unintentionally through prevailing norms, rules, policies, and behaviours.

**Example of Indirect Discrimination**
A job advertisement specifies that applicants need to be native English speakers. This would automatically disadvantage candidates of different nationalities regardless of their levels of fluency or language training.

These three facets of anti-discrimination and equality law (harassment, direct discrimination, and indirect discrimination) have significantly shaped contemporary approaches to AI Fairness.[40] [41] [42] [43] Indeed, attempts to put the principle of AI Fairness into practice have largely converged around the priority to do no discriminatory harm to affected people along each of these three vectors of potential injury.[44] It is thus helpful to think about basic AI Fairness considerations as tracking three corresponding questions (an example of an associated discriminatory harm is provided alongside each):

## Question 1

**How could the use of the AI system we are planning to build or acquire—or the policies, decisions, and processes behind its design, development, and deployment—lead to the discriminatory harassment of impacted individuals?**

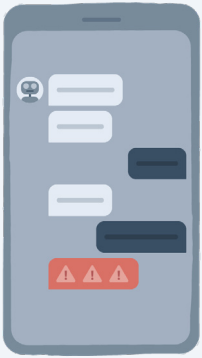

### Example

An AI-enabled customer support chatbot is built from a large language model pre-trained on billions of data points scraped from the internet. The model is then customised to provide tailored responses to customer questions about the provision of a public service. After the system goes live, the project team discovers that when certain customer names, which are indicative of protected classes, are entered, the chatbot generates racist and sexist text responses to customer inquiries.

## Question 2

**How could the use of the AI system we are planning to build or acquire—or the policies, decisions, and processes behind its design, development, and deployment—lead to the disproportionate adverse treatment of impacted individuals from protected groups on the basis of their protected characteristics?**

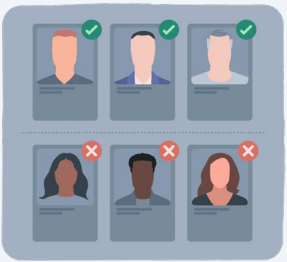

### Example

An AI system used to filter job applications in a recruitment process is trained on historic data that contains details about the characteristics of successful candidates over the past several years. White male applicants were predominantly hired over this time. As a result, the system learns to infer the likelihood of success based on proxy features connected to the protected characteristics of race and sex.[45] It consequently filters out ethnic minority and non-male job candidates from the applicant pool.[46]

## Question 3

**How could the use of the AI system we are planning to build or acquire—or the policies, choices, and processes behind its design, development, and deployment—lead to indirect discrimination against impacted individuals from protected groups?**

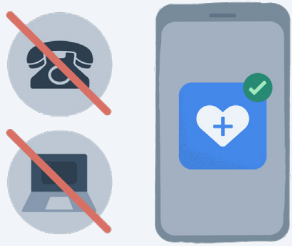

**Example**

An AI medical diagnosis system is built and made available to all individuals registered in a national health system through a smart phone application. The barriers to healthcare access faced by some residents are not sufficiently considered and, after the app is launched, it was discovered that the service disproportionately favours younger, more digitally literate, and more affluent individuals, disadvantaging the elderly, less digitally literate individuals and people who do not have access to smartphone technologies and internet connections.[47] [48]

It is important to note, regarding this final question on indirect discrimination, that the Public Sector Equality Duty mandates considerations both of how to 'reduce the inequalities of outcome which result from socio-economic disadvantage' and of how to advance equality of opportunity and other substantive forms of equality. This means that your approach to putting the principle of AI Fairness into practice must includes **social justice considerations** (discussed on page 20) considerations that concentrate on how the production and use of AI technologies can address and rectify structural inequalities and institutionalised patterns of inequity and discrimination rather than reinforce or exacerbate them.

We must consequently take a multi-pronged approach to AI Fairness that integrates:

1. **formal approaches to non-discrimination and equality** (which focus primarily on consistent and impartial application of rules and equal treatment before the law); and

2. **substantive and transformative approaches** (which focus on equalising the distribution of opportunities and outcomes and on the fundamental importance of addressing the material pre-conditions and structural changes needed for the universal realisation of equitable social arrangements).

Key Concepts   Introduction to Fairness                                                      16

> This section was written by the Equality and Human Rights Commission.

# The Public Sector Equality Duty (PSED) in the Equality Act 2010[49]

Under the PSED, if your organisation is planning to use a specific AI technology to deliver one of its functions or services, it must:

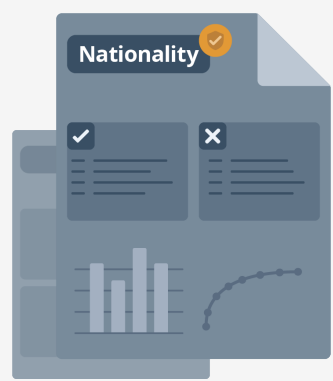

1. **think about its potential impact (negative and positive)** on people with protected characteristics under the Equality Act 2010[50] **before** going ahead with it; and

2. **monitor its actual impact during and after implementation.** The latter is essential to satisfy the **ongoing nature of the PSED.**

These are **legal requirements** that apply to **all public bodies in Great Britain** as well as to **any organisation that has been contracted out by a public body** to develop or use AI technologies to deliver a public function.

### The Importance of Data and Evidence

To assess the potential and actual impact of using a specific AI technology on people with protected characteristics, organisations will need to **gather relevant data and evidence**. For example, from:

- Research and engagement with other organisations that have experience using similar AI technology.

- Relevant case law or concerns raised by regulators or in the media.

- Service users and community groups who may be impacted by the use of AI technology.

- Individual concerns and complaints raised by members of the public.

**Collecting and analysing data is key**. It is the only way for your organisation to understand and monitor whether:

- the **intended benefits** of using AI are being realised;

- any **negative impacts** your organisation anticipated have effectively been mitigated; and

- any **new or unforeseen negative effects on people with protected characteristics** have emerged.

## The Need to Consider Outcomes Carefully

If, after considering data and evidence, your organisation anticipates that using a specific AI technology is likely to lead to unlawful discrimination under the Equality Act 2010, it should not use it. If your organisation assesses that using such AI technology could lead to putting some groups at a particular disadvantage in accessing services and opportunities, it will need to consider how this negative impact could be mitigated, including through adapting the way it uses AI.
If when monitoring the actual impact of using a specific AI technology, your organisation assesses that it has led to unforeseen negative consequences, it will have to consider the best way forward promptly and effectively. This may include stopping using the AI technology, or adapting the way it is used to mitigate negative impact.

You can find more information on how to consider the impact of using AI technologies on people with protected characteristics under the Equality Act 2010 as well as practical examples on the [Equality and Human Rights Commission's website](#).

# Discriminatory Non-Harm

There are different ways to characterise or define fairness in the design and use of AI systems. However, you should consider the principle of discriminatory non-harm as a minimum required threshold of fairness. This principle directs us to do no harm to others through direct or indirect discrimination or through discriminatory harassment linked to a protected characteristic that violates the dignity of impacted individuals, degrades their identity, or creates a humiliating or offensive environment for them.

> **KEY CONCEPT**
>
> **Principle of Discriminatory Non-Harm (Do No Discriminatory Harm)**
>
> The producers and users of AI systems should prioritise the identification and mitigation of biases and discriminatory influences, which could lead to direct or indirect discrimination or discriminatory harassment. This entails an end-to-end focus on how unfair biases and discriminatory influences could arise:
>
> 1. in the processes behind the design, development, and deployment of these systems;
> 2. in the outcomes produced by their implementation; and
> 3. in the wider economic, legal, cultural, and political structures or institutions in which the AI project lifecycle is embedded—and in the policies, norms, and procedures through which these structures and institutions influence actions and decisions throughout the broader AI innovation ecosystem.
>
> Developers and users of AI systems should, in this respect, acknowledge and address discriminatory patterns that may originate in the data used to train, test, and validate the system and in the model architectures that generate system outputs. Model architectures include, for instance, variables, parameters, and inferences.
>
> Beyond this, the principle of discriminatory non-harm implies that AI project teams ensure, more broadly, that their research, innovation, and implementation practices are undertaken in a responsible and ethical manner, keeping in mind the historical tendency that deficiencies in the deployment and operation of faulty systems often disparately impact protected, underrepresented, or disadvantaged groups.
>
> The principle of discriminatory non-harm applies to any AI system that processes social or demographic data (i.e. data pertaining to features of human subjects, population- and group-level traits and characteristics, or patterns of human activity and behaviour). However, the principle applies equally to AI systems that process bio-physical or biomedical data. In this case, imbalanced datasets, selection biases, or measurement errors could have discriminatory effects on impacted individuals and communities—for instance, where a demographic group's lack of representation in a biomedical dataset (e.g. one used to train a diagnostic prediction model) means that the trained system performs poorly for that group relative to others that are better represented in the data.

> **What is Social Justice?**
>
> Social justice concerns inform considerations around discriminatory harms. In particular, considerations around indirect discrimination and how to rectify wrongs arising from it. Social justice is a commitment to the achievement of a society that is:
>
> - equitable;
> - fair; and
> - capable of confronting the root causes of injustice.
>
> In an equitable and fair society, all individuals are recognised as worthy of equal moral standing. They are able to realise the full assemblage of fundamental rights, opportunities, and positions.
>
> In a socially just world, every person has access to the necessary resources or materials to participate fully in work life, social life, and creative life. For instance, through the provision of proper education, adequate living and working conditions, general safety, social security, and other means of realising maximal health and wellbeing.
>
> Social justice also entails the advancement of diversity and participatory parity. It entails a pluralistically informed recognition of identity and cultural difference. Struggles for social justice typically include accounting for historical and structural injustice coupled to demands for reparations and other means of restoring rights, opportunities, and resources to those who have been denied them or otherwise harmed.

Prioritising discriminatory non-harm implies that the producers and users of AI systems ensure that the decisions and behaviours of their models do not:

1. treat impacted individuals adversely based on their membership in some protected class or socioeconomic group; or

2. generate discriminatory or inequitable impacts that unfairly disadvantage members of some protected class or socioeconomic group in comparison with others who are not members of that class or group.

This can be seen as a proportional approach to bias mitigation because it sets a baseline for fair AI systems, while creating conditions for AI teams to strive towards an ideal for fair outcomes for all people as moral equals.

Finally, the scope of the principle means that, beyond project teams, any individuals, organisations, or departments who are procuring AI systems must ensure that the vendors of such systems can demonstrate the mitigation of potential biases and discriminatory influences in the processes behind their production and in their outputs.

**What is the Difference Between Output and Outcome Used in This Workbook?**

Output refers to the immediate or direct results or predictions generated by the AI system based on the input data and the model used.

Outcome refers to the overall result or effect of the AI system's output. In the context of clinical decisions, for instance, the outcome refers to the treatment path that the patient will take, having considered suggestions by the clinician, who may or may have not followed the AI system's output. The outcome can also refer to the result of an intervention or treatment, and its effect on the patient's health and wellbeing.

# AI Fairness as a Contextual and Multivalent Concept

AI Fairness must be treated as a contextual and multivalent concept. It manifests in a variety of ways and in a variety of social, technical, and sociotechnical environments. It must, accordingly, be understood in and differentiated by the specific settings in which it arises. Our operating notion of AI Fairness should distinguish between the kinds of fairness concerns that surface in:

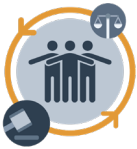

1. **The Context of the Social World that Precedes and Informs Approaches to Fairness in AI/ML Innovation Activities**

    i.e. in general normative and ethical notions of fairness, equity, and social justice, in human rights laws related to equality, non-discrimination, and inclusion, and in anti-discrimination laws and equality statute.

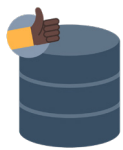

2. **The Dataset**

    i.e. in criteria of fairness and equity that are applied to responsibly collected and maintained datasets.

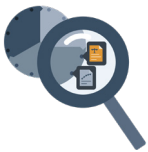

3. **The Design, Development, and Deployment Context**

    i.e. in criteria of fairness and equity that are applied:

    a. in setting the research agendas and policy objectives that guide decisions made about where, when, and how to use AI systems; and

    b. in actual model design and development and system implementation environments, as well as in the technical instrumentalisation of formal fairness metrics that allocate error rates and the distribution of outcomes through the retooling of model architectures and parameters.

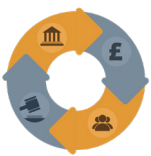

4. **The Ecosystem Context**

    i.e. in criteria of fairness and equity that are applied to the wider economic, legal, cultural, and political structures or institutions in which the AI project lifecycle is embedded—and to the policies, norms, and procedures through which these structures and institution influence actions and decisions throughout the AI innovation ecosystem.

Each of these contexts will generate different sets of fairness concerns. In applying the principle of discriminatory non-harm to the AI project lifecycle we will accordingly break down the principle of fairness into six subcategories that correspond to their relevant practical contexts:

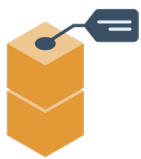

### Data Fairness

The AI system is trained and tested on datasets that are properly representative, fit-for-purpose, relevant, accurately measured, and generalisable.

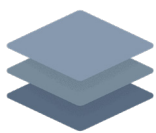

### Application Fairness

The policy objectives and agenda-setting priorities that steer the design, development, and deployment of an AI system (and the decisions made about where, when, and how to use it) do not create or exacerbate inequity, structural discrimination, or systemic injustices. They are also acceptable to and line up with the aims, expectations, and sense of justice possessed by impacted people.

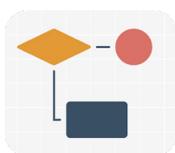

### Model Design and Development Fairness

The AI system has a model architecture that does not include target variables, features, processes, or analytical structures (correlations, interactions, and inferences) which are discriminatory, unreasonable, morally objectionable, or unjustifiable or that encode social and historical patterns of discrimination.

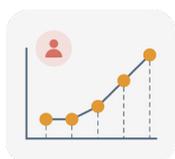

### Metric-Based Fairness

Lawful, clearly defined, and justifiable formal metrics of fairness have been operationalised in the AI system. They have been made transparently accessible to relevant stakeholders and impacted people.

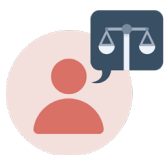

### System Implementation Fairness

The AI system is deployed by users sufficiently trained to implement it. They have an appropriate understanding of its limitations and strengths and deploy it in a bias-aware manner that gives due regard to the unique circumstances of affected individuals.

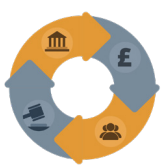

### Ecosystem Fairness

The wider economic, legal, cultural, and political structures or institutions in which the AI project lifecycle is embedded do not steer AI research and innovation agendas in ways that entrench or amplify asymmetrical and discriminatory power dynamics or that generate inequitable outcomes for protected, marginalised, vulnerable, or disadvantaged social groups. Nor do the policies, norms, and procedures through which such structures and institutions influence actions and decisions throughout the AI innovation ecosystem.

# Data Fairness

Responsible data acquisition, handling, and management is a necessary component of fairness. If the results of your AI project are generated by biased, compromised, or skewed datasets, affected stakeholders will not be adequately protected from discriminatory harm. Your project team should keep in mind the following key elements of Data Fairness:

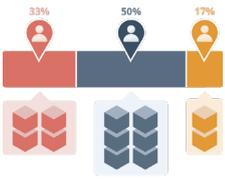

### 1. Representativeness

Depending on the context, either underrepresentation or overrepresentation of disadvantaged or legally protected groups in the data sample may lead to the systematic disadvantaging of vulnerable or marginalised stakeholders in the outcomes of the trained model.

To avoid such kinds of sampling bias, domain expertise will be crucial to assess the fit between the data collected or procured and the underlying population to be modelled. Technical team members should, if possible, offer means of remediation to correct for representational flaws in the sampling.

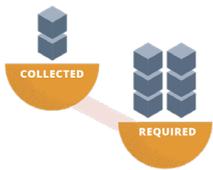

### 2. Fit-For-Purpose and Sufficiency

The quantity of data collected or procured has a significant impact on the accuracy and reasonableness of the outputs of a trained model. A data sample not large enough to represent with sufficient richness the significant or qualifying attributes of the members of a population to be classified may lead to unfair outcomes. Insufficient datasets may not equitably reflect the qualities that should rationally be weighed in producing a justified outcome that is consistent with the desired purpose of the AI system.

An important question to consider in the data collection and procurement process is: Will the amount of data collected be sufficient for the intended purpose of the project? Members of the project team with technical and policy competences should collaborate to determine if the data quantity is, in this respect, sufficient and fit-for-purpose.

> This section was written by the Information Commissioner's Office.
>
> **Fairness in UK Data Protection and AI**
>
> Fairness is a foundational principle of the UK data protection regulation. It aims to protect people's rights and freedoms in relation to the processing of their personal data. The framework includes the UK General Data Protection Regulation ([UK GDPR](#)) and the Data Protection Act 2018 (DPA 2018). If you use personal data during AI development or deployment you need to comply with data protection.
>
> The data protection legislation is overseen by the Information Commissioner's Office ([ICO](#)), which is the UK's independent data protection authority. It has produced [a suite of products on AI](#) to assist developers and users of AI systems.
>
> As part of complying with data protection you will need to comply with its fairness principle. In simple terms, fairness in data protection means that organisations should only process personal data in ways that people would reasonably expect and not use it in any way that could have unjustified adverse effects on them. Organisations should not process personal data in ways that are unduly detrimental, unexpected, or misleading to the individuals concerned.
>
> The ICO has produced foundational [Guidance on AI and Data Protection.](#) Among other things, it explains how to interpret data protection's fairness principle in the context of developing and using artificial intelligence or machine learning.

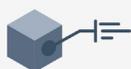
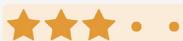

### 3. Source Integrity and Measurement Accuracy

Effective bias mitigation begins at the very start of data extraction and collection processes. Both the sources and instruments of measurement may introduce discriminatory factors into a dataset. When incorporated as inputs in the training data, biased prior human decisions and judgments—such as prejudiced scoring, ranking, interview-data or evaluation—will become the 'ground truth' of the model and replicate the bias in the outputs of the system.

In order to secure discriminatory non-harm, you must do your best to make sure your data sample has optimal source integrity. This involves securing or confirming that the data gathering processes involved suitable, reliable, and impartial sources of measurement and sound methods of collection.

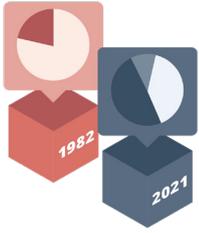

### 4. Timeliness and Recency

If your datasets include outdated data then changes in the underlying data distribution may adversely affect the generalisability of your trained model. Provided these distributional drifts reflect changing social relationship or group dynamics, this loss of accuracy with regard to the actual characteristics of the underlying population may introduce bias into your AI system.

In preventing discriminatory outcomes, you should scrutinise the timeliness and recency of all elements of the data that constitute your datasets.

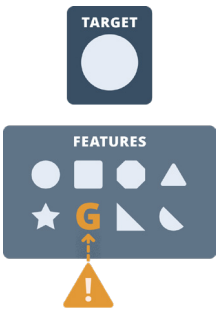

### 5. Relevance, Appropriateness, and Domain Knowledge

Without a solid understanding and utilisation of the most appropriate sources and types of data, AI systems may fail to be robust and bias-mitigating.

Solid domain knowledge of the underlying population distribution and of the predictive or classificatory goal of the project is instrumental for choosing optimally relevant measurement inputs that contribute to the reasonable determination of the defined solution. You should make sure that domain experts collaborate closely with your technical team to assist in the determination of the optimally appropriate categories and sources of measurement.

To ensure the uptake of best practices for responsible data acquisition, handling, and management across your AI project delivery workflow, you should start the creation of a Data Factsheet at the initial stage of your project. This factsheet should be maintained diligently throughout the design and implementation lifecycle. This will help secure optimal data quality, comply with regulatory obligations such as data protection's [accuracy principle](), deliberate bias-mitigation aware practices, and optimal auditability. The Data Factsheet should include:

- A comprehensive record of data provenance, procurement, preprocessing, lineage, storage, and security

- Qualitative input from team members about determinations made with regard to:
    - Data representativeness
    - Data sufficiency
    - Source integrity
    - Data timeliness
    - Data relevance
    - Training/testing/validating splits
    - Unforeseen data issues encountered across the workflow

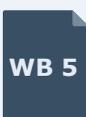 **WB 5** More details about the processes that need to be carried out to create a Data Factsheet are included in **Workbook 5: Responsible Data Stewardship in Practice.**

# Application Fairness

Fairness considerations should enter into your AI project at the earliest point in the design stage of the lifecycle. This is because, the overall fairness of an AI system is significantly determined by the objectives, goals, and policy choices that lie behind initial decisions to dedicate time and resources to its design, development, and deployment.

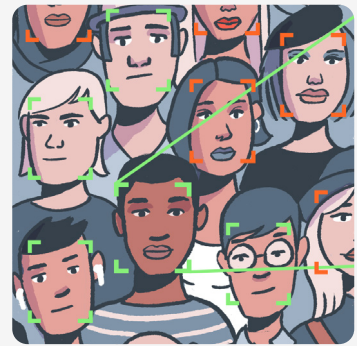

For example, the choice made to build a biometric identification system, which uses live facial recognition technology to identify criminal suspects at public events, may be motivated by the objective to increase public safety and security. However, many members of the public may find this use of AI technology unreasonable, disproportionate, and potentially discriminatory. In particular, members of communities historically targeted by disproportionate levels of surveillance from law enforcement may be especially concerned about the potential for abuse and harm.

Appropriate fairness and equity considerations should, in this case, occur at the horizon scanning and project planning stage (e.g. as part of a stakeholder impact assessment process that includes engagement with potentially affected individuals and communities). Aligning the goals of a project team with the reasonable expectations and potential equity concerns of those affected is a key component of the fairness of the project.

Application Fairness, therefore, entails that the objectives of an AI project are nondiscriminatory and are acceptable to and line up with the aims, expectations, and sense of justice of those affected.[51] [52] [53] [54] [55] [56] As such, whether the decision to develop and use of an AI technology can be described as 'fairness-aware' depends upon ethical considerations that are external and prior to considerations of the technical feasibility of building an accurate system, or the practical feasibility of accessing, collecting, or acquiring enough and the right kind of data.

Beyond this priority of assuring the equity and ethical permissibility of project goals, Application Fairness requires additional considerations in framing decisions made at the horizon scanning and project scoping or planning stage:

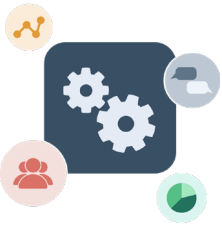

## 1. Equity Considerations Surrounding Real-World Context of the Issue to Be Solved

When your project team is assessing the fairness of using an AI solution to address a particular issue, it is important to consider how equity considerations extend beyond the statistical and sociotechnical contexts of designing, developing, and deploying the system. Applied concepts of AI Fairness should not be treated in a technology-centred way, as originating exclusively from the design and use of any particular AI system. Nor should AI Fairness be treated as an abstraction that can be engineered into an AI system through technical or mathematical retooling (e.g. by operationalising formal fairness criteria).[57] [58] [59] [60] When designers of AI systems limit their understanding of the scope of 'fairness' to these two dimensions, it can constrain their perspectives in such a way that they erroneously treat only the patterns of bias and discrimination that arise in AI innovation practices or those that can be measured, formalised, or statistically understood as indicators of inequity.

Rather, fairness considerations should be grounded in a human-centred approach, which includes critical examination of the wider social and economic patterns of disadvantage, injustice, and discrimination that arise in the real-world contexts surrounding the issue being addressed. Such considerations should include an exploration of how such patterns of inequity may lead to the unequal distribution of the risks and adverse impacts of the AI system, as well as an unequl access to its potential benefits.

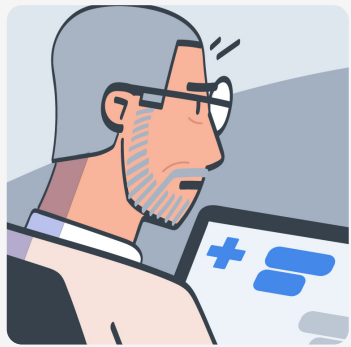

For instance, while the development of an AI chatbot to replace a human-serviced medical helpline may provide effective healthcare guidance for some individuals, it could have disparate adverse impacts on others, Some vulnerable elderly populations or socioeconomically deprived groups may face barriers to accessing and using the app. Here, consideration of the real-world contexts surrounding the issue being addressed allows for an awareness of the social and economic conditions that could impair the fairness of the application.

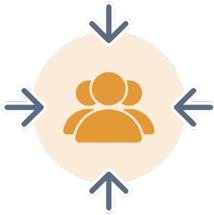

## 2. Equity Considerations Surrounding the Group Targeted by the AI Innovation Intervention

AI applications that make predictions about or classify people target specific groups within the wider population. For instance, a résumé filtering system used to select desirable candidates in a recruitment process will draw from a pool of job applicants. The job applicants constitute a subgroup within the broader population. Potential equity issues may arise here. This is because the selection of subpopulations sampled by AI applications is non-random. Instead, the sample selection may reflect particular social patterns, structures, and path dependencies that are unfair or discriminatory.[61] The sample used for the résumé filtering system may reflect long-term hiring patterns, where a disproportionate number of male job candidates from privileged educational backgrounds have been actively recruited. Such practices have historically excluded people from other gender identities, and socioeconomic and educational backgrounds. The pattern of inequity surfaces, in this instance, not within the sampled subset of the population. Rather, it surfaces in the way that discriminatory social structures have led to the selection of a certain group of individuals into that subset of job applicants.[62][63]

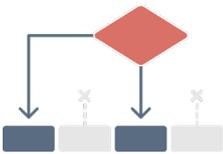

## 3. Equity Considerations Surrounding the Way that the Model's Output Shapes the Range of Possible Decision-Outcomes

AI applications that assist human decision-making shape and delimit the range of possible outcomes for the problem areas they address.[64] For instance, a predictive risk model used in children's social care may generate an output that directly affects the choices available to a social worker. The model's target is the identification of at-risk children. This may lead to social care decisions that focus narrowly on whether to take a child into care. The focus on negative outcomes could restrict the range of viable choices open to the social worker. It may de-emphasise the potential availability of other strengths-based approaches. For example, stewarding positive family functioning through social supports and identifying and promoting protective factors. Restricting the range of possible actions may potentially close off alternative decision-making paths in the social care environment.

# Model Design and Development Fairness

Since human beings have a hand in all stages of the construction of AI systems, fairness-aware design must take precautions across the AI project workflow to prevent bias from having a discriminatory influence.

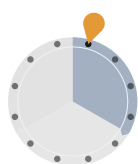

**Design Phase**

## Problem Formulation

At the initial stage of Problem Formulation and outcome definition, technical and non-technical members of your team should work together to translate project goals into measurable targets. This will involve the use of both domain knowledge and technical understanding to define what is being optimised in a formalisable way and to translate the project's objective into a target variable or measurable proxy, which operates as a statistically actionable rendering of the defined outcome.

At each of these points, your team will make choices about the design of the algorithmic system that may introduce structural biases which ultimately lead to discriminatory harm. Your team must take special care here to identify affected stakeholders and to consider how vulnerable groups might be negatively impacted by the specification of outcome variables and proxies. You must also pay attention to the question of whether these specifications are reasonable and justifiable given the general purpose of the project. Lastly, your team must consider the potential impacts that the outcomes of the system's use will have on the individuals and communities involved.

These challenges of fairness-aware design at the Problem Formulation stage show the need for making diversity and inclusive participation a priority from the start of the AI project lifecycle. This involves both the collaboration of the entire team and the attainment of stakeholder input about the acceptability of the project plan. This also entails collaborative deliberation across the project team and beyond about the ethical impacts of the design choices made.

| Case Study | Problem Formulation |

Hospital administrators in a country with a privatised medical system decided to develop an AI model to refer patients to specialised healthcare programmes such as intensive one-on-one services, home visits, and prioritised appointments. The model would determine the scale of healthcare intervention for each patient (whether the patient was granted specialised care or not) by scoring their health need. The model's target variable was decided to be health need, and it was determined that the measurable proxy for this would be the patient's history of healthcare costs.

However, when this model came into use, its discriminatory effects soon became apparent. By using previous medical costs as a proxy for health need, the trained model generated lower health need scores for patients from racial minority groups than for equally unhealthy patients in the racial majority. A couple of factors indicating health inequity were behind this.

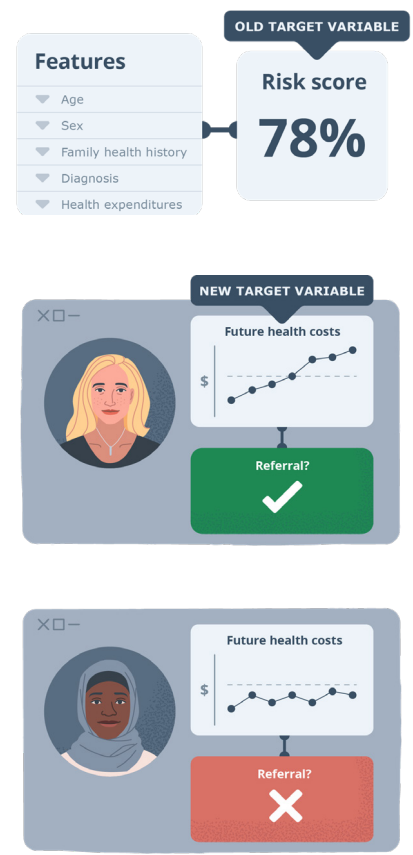

First, because of associations between socioeconomic status and race, minority ethnic groups have traditionally faced barriers when accessing healthcare, such as an inability to take time off work or to find viable transportation to medical facilities. Second, having endured historical maltreatment in the provision of medical services, some communities have reduced trust in the healthcare system, which directly affects their level of engagement. This means that affected patients (even the insured) are less likely to run up the same level of healthcare costs as majority group patients of greater advantage.

This is a good example of a lack of fairness-aware model design at the Problem Formulation stage. The model designers identified healthcare cost as a proxy for health need without considering the influences of legacies of inequality and discrimination on that measurement. As a consequence, the users of the model systemically discriminated against members of a disadvantaged group, who were denied healthcare interventions even when they were in poorer health than other beneficiaries in the majority, who received care.

*This case study was based on the work of Obermeyer, Z., Powers, B., Vogeli, C., & Mullainathan, S. (2019). Dissecting racial bias in an algorithm used to manage the health of populations.*[65]

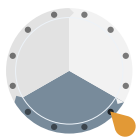

**Development Phase**

# Preprocessing & Feature Engineering

Human judgement enters into the process of algorithmic system construction at the stage of labelling, annotating, and organising the training data to be utilised in building the model. Choices made about how to classify, categorise, and structure raw inputs must be taken in a fairness-aware manner. The project team needs to give due consideration to the social contexts that may introduce bias into such acts of catergorisation and classification. It should put in place similar fairness-aware processes to review automated or outsourced classifications and categorisations. Likewise, efforts should be made to attach solid contextual information and ample metadata to the datasets. This allows downstream analyses of data processing to have access to properties of concern in bias mitigation.

**Article 22 of the UK GDPR**[73]

This article gives data subjects the right not to be subject to a decision that is totally automated that have a legal or similarly significant effect on them. Fairness-aware processes should be put in place to review automated or outsourced classifications. In particular, review in the context of those that fall under the purview of Article 22 of the UK GDPR.

At this stage, your team will select the attributes or features that will serve as input variables for your model. This involves deciding what information may or may not be relevant or rationally required to make an accurate and unbiased classification or prediction. Moreover, the tasks of aggregating, extracting, or decomposing attributes from datasets may introduce human appraisals that have biasing effects.

Human decisions about how to group or disaggregate input features (e.g. how to carve up categories of gender or ethnic groups) or about which input features to exclude altogether (e.g. leaving out deprivation indicators in a predictive model for clinical diagnostics) can have significant downstream influences on the fairness and equity of an AI system. This applies even when algorithmic techniques are employed to extract and engineer features, or support the selection of features (e.g. to optimise predictive power).

For this reason, discrimination awareness should play a large role at this stage of the AI model-building workflow as should domain knowledge and policy expertise. Your team should proceed in the model development stage aware that choices made about grouping or separating and including or excluding features as well as more general judgements about the comprehensiveness or coarseness of the total set of features may have significant consequences for historically marginalised, vulnerable, or protected groups.

| Case Study | Preprocessing & Feature Engineering |

Human Papillomavirus causes 99% of cervical cancers.[66] Cervical screenings are critical for identifying and treating abnormalities related to this virus before they develop. An AI model has been developed to support a public health programme aiming to maximise coverage and uptake of screenings among the eligible populations. This system facilitates patients' access to preventative care and early intervention.

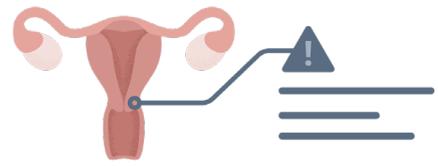

The database used to build the model was organised into classifications including: age, sex, ethnicity, sexual orientation, address, health history, conditions, medicines, and lifestyle information. During the Data Preprocessing & Feature Engineering stage, data was formatted, and the features used for the model to identify eligible patients were selected. Because any resident registered to a General Practice (GP) who has a cervix is eligible to receive cervical screenings,[67] [68] the features of age, sex, address, and GP registration were selected. The model identifies individuals whose medical records are labelled as 'female' and indicate an age range between 25 and 64.

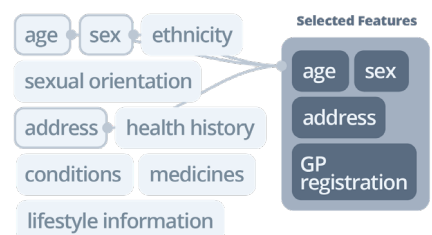

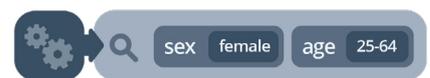

However, the medical database used did not include categories such as 'sex recorded at birth', 'legal sex', or 'gender'. This would have enabled the identification of individuals who were eligible for screening but had changed the sex classification in their medical record from female to male. By not accounting for these categories during the Data Preprocessing & Feature Engineering stage (i.e. by not deriving these categories from patient data) this model could not identify eligible transgender patients. This meant that an adverse effect of using the system was the creation of a dynamic of unequal access to preventative care.[69] [70] To receive this sort of care, eligible transgender patients would have to individually request screenings from their GPs, presenting significant barriers to access when considering that cervical screenings are often associated with female patients.[71] Bias within the Data Preprocessing & Feature Engineering stage of developing this cervical screening uptake model resulted in a replication and amplification of discriminatory healthcare processes that contribute to disparate and potentially disastrous health outcomes for transgender populations.[72]

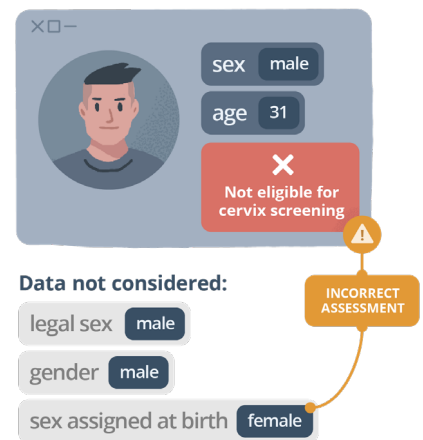

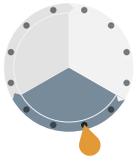

**Development Phase**

# Model Selection & Training

The Model Selection & Training stage determines the model type and structure. It also applies a training dataset to the chosen model. In some projects, this will involve the selection of multiple prospective models. The purpose is to compare the models based on some performance metric, such as accuracy or sensitivity. The set of relevant models is likely to have been highly constrained by many of the issues dealt with in previous stages (e.g. available resources and skills, Problem Formulation). For instance, the problem may demand a supervised learning algorithm instead of an unsupervised learning algorithm.

Fairness and equity issues can surface in Model Selection & Training processes in at least two ways. First, they can arise when the choice between algorithms has implications for explainability. For instance, it may be the case that there are better performing models in the pool of available options but which are less interpretable than others. This difference becomes significant when the processing of social or demographic data increases the risk that biases or discriminatory proxies lurk in the algorithmic architecture. Model interpretability can increase the likelihood of detecting and redressing such discriminatory elements.

Second, fairness and equity issues can arise when the choice between algorithms has implications for the differential performance of the final model for subgroups of the population. For instance, where several different learning algorithms are simultaneously trained and tested, one of the resulting models could have the highest overall level of accuracy while, at the same time, being less accurate than others in the way it performs for one or more marginalised subgroups. In cases like this, technical members of your project team should proceed with attentiveness to mitigating any possible discriminatory effects of choosing one model over another and should consult members of the wider team—and impacted stakeholders, where possible—about the acceptability of any trade-offs between overall model accuracy and differential performance.

The process of tuning hyperparameters, setting metrics, and resampling data during the training phase also involves human choices that may have fairness and equity consequences in the trained model. For instance, the way your project team determines the training-testing split of the dataset can have a considerable impact on the need for external validation to ensure that the model's performance "in the wild" meets reasonable expectations. Therefore, your technical team should proceed with an attentiveness to bias risks, and continual iterations of peer review and project team consultation should be encouraged to ensure that choices made in adjusting the dials, parameters, and metrics of the model are in line with bias mitigation and discriminatory non-harm goals.

**Case Study** — Model Selection & Training

An AI model was developed by a private medical technology company for use in radiology and medical image scanning. The model is based on a deep neural network which carries out complex rounds of feature representation on the input data (i.e. medical scans) to predict the medical diagnosis.

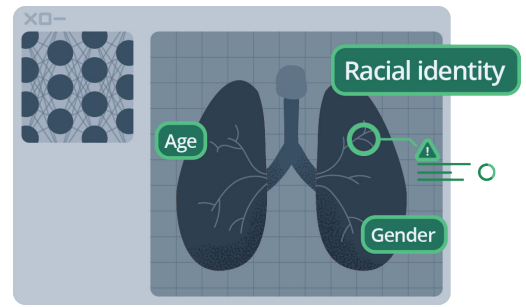

To reduce the burden on radiologists and improve efficiency, a hospital network deployed this AI model for use in several clinics and diagnostic centres across the country. It was soon found that the model was able to accurately determine racial identity, alongside age and gender, from the medical images alone, a task that expert radiologists at the hospitals have been unable to do.[74] [75]

This raised immediate questions about why the system was picking up racial identity and whether this inference was occurring for reason of differences in the quality of imaging technologies or techniques that could be indicative of broader biases and discriminatory influences. For instance, differences in the quality of these imaging technologies and techniques could track differences in the resourcing of care which are correlated with socioeconomic status and race.[76] [77] This would create a situation where the system would predict racial characteristics based on a lurking discriminatory variable.

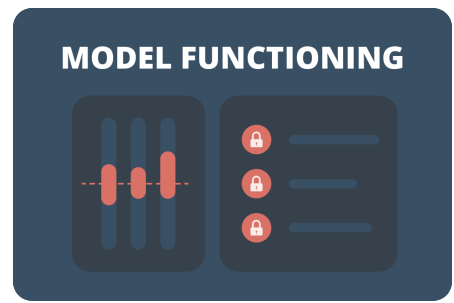

**MODEL FUNCTIONING**

However, the introduction of the model raised additional concerns over its 'black-box' nature as the opaque model had limited transparency and interpretability.[78] Clinicians faced challenges of both the intrinsic opacity (i.e. understanding the inner functioning of the model) and the extrinsic opacity (i.e. when their access to the model functioning was subject to proprietary restrictions by the company that developed the model). This, in turn, deteriorated clinician-patient relationships which are often built on trust, reassurance, and dialogue.[79] [80] The correlations between race and medical imaging required further justification at medical board hearings. Hospital staff were unable to ascertain the cause of the associations, if the relations may be spurious, or if other biases may have been baked into other stages of the model lifecycle.[81] Later, regulatory governance was called to action because of downstream risks. For instance, if the model picked up on racial features more than features of the disease itself, the model could learn to output harmful associations based on race than veritable disease vectors.

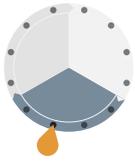

**Development Phase**

# Model Testing & Validation

The process of tuning hyperparameters, setting metrics, and resampling data at the Model Testing and Validation phase involves human choices that may have fairness and equity consequences in the trained model.

## Evaluating and Validating Model Structures

Model Design and Development Fairness requires checking the trained model for lurking or hidden proxies for discriminatory features that may act as significant factors in its output. If the model includes such proxies, it may lead to unfair treatment of a sensitive group. This may lead to implicit 'redlining'.[82] [83]

Designers must additionally scrutinise the moral justifiability of the significant correlations and inferences that are determined by the model's learning mechanisms themselves. When processing social or demographic data related to human features, it's important to use machine learning models that can be assessed to ensure they don't result in discriminatory outcomes. If a model is too complex to be evaluated in this way, it should be avoided. In cases where this is not possible, a different, more transparent and explainable model or portfolio of models should be chosen.

Model structures must also be confirmed to be procedurally fair, in a strict technical sense. This means that any rule or procedure employed in the processing of data by an algorithmic system should be consistently and uniformly applied to every decision subject whose information is being processed by that system. Your team should be able to certify that any relevant rule or procedure is applied universally and uniformly to all relevant individuals.

Implementers of the system should be able to replicate any output the system generates by following the same rules and procedures used for the same inputs. This secures the equal procedural treatment of decision subjects. It also ensures that no changes are made to the system that could disadvantage a specific person. It's important to note that this applies to deterministic algorithms with fixed parameters, but attention should be paid to procedural fairness issues with dynamic learning algorithms, which can evolve and generate different outputs over time for the same inputs. Where the system is dynamic and continuously "learning", fairness awareness should involve considerations of the equity consequences of applying the different rules of the evolving model to the same or similar people over time.

**Case Study** — Model Testing & Validation

Long wait times for emergency treatment in hospitals pose risks to patients requiring urgent care. At the same time, discharging existing patients to make their beds available also poses risks if patients are released too soon. This leaves doctors with a high-stakes dilemma where they must balance the needs of both new and existing patients while optimising the hospital's available resources.

An AI application was developed to support doctors in deciding when patients are safe to be discharged. The model predicts patients' risk of hospital emergency re-admissions within 30 days of discharge by considering patient features such as vital signs (blood pressure, heart rate, temperature, oxygen levels), and data from their Electronic Health Records (EHRs) (including current illness, test results, medication history, other diagnoses, and previous hospital admissions). The model presents doctors with a patient score from 0 to 20, with 20 representing the highest level of risk and 0 indicating that the patient is likely to be safe and well after discharge. Doctors can use this information to identify which patients are safe to be sent home.

However, this predictive risk model raises fairness and equity concerns regarding the treatment of socioeconomically deprived subpopulations.[84] Due to inequalities in primary healthcare provision (e.g. lack of cultural competency, stigmatisation, receiving less appropriate care such as lower rates of medication and diagnostic testing) and barriers to access (e.g. inability to take time off from work or find childcare), socioeconomically disadvantaged patients are less likely to have comprehensive or lengthy medical records.[85]

By using various features in EHRs that are indicative of the provision of and access to primary healthcare as a significant factor in the prediction of patient risk without correcting for health inequalities, the AI application had incorporated a discriminatory analytical structure. The project team did not evaluate the differential performance of the model between subgroups and failed to identify disparate impact before deployment. The absence of a substantial medical history and a record of healthcare maintenance and intervention among members of socioeconomically disadvantaged groups operated as a lurking discriminatory proxy for their deprived status, and the presence of the same for materially advantaged patients meant that they would be scored as having a greater health risk even if they were potentially healthier and more medically resilient.[86] This led to the former having a greater chance of being released too soon and being vulnerable to health complications, and the latter being prioritised for continued treatment.

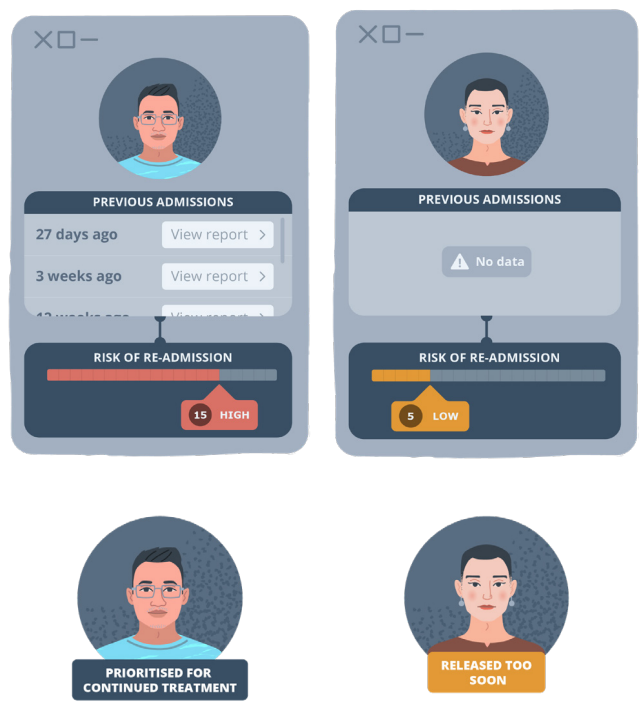

# Metric-Based Fairness

As part of ensuring diligent fairness considerations, well-informed thought must be put into how project teams define and measure the formal metrics of fairness that can be operationalised into the AI systems.

Metric-Based Fairness involves the mathematical mechanisms that can be incorporated into an AI system to allocate the distribution of outcomes and error rates for members of relevant groups (e.g. groups with protected or sensitive characteristics). The distribution of error rates refers to the differential performance of an AI system (e.g. it accuracy, precision, or sensitivity) for different demographic groups. The distribution of outputs refers to the way favourable or unfavourable predictions or classifications are spread across different demographic groups (for example, the distribution of job candidates who are predicted to be viable for an available position and thus filtered into a short list for interviews).

When establishing an approach to Metric-Based Fairness, your project team will be confronting challenging issues. For instance, they may have to justify why the distribution of model outputs or the distribution of error rates and performance indicators (like precision or sensitivity) is different across different groups.

There is a great diversity of beliefs as to what makes the outputs of AI systems allocatively fair. Broadly speaking, fairness metrics fall within two approaches: group and individual fairness.

---

**KEY CONCEPT**

**Group Fairness**

This approach focuses on ensuring fairness across relevant groups. It aims to define a metric that is used to compare model performance across relevant groups.

---

**KEY CONCEPT**

**Individual Fairness**

This approach focuses on ensuring fairness at the level of the individual. It aims to ensure that individuals with similar relevant qualifications receive similar system outcomes.

---

Within these broad approaches, you will find different definitions of Metric-Based Fairness. You will find below a summary table of several of the main definitions that have been integrated by researchers into formal models. Each fairness metric prioritises different principles.

# Selected fairness metrics

**Group Fairness**

### Demographic/Statistical Parity[87]

An outcome is fair if each group in the selected set receives benefit in equal or similar proportions. This means that there is no correlation between a sensitive or protected attribute and the result of the allocation. This approach is intended to prevent disparate impact, where an algorithmic process disproportionately harms people who are part of disadvantaged or protected groups.[88]

**Group Fairness**

### Equalised Odds

An outcome is fair if false positive and true positive rates are equal across groups. In other words, both the probability of incorrect positive predictions and the probability of correct positive predictions should be the same across protected and privileged groups. This approach is motivated by the position that sensitive groups and advantaged groups should have similar error rates in outcomes of algorithmic decisions.

**Group Fairness**

### True Positive Rate Parity

An outcome is fair if the 'true positive' rates of an algorithmic prediction or classification are equal across groups. A 'true positive' refers to a result that correctly identifies the presence of a condition or attribute that the system is designed to detect. The true positive rate parity approach is intended to align the goals of bias mitigation and accuracy by ensuring that the accuracy of the model is equivalent between relevant population subgroups. This method is also referred to as 'equal opportunity' fairness. It aims to secure equalised odds of an advantageous outcome for qualified individuals in a given population regardless of the protected or disadvantaged groups of which they are members.

Group Fairness

**Positive Predictive Value Parity**

An outcome is fair if the rates of positive predictive value (the fraction of correctly predicted positive cases out of all predicted positive cases) are equal across sensitive and advantaged groups. Outcome fairness is defined here in terms of a parity of precision, where the probability of members from different groups actually having the quality they are predicted to have is the same across groups.

Individual Fairness

**Individual Fairness**

An outcome is fair if it treats individuals with similar relevant qualifications similarly. This approach relies on the establishment of a similarity metric that shows the degree to which pairs of individuals are alike with regard to a specific task.

Individual Fairness

**Counterfactual Fairness**

An outcome is fair if an automated decision made about an individual belonging to a sensitive group would have been the same were that individual was a member of a different group in a closest possible alternative (or counterfactual) world.[89] Like the individual fairness approach, this method of defining fairness focuses on the specific circumstances of an affected decision subject, but, by using the tools of contrastive explanation, it moves beyond individual fairness insofar as it brings out the causal influences behind the algorithmic output. It also presents the possibility of offering the subject of an automated decision knowledge of what factors, if changed, could have influenced a different outcome. This could provide them with actionable recourse to change an unfavourable decision.

**Example: Diagnosis System**

An AI system is being developed to help clinicians analyse electrocardiogram signals. When deployed, this system will assist doctors in diagnosing cardiac conditions, including heart failure. High blood pressure, diabetes, obesity, and smoking can contribute to the development of heart failure and other cardiac conditions. These risk factors are linked to other socioeconomic determinants of health such as income, occupation, housing conditions, and levels of social support. Because of historical and structural disparities, some demographic groups are more likely to experience multiple contributing factors to heart failure.

This combination of the complex range of risk factors and the linkage of certain of these factors to disadvantaged demographic groups increases the chances that the diagnostic AI system could perform better for some groups than for others. This is because the data used to train the system might not capture some of the risk factors that are more commonly connected with disadvantaged or marginalised groups. Conversely, it may overrepresent risk factors that are more prevalent in advantaged and thus better measured groups. The project team needs to define their approach to Metric-Based Fairness.

A group fairness approach will ensure that the predictions of the AI system are consistent across demographic groups. It can be evaluated using different fairness metrics. For instance:

- **True Positive Rate Parity**

  The AI system may deliver the same rate of correct diagnoses across demographic groups. This means that the accuracy of the model for positive predictions would be equivalent across disadvantaged and advantaged groups.

- **Equalised Odds**

  The AI system may deliver the same rates of accurate diagnoses and inaccurate diagnoses across demographic groups. This means that disadvantaged and advantaged demographic groups would have the same probability of receiving true and false positive predictions.

- **Counterfactual Fairness**

  The AI system would perform the same for individuals who share all the same characteristics except for membership in a sensitive or protected group. This means that the system would deliver the same diagnosis for two patients that are similar aside from belonging to different demographic groups.

Regardless of the diversity of the fairness definitions you have to choose from, your project team must ensure that decisions made regarding formal fairness metrics are lawful and conform to governing equality, non-discrimination, data protection, and human rights laws. Where appropriate, relevant experts should be consulted to confirm the legality of such choices. Moreover, your team's determination of Metric-Based Fairness should heavily depend on the specific use case being considered. Each definition has strengths that may be more or less suitable in different scenarios, so it is important to consider the practical consequences of each metric in the particular context of its potential deployment.

**Further Technical and Sociotechnical Considerations**

Note that different fairness-aware methods involve different types of technical interventions at the preprocessing, modelling, or post-processing stages of production and use. Incorporating and testing for each metric type requires different technical specifications to be met, and so this also needs to be considered. In [Appendix A](Appendix A), we set out some of the prominent algorithmic fairness techniques that may be implemented from the Data Preprocessing step of the workflow to the System Deployment phase to support Metric-Based Fairness (note that this is a rapidly developing field, so your technical team should keep updated about further advances).

It is also important to consider here that, regardless of the technical approach to Metric-Based Fairness that is taken, there are unavoidable trade-offs and inconsistencies between mathematical definitions of Metric-Based Fairness that must be weighed in determining which of them are best fit for your use case. For instance, the desire for equalised odds (error rate balance) across subgroups can clash with the desire for equalised model calibration (correct predictions of gradated outcomes) or parity of positive predictive values across subgroups. The complex nature of these considerations again highlights that determining your fairness definition should be a cooperative and multidisciplinary effort across the project team.

Take note, though, that these technical approaches have limited scope in terms of the bigger picture issues of application and design fairness that we have already stressed. Moreover, metric-based approaches face other practical and technical barriers. For instance, to carry out group comparisons, formal approaches to fairness require access to data about sensitive/protected attributes as well as accurate demographic information about the underlying population distribution (both of which may often be unavailable or unreliable). Furthermore, the work of identifying sensitive/protected attributes may pose additional risks of bias. Lastly, metric-based approaches also face challenges in the way they handle combinations of protected or sensitive characteristics that may amplify discriminatory treatment. These have been referred to as intersectional attributes (e.g. the combination of gender and race characteristics), and they must also be integrated into fairness and equity considerations.

# System Implementation Fairness

When your project team is approaching the stage before the final release, you should begin to build out your plan for implementation training and support. This plan should include adequate preparation for the responsible and unbiased deployment of the AI system by its on-the-ground users. Automated decision-support systems present novel risks of bias and misapplication at the point of delivery. Special attention should be paid to preventing harmful or discriminatory outcomes at this critical juncture of the AI project lifecycle.
If those systems are solely automated and lead to legal or similarly significant impacts on individuals you will be required to comply with Article 22 of the UK GDPR. You will be indeed expected to put in place technical and organisational measures to mitigate unfair discrimination based on Recital 71. In order to design an optimal regime of implementer training and support, you should pay special attention to the unique pitfalls of bias-in-use to which the deployment of AI technologies give rise. These can be loosely classified as **Decision-Automation Bias** and **Automation-Distrust Bias**.

## 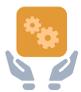 Decision-Automation Bias (Overreliance)

Deployers of AI systems may tend to become hampered in their critical judgement, rational agency, and situational awareness. This happens because they believe that the AI system is objective, neutral, certain, or superior to humans. When this occurs, implementers may rely too much on the system or miss important faults, errors, or problems that arise over the course of its use. Implementers can become complacent and overly deferent to its directions and cues.

# Case study: Decision-Automation Bias (Overreliance)

An AI application based on a deep neural network algorithm was developed to assist pathologists in detecting the presence or absence of tumours in histopathological images (microscopic images of body tissue). The AI system would improve the performance in predictions of difficult borderline cases. It would also speed up the analysis of low-risk instances that were considered clear and uncontroversial. Over the course of the trialling of the application, it was discovered that several clinicians relied too much on the results generated by the AI system. This occurred even when there was some evidence of the presence of tumours but the system predicted tumour absence with high confidence.

It was later discovered that the application's architecture was faulty, making the clinicians' overreliance on the system's classifications a significant issue. Specifically, the deep neural network was incorrectly using the presence of fat tissue in the pathology slides as a signal of the absence of tumours. This is an indication of defectiveness in the trained model. To combat the tendency of clinicians to overrely on the results of the application—even in instances of evidence that contradicted system outputs—the AI project team instituted a rigorous training regime. The new training regime reinforced the importance of continuous critical intervention and situational awareness across System Implementation processes.

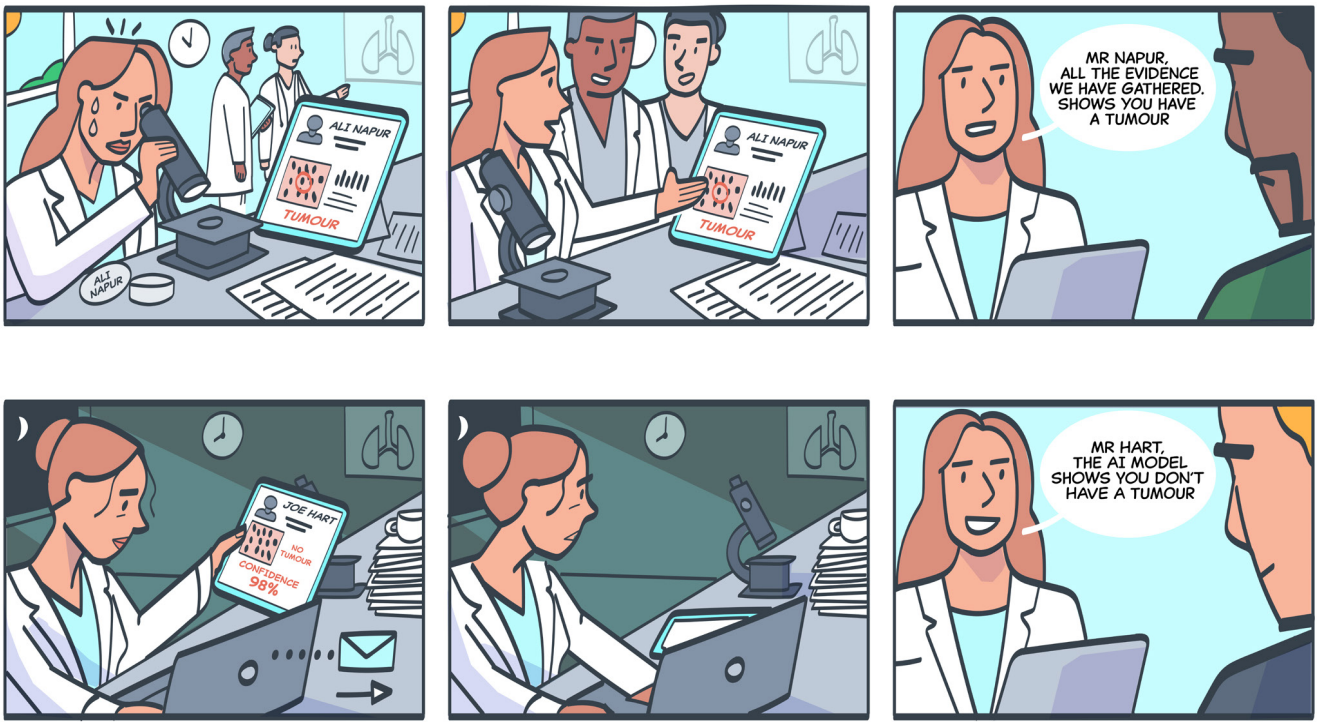

# Decision-Automation Bias (Overcompliance)

Decision-Automation Bias may also lead to overcompliance or errors of commission. This occurs when implementers defer to the perceived infallibility of the system and thereby become unable to detect problems emerging from its use for reason of a failure to hold the results against available information. Both overreliance and overcompliance may lead to what is known as out-of-loop syndrome. The degradation of the role of human reason and the de-skilling of critical thinking hampers their ability to responsibly complete the tasks that have been automated. This condition may bring about a loss of the ability to respond to system failure. It may also lead both to safety hazards and to dangers of discriminatory harm. To combat risks of Decision-Automation Bias, you should operationalise strong regimes of accountability at the site of user deployment to steer human decision-agents to act on the basis of good reasons, solid inferences, and critical judgment.

**Case study** — Decision-Automation Bias (Overcompliance)

An AI application based on a natural language processing algorithm was launched to support the early detection of Alzheimer's disease. The model used patients' verbal responses to a series of prompts. It assessed their language patterns to detect cognitive impairment. The AI application was used within GPs' screening processes, which also includes memory tests, concentration tests, and spatial orientation tests. GPs evaluated the tests' results alongside the model's predictions to decide whether to refer patients for further specialist examination. The model helped to improve physicians' accuracy in early screenings by 40%.

However, the model was developed using recordings of native English speakers primarily from a geographic region with a distinctive speaking style. When deployed nationally in diverse public healthcare contexts, which contained a range of speaking styles, accents, and language patterns, it was prone to false positives. The model diagnosed patients who were found not to have Alzheimer's disease upon further examination. When investigated, it was noted that in many false-positive cases, doctors overcomplied with the application's outputs. They over-rode their correct assessments in favour of the model's predictions. This occurred even where their judgement had been correctly informed by evidence of accurate memory, strong concentration, and appropriate spatial orientation.

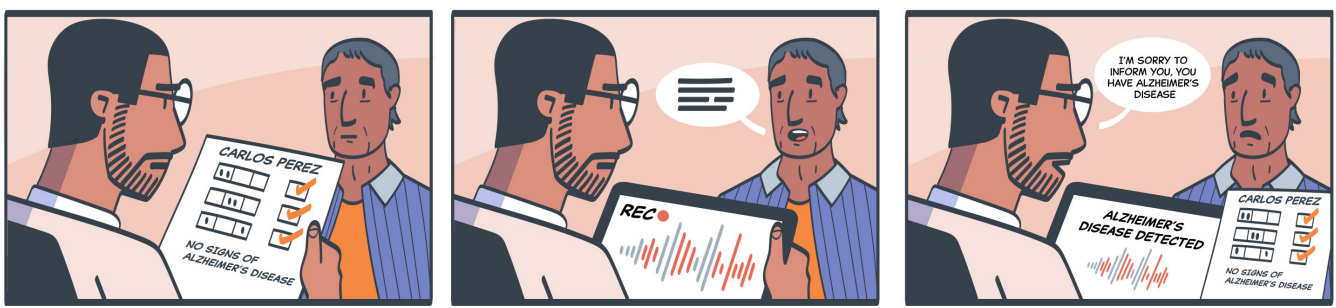

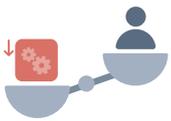

# Automation-Distrust Bias

At the other extreme, users of an automated decision-support system may tend to disregard its salient contributions to evidence-based reasoning either as a result of their distrust or scepticism about AI technologies in general or as a result of their overprioritisation of the importance of prudence, common sense, and human expertise. An aversion to the non-human and amoral character of automated systems may also influence decision subjects' hesitation to consult these technologies in high-impact contexts such as healthcare, transportation, and law.

**Case study**  **Automation-Distrust Bias**

Radiomics is an emerging application of AI used to support medical diagnoses. Radiomic systems extract features from digital images and provide doctors with information about digital images that may be undetected by their naked eye. This data, combined with an array of other clinical information, is particularly helpful for aiding doctors in the early detection and treatment of diseases, when clinicians may visually overlook the presence of a condition.

A radiomics-based AI application was built to assist doctors in the early diagnosis of breast cancer. The system extracts data from mammogram screenings, processes it along with other data from patients' Electronic Health Records and genomic profiles, and delivers a prediction regarding patients' probability of developing breast cancer. When the application was being trialled, it was discovered that some clinicians tended to reject high-probability diagnoses of very early-stage forms of breast cancer, where they could not visually or clinically confirm the presence of cancerous cells. These doctors reported that they prioritised their professional judgement over opaque algorithms and a lack of clinical evidence. To mitigate the possibilities that this pattern of clinical decision-making was being influenced by automation-distrust bias, the AI project team developed a training curriculum that detailed the application's technical specifications, the measures taken to ensure its safety, reliability, and robustness, and the advantages and challenges of using this type of technology in oncological settings.

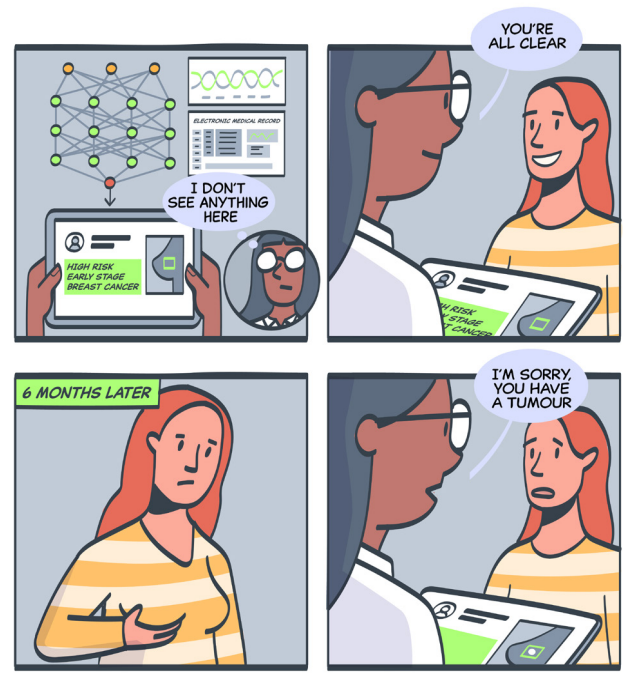

# Taking Account of the Context of Impacted Individuals in System Implementation

In cases where you are using a decision-support AI system that draws on statistical inferences to determine outcomes which affect individual persons, fair implementation processes should include considerations of the specific context of each impacted individual. For instance, when you are using a predictive risk model that helps an adult social care worker determine an optimal path to caring for an elderly patient, you should consider their specific context. Any application of this kind of system's recommendation will be based on statistical generalisations. These generalisations pick up relationships between the decision recipient's input data and patterns or trends that the AI model has extracted from the underlying distribution of that model's original dataset. Such generalisations will be predicated on inferences about a decision subject's future behaviours (or outcomes related to them) based on populational-level correlations. These predictions are based on the historical characteristics and attributes of the members of groups to which that person belongs rather than on the specific qualities of the person themself.

Fair and equitable treatment of decision subjects entails that their unique life circumstances and individual contexts be taken into account in decision-making processes that are supported by AI-enabled statistical generalisations. For this reason, you should train your implementers to think contextually and holistically about how these statistical generalisations apply to the specific situation of the decision recipient. This training should involve preparing implementers to work with an active awareness of the sociotechnical aspect of implementing AI decision-assistance technologies from an integrative and human-centred point of view. You should train implementers to apply the statistical results to each particular case with appropriate context sensitivity and 'big picture' sensibility. This means that the respect they show to the dignity and uniqueness of decision subjects can be supported by interpretive understanding, reasonableness, and empathy.

# Putting System Implementation Fairness Into Practice

To secure and safeguard fair implementation of AI systems by users well-trained to utilise the algorithmic outputs as tools for making evidence-based judgements, you should consider the following measures:

1. Training of implementers should include the conveyance of basic knowledge about the statistical and probabilistic character of machine learning and about the limitations of AI and automated decision-support technologies. This training should avoid any anthropomorphic (or human-like) portrayals of AI systems and should encourage users to view the benefits and risks of deploying these systems in terms of their role in assisting human judgment rather than replacing it.

2. Forethought should be given in the design of the user-system interface about human factors and about possibilities for Implementation Biases. The systems should be designed to enable processes that encourage active user judgment and situational awareness. The interface between the user and the system should be designed to make clear and accessible to the user the system's rationale.

3. Training of implementers should include a pre-emptive exploration of the cognitive and judgmental biases that may occur across deployment contexts. This should be done in a use-case based manner that highlights the particular misjudgements that may occur when people weigh statistical evidence. Examples of the latter may include overconfidence in prediction based on the historical consistency of data, illusions that any clustering of data points necessarily indicates significant insights, and discounting of societal patterns that exist beyond the statistical results. More information on related cognitive biases can be found in Appendix B.

Key Concepts    System Implementation Fairness    48

# Ecosystem Fairness

The AI project lifecycle does not exist in isolation from, or independent of, the wider social system of economic, legal, cultural, and political structures and institutions in which the production and use of AI systems take place. Rather, because the project lifecycle is embedded in these structures and institutions, the policies, norms, and procedures through which such structures and institutions influence human action also influence the AI project lifecycle itself.

Sociohistorically rooted inequities and biases at this ecosystem level can therefore steer or shape AI research and innovation agendas in ways that can generate inequitable outcomes for protected, marginalised, vulnerable, or disadvantaged social groups. Such ecosystem-level inequities and biases may originate in and further reinforce asymmetrical power structures, unfair market dynamics, and skewed research funding schemes that favour or bring disproportionate benefit to those in the majority, or those who wield disproportionate power in society, at the cost of those who are disparately impacted by the discriminatory outcomes of the design, development, and use of AI technologies.

Ecosystem-level inequities can occur across a wide range of AI research and innovation contexts:

For instance, when AI-enabled health interventions such as mobile-phone-based symptom checker apps or remote AI-assisted medical triaging or monitoring are designed without regard for the barriers to access faced by protected, vulnerable, or disadvantaged groups, they will disproportionately benefit users from other, more advantaged groups.

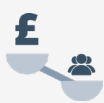
Likewise, where funding of the development of AI technologies, which significantly affect the public interest, is concentrated in the hands of firms or vendors who singly pursue commercial or financial gain, this may result in:

- exclusionary research and innovation environments;
- AI systems that are built without due regard for broad fairness and equity impacts; and
- deficiencies in the development and deployment of wide-scale, publicly beneficial technology infrastructure.

Moreover, widespread structural and institutional barriers to diversity and inclusion can create homogeneity on AI project teams (and among organisational policy owners) that has consequential fairness impacts on application decisions, resource allocation choices, and system deployment strategies.

The concept of Ecosystem Fairness highlights the importance of mitigating the range of inequity-generating path dependencies that originate at the ecosystem level and that are often neglected by or omitted from analyses of AI project lifecycles. **Ecosystem Fairness, therefore, focuses on rectifying the social structures and institutions that engender indirect discrimination. It also focuses on addressing the structural and institutional changes needed for the corrective modification of social patterns that produce discriminatory impacts and socioeconomic disadvantage.** In this way, Ecosystem Fairness involves the transformation of unjust economic, legal, cultural, and political structures or institutions with the aim of the universal realisation of equitable social arrangements.

# Part Two: Putting Fairness Into Practice

Up to this point, we have explored the complexities involved in defining fairness and offered a way of understanding the different valences of fairness that is practice-based and contextually responsive. We have also offered an overview of the principle of discriminatory non-harm (do no discriminatory harm) to help establish normative criteria that can inform the fairness-aware design, development, and deployment of AI systems.

We are now well-positioned to delve into how best to put the principle of fairness into practice. This will involve both focusing on end-to-end processes of bias mitigation and, where applicable, making explicit the formal fairness definitions chosen and technically operationalised in the model and its deployment.

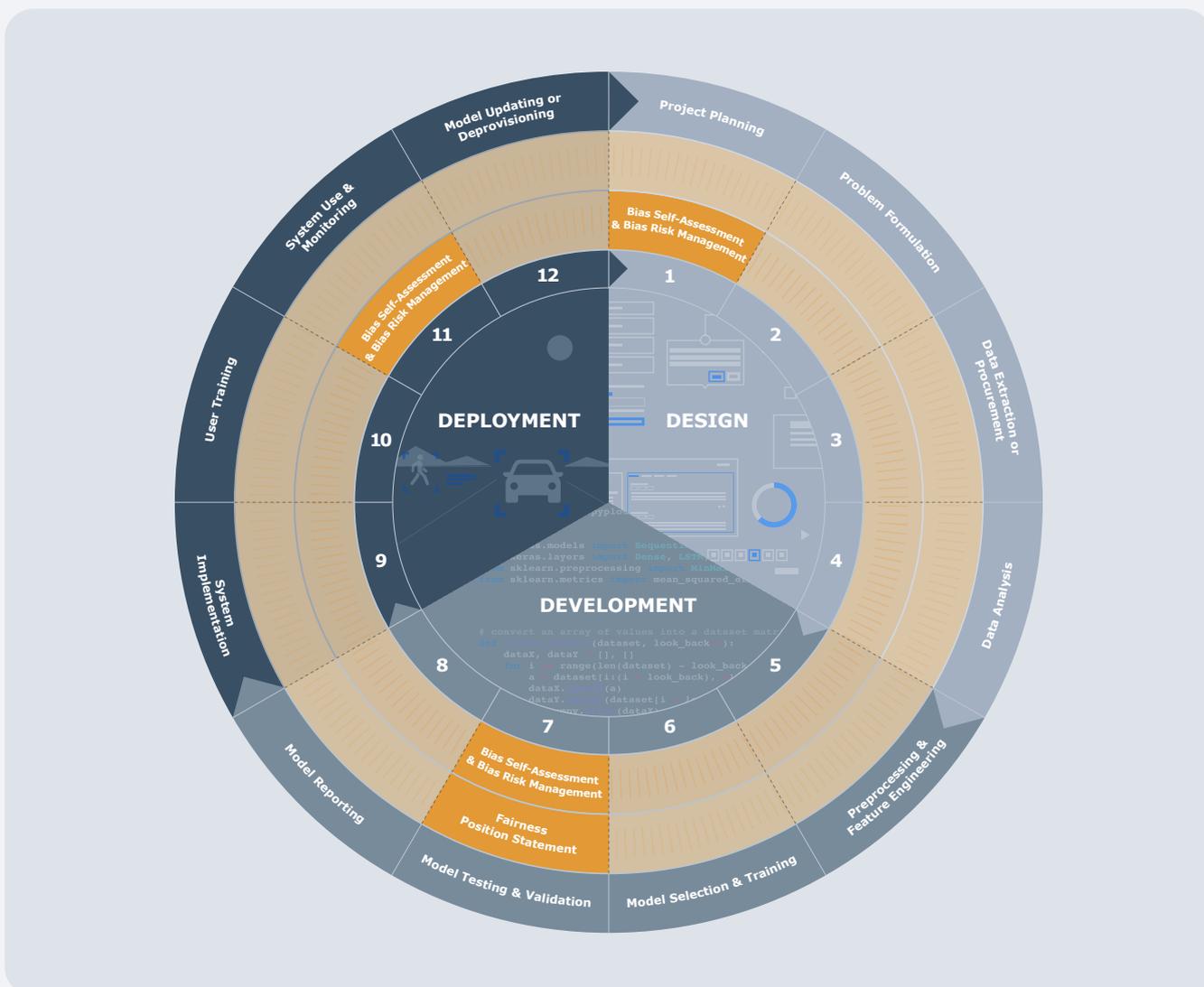

# Bias Self-Assessment and Bias Risk Management

Considering fairness-aware design and implementation from a workflow perspective will assist you in pinpointing risks of bias or downstream discrimination and streamlining possible solutions in a proactive, pre-emptive, and anticipatory way. At each stage of the AI project lifecycle, you and the relevant members of your team should carry out a collaborative Bias Self-Assessment and Risk Mitigation Plan. This plan is to be initially completed during the Project Planning step of the AI lifecycle, informed by the PS Report and a consideration of biases that are relevant to each project stage. See Appendix B (page 79) for a taxonomy of unfair biases across the AI innovation lifecycle. This plan is to be revisited during each relevant lifecycle stage and updated after completing each iteration of your Stakeholder Impact Assessment, each time identifying and responding to emerging biases through mitigation activities, and providing up-to-date documentation on the project's management of the principle of Fairness.

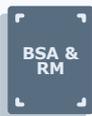

## Bias Self-Assessment and Bias Risk Management Template for *Project Name*

Date completed: ..........................   Team members involved: ....................................................................... .

| AI Lifecycle Stage | Bias | Category | Risk Mitigation Action |
|---|---|---|---|
| Project Planning | Historical Bias | World Bias | *Conduct Stakeholder Impact Assessment in partnership with impacted communities to ensure that the real-world context of the project, including the social and economic conditions of stakeholders are accounted for in project objectives and that these objectives are acceptable within stakeholders' sense of justice.* |
| Problem Formulation | | | |
| Data Extraction or Procurement | | | |
| Data Analysis | | | |

Conducting a self-assessment consists of the following steps (repeated iteratively across the project lifecycle):

### Step 1

**Familiarise yourself with biases that are relevant to each project stage** by reviewing the biases that map onto the Taxonomy of Biases Across AI Lifecycle outlined in [Appendix B (page 79)](#).

### Step 2

**Reflect on how your particular AI project might be vulnerable to biases that may arise at each stage**, and identify biases that may be present across your project workflow.

### Step 3

**Determine and document bias risk mitigation actions** you will implement to correct any biases that have been identified and strengthen specific stages in the workflow that have possible discriminatory consequences.

The Fairness Self-Assessment and Mitigation Plan template will help you and your team go through steps 2 and 3. Having familarised yourself with social, statistical, and cognitive biases across the AI project workflow, the template will allow you to document potential biases you've identified for your project, as well as risk mitigation actions you will implement to address them.

# Fairness Position Statement

In our discussion of Metric-Based Fairness above, we explored how there are many different ways to formally define how to allocate fairness of a system's outputs and error rates. Some of these ways may contradict each other or be inconsistent. A Fairness Position Statement provides an opportunity for AI project teams to be transparent about both their choices of formal fairness metrics and the rationale behind those choices.

During the design and development of your AI system, your project team will need to establish which formal fairness definitions to apply by thoroughly considering the use case appropriateness as well as technical feasibility of the relevant fairness metrics. When decisions about fairness criteria have been finalised, your project team should prepare a Fairness Position Statement (FPS), in which the Metric-Based Fairness criteria being employed in the model is made explicit and explained in plain and nontechnical language. This should include a clear explanation of the rationale behind the choices made. The FPS should then be made publicly available for review by all affected stakeholders.

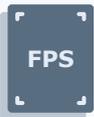

## Fairness Position Statement for Project Name

Date completed: ......................

Team members involved:
........................................................
........................................................

**Established Fairness Metrics:**

**Explanation of Choice and Rationale:**

# AI Fairness in Practice

# Activities

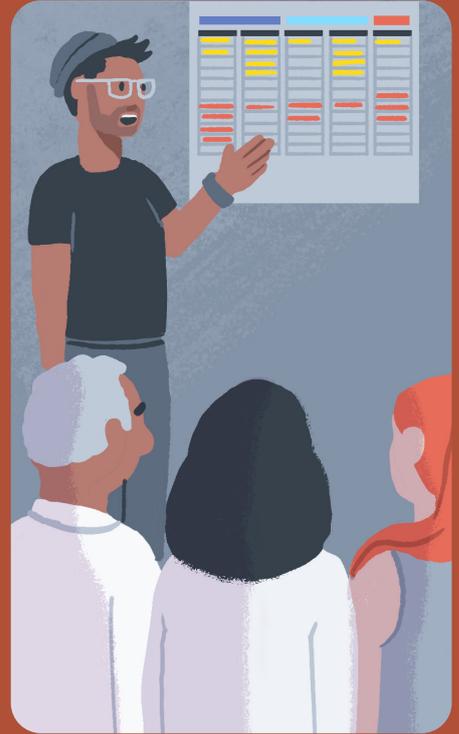

# Activities Overview

In the previous sections of this workbook, we have presented an introduction to the core concepts of AI Fairness. In this section we provide concrete tools for applying these concepts in practice. The purpose of the AI Fairness in Practice activities is to help participants understand how the principle of Fairness may be operationalised within the design and development of AI systems. Participants will gain an understanding of this principle in practice by engaging with a variety of case studies related to AI in healthcare. They will address different forms of bias that may occur throughout the workflow by addressing them through group activities.

We offer a collaborative workshop format for team learning and discussion about the concepts and activities presented in the workbook. To run this workshop with your team, you will need to access the resources provided in the link below. This includes a Miro board with case studies and activities to work through.

Workshop resources for **AI Fairness in Practice:**
turing.ac.uk/aieg-4-activities

**A Note on Activity Case Studies**

Case studies within the Activities sections of the AI Ethics and Governance in Practice workbook series offer only basic information to guide reflective and deliberative activities. If activity participants find that they do not have sufficient information to address an issue that arises during deliberation, they should try to come up with something reasonable that fits the context of their case study.

> **Note for Facilitators**
>
> In this section, you will find the participant and facilitator instructions required for delivering activities corresponding to this workbook. Where appropriate, we have included Considerations to help you navigate some of the more challenging activities.
>
> Activities presented in this workbook can be combined to put together a capacity-building workshop or serve as stand-alone resources. Each activity corresponds to a section within the Key Concepts in this workbook, which are detailed on the following page.

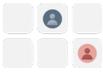

### Considering Application and Data Bia

Practise analysing real-world patterns of bias and discrimination that may produce data biases.

**Corresponding Sections**

- → Data Fairness (page 24)
- → Application Fairness (page 27)
- → Part Two: Putting Fairness into Practice (page 51)
- → Appendix B: Mapping Biases Across the AI Project Lifecycle (page 79)
  - → World Biases (page 84)
  - → Data Diases (page 86)

### Design Bias Reports

Practise assessing how bias may play out throughout the different stages of designing AI system.

**Corresponding Sections**

- → Model Design and Development Fairness (page 30)
- → Part Two: Putting Fairness into Practice (page 51)
- → Appendix B: Mapping Biases Across the AI Project Lifecycle (page 79)
  - → Design Biases (page 89)

### Defining Metric-Based Fairness

Practise selecting definitions of fairness that fit specific use cases for which outcomes are being considered.

**Corresponding Sections**

- → Metric-Based Fairness (page 38)
- → Part Two: Putting Fairness into Practice (page 51)

### Redressing System Implementation Bias

Practise identifying and redressing different forms of Implementation Bias.

**Corresponding Sections**

- → System Implementation Fairness (page 43)
- → Part Two: Putting Fairness into Practice (page 51)

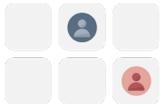  90 mins | Participant Instructions

# Considering Application and Data Bias

**Objective**

The purpose of this activity is to practise examining patterns of real-world disadvantage, injustice, and discrimination that may contribute to biased data.

**Activity Context**

**Real-World Patterns of Health Inequality and Discrimination**

The health consequences of a treatment of a medical condition or an interaction with the healthcare system vary among individuals and between different population groups. These *differences in health outcomes* are known as **health inequalities**. Interacting factors of health inequality include:

- widespread disparities in living and working conditions;
- differential access to and quality of healthcare;
- systemic racism; and
- and other deep-seated patterns of discrimination.[90]

Factors such as systemic racism, marginalisation, and structural inequality contribute to disproportionate vulnerability to disease and poor health outcomes in disadvantages communities. These *unfair and avoidable systematic differences in health* among groups in society is what we know as health inequities.

AI systems are a relevant part of the health informatics toolkit to fight diseases. But AI is well known to be susceptible to biases that can entrench the design, development, and deployment process. Uncritically deploying AI in healthcare risks amplifying adverse effects on vulnerable groups, exacerbating **health inequity**.[91]

**Health Discrimination in Datasets**

AI technologies rely on large datasets. When patterns of health inequality affect those datasets, the algorithmic models they generate will reproduce inequities.

**Patterns of Health Inequality and Discrimination that Contribute to Data Bias**

- **Discriminatory Healthcare Processes and Clinical Decision-Making**
  In clinical and public health settings, electronic health records, case notes, training curriculums, clinical trials, academic studies, and public health monitoring records are the basis of AI models. Biased judgment and decision-making, as well as discriminatory healthcare processes, policies, and governance regimens can affect these sources. Datasets deriving from them thus reflect complex and historically situated practices, norms, and attitudes.

- AI models that draw inferences from such medical data for diagnosis or prognosis might incorporate the biases of previous inequitable practices. As a result, the use of models trained on these datasets could reinforce or amplify discriminatory structures.[92]

- **Unequal Access and Resource Allocation**
  Datasets composed of electronic health records, genome databases, and biobanks often undersample those who have irregular or limited access to the healthcare system, such as minoritised ethnicities, immigrants, and socioeconomically disadvantaged groups.[93] Similarly, the increased use of digital technologies, like smartphones, for health monitoring (e.g. through symptom tracking apps) also creates potential for biased datasets as they exclude those without digital access.[94] The resources needed to ensure satisfactory dataset quality and integrity is at times limited to digitally mature hospitals in well-off neighbourhoods, that disproportionately serve a privileged segment of a population to the exclusion of others. Where data from electronic health records resulting from these contexts contribute to the composition of AI training data, problems surrounding discriminatory effects arise.

## Part One: Privilege Walk  ⏱ 45 mins

1. Individually, take a moment to read over the activity context.

2. Each member of your team will be assigned a stakeholder profile from the **Considering Real-World Contexts** section.

3. Your facilitator will guide the team through a series of questions related to experiences in healthcare. Team members are to step into the shoes of their assigned profile and answer these questions based on the answers they would be **likely to give**.

   - Profiles will move forwards or backwards. For each question, the facilitator will specify the direction of movement depending on the answers the profiles would be likely to give.

   - Profiles will stay in place if:

     - a question doesn't apply to your stakeholder;

     - you think your character would be equally positively and negatively affected; or

     - you think you do not have enough information about the character and an answer would require you to make assumptions about them.

   - Respond to these questions to the best of your ability. If there are any questions for which you don't know the answer, feel free to not answer.

4. After all questions have been answered, have a group discussion about how the likely experiences of profiles may exemplify patterns of inequality and discrimination in healthcare experienced by the identity groups they belong in.

   - Your co-facilitator will take notes about this discussion in the **Real-World Patterns** section.

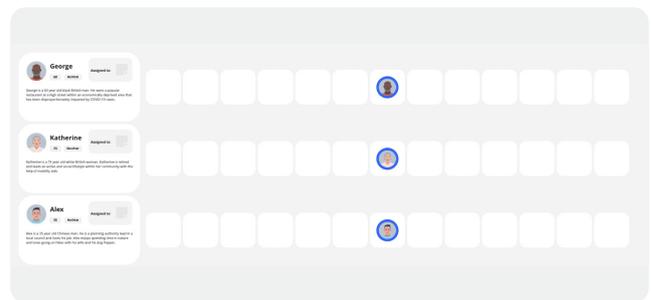

**Considering Real-World Contexts**

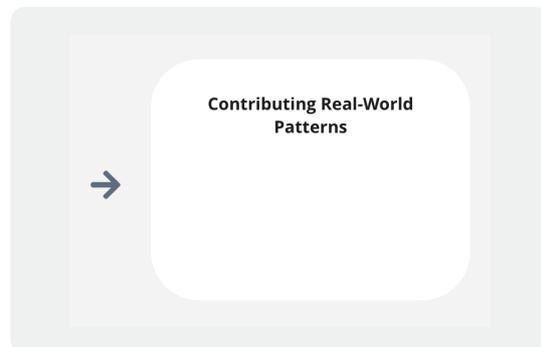

**Contributing Real-World Patterns**

**Part Two: Data Fairness Charades**  45 mins

1. In this activity, your team will be divided into groups. You will work together to understand how real-world patterns of inequality, compiled with a lack of consideration for elements of Data Fairness, can yield to AI systems that do not protect stakeholders from discriminatory harm.

2. Each team will be assigned a case study card containing an example of an AI system.

3. One team member will read the card out loud. Your group will think carefully and discuss which element of Data Fairness this card is associated with.

4. When time has passed, reconvene as a team.

5. In turns, each group will share the element of Data Fairness they believe their assigned card is associated with. The facilitator will disclose the correct answer and respond to any questions the groups may have about their case studies.

6. Take a moment to individually review the team notes on the **Real-World Patterns** section, as you will be next encouraged to connect these notes and overall discussions on real-world patterns of inequality affecting the stakeholder profiles from Part One of the activity with the instances of Data Fairness of Part Two.

7. Have a group discussion about how real-world patterns of inequality may have contributed to each instance of Data Fairness. Consider the questions:

   a. *What groups of stakeholders were not protected from discriminatory harm in these case studies?*

   b. *What real-world patterns of inequality and discrimination in healthcare experienced by these groups of stakeholders may have contributed to biased datasets? How do they relate to the elements of Data Fairness illustrated by the case studies?*

   c. *How might consideration for the key element of Data Fairness associated with these case studies have yielded different outcomes?*

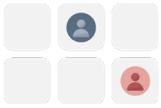

⏱ 90 mins | Facilitator Instructions

# Considering Application and Data Bias

**Part One: Privilege Walk** ⏱ 45 mins

1. Give the team 8 minutes to read over the activity context. Ask the team if they have any questions.

2. Next, give the team some minutes to read over the instructions of the Privilege walk activity. Ask the team if they have any questions.

3.

4. Assign a stakeholder profile to each team member.

5. Conduct the activity by asking the group the questions on the following page:

**If Delivering Physically**

- If there are more profiles than team members, assign profiles in the order they are on the board. There will be leftover profiles.
- If there are more team members than profiles, assign individual profiles to multiple team members.
- Ask the team to line up on one side of the room. They will be moving forwards and backwards towards the other side of the room depending on their responses to the questions. If a question does not apply to their profile, participants think they would need more information about the character to answer, or participants think it is likely their character would be equally positively and negatively affected, they are to stay in place.

**If Delivering Digitally**

- If there are more profiles than team members, let participants choose profile themselves.
- If there are more team members than profiles, duplicate profiles, creating additional rows on the board, assigning the same profiles to more than one team member.
- Direct the team towards the **Considering Real-World Contexts** section.
  - They will be moving the icon representing their profile either right (forward) or left (backwards).
  - If a question does not apply to their profile, participants think they would need more information about the character to answer, or participants think it is likely their character would be equally positively and negatively affected, they are to stay in place.

## Unequal Access and Resource Allocation

a. *Is this person likely to feel represented by their healthcare providers (e.g. physician, nurse, or other health professionals)?*
If yes: → Move forward

b. *Might fear of discrimination or uncomfortable experiences related to their identity lead to this person avoiding healthcare services?*
If yes: ← Move backward

c. *Is this person likely to easily access digital health services?*
If yes: → Move forward

d. *Is it likely that this person has access to well-funded public healthcare provision, or can afford private healthcare?*
If yes: → Move forward

## Discriminatory Healthcare Processes

a. *Might this person's circumstances present them with challenges in providing documentation (i.e. proof of address, identification) for registering to a GP practice?*
If yes: ← Move backward

b. *Is it possible that healthcare needs that are specific to this person's characteristics are not covered by most health insurances or underfunded in public health services?*
If yes: ← Move backward

c. *Are most medical providers widely trained in health considerations related to this person's identity or background?*
If yes: → Move forward

## Biased Clinical Decision-Making

a. *Are healthcare providers likely aware of any religious or cultural needs that this person might have?*
If yes: → Move forward

b. *Is it possible that this person is subject to stereotypes when receiving care which might lead to assumptions around their behavioural health risks (i.e. smoking, drinking, using drugs, engaging in risky activity)?*
If yes: ← Move backward

c. *Is it possible that this person might fear stigmatisation on behalf of healthcare providers?*
If yes: ← Move backward

d. *Might this person be stereotyped as a social threat such as a "benefits scrounger" or "health tourist", or be associated with the spread of specific diseases?*
If yes: ← Move backward

## Disparities in Living Conditions

a. *Is it possible that this person has been discouraged from a health promoting activity such as sports or community life because of race, class, ethnicity, gender, disability, or sexual orientation?*
If yes: ← Move backward

b. *Might this person's context or specific circumstances (i.e. living environment, job) contribute to unhealthy outcomes?*
If yes: ← Move backward

c. *Is this person likely comfortable being in public without fear of assault, profiling, or violence?*
If yes: ← Move backward

d. *Could this person's life experiences related to their identity contribute to issues with mental health?*
If yes: ← Move backward

5. After all questions have been answered, have a group discussion about how the likely experiences of profiles may reflect patterns experienced by the identity groups they belong in. Consider the questions:

    - Were there any questions that seemed particularly pertinent to team members' assigned profiles?

    - What instances of inequality did you notice through this activity that falls under either unequal access and resource allocation or discriminatory healthcare processes and clinical decision-making?

    - How does identity-based advantages affect patients' experiences in healthcare?

    - How do multiple forms of inequality or disadvantage contribute to the end result of the activity? What does this mean for the healthcare experiences of patients with intersecting characteristics?

    - **Co-facilitator:** write the team observations in notes, placing them within their appropriate section in the **Contributing Real-World Patterns** section.

        - In addition to the suggested questions, you may also consider asking the team what other instances of inequality likely experienced by the stakeholder profiles relate to unequal access and resource allocation or discriminatory healthcare processes and clinical decision-making.

**Part Two: Data Fairness Charades**  45 mins

1. Give the team a moment to read over the activity instructions, and review the different elements of Data Fairness outlined on the board, answering any questions.

2. Divide the team into groups. Each group will be assigned a case study of an example of discriminatory AI system. They will have to discuss which element of Data Fairness their card is associated with.

3. Next, set a timer for the assigned minutes for this activity.

   **If Delivering Physically**
   - Each group will be given a card containing a case study.

   **If Delivering Digitally**
   - Direct the team towards the Data Fairness charades section of the board, specifying the card assigned to each group.

4. When the assigned time has passed, ask participants to reconvene as a team.

5. Go through each group's answer, using the considerations section of this activity to correct any wrong answers. Allow each group to discuss their rationale behind their answer.

6. Next, give the team a moment to review the cards and the notes on the **Contributing Real-World Patterns** section.

7. Lead a group discussion, considering the questions on the participant instructions.

   - **Co-facilitator:** Take notes about the group discussion, writing them in the **Contributing Real-World Patterns** section.

| Case Study Card | Relevant Element of Data Fairness | Contributing Real-World Pattern in Healthcare |
|---|---|---|
| **Card 1**<br><br>An AI system used to diagnose patients with Obsessive Compulsive Disorder (OCD) based on their medical records is developed using records from users of clinical services for OCD. Uptake of such services is low amongst minority ethnic groups, reflecting in available records.[95] The system therefore yields suboptimal diagnoses for members of these communities.[96] | Representativeness | Unequal access and resource allocation |
| **Card 2**<br><br>An AI system used to detect cardiovascular conditions across the UK is developed using several decades of patients' medical records from a small but well funded hospital that recently turned paper records into electronic health records (EHRs). The hospital is located in a predominantly white, suburban area. The resulting system is prone to error when diagnosing black and other minority ethnic people, as well as white patients from urban areas.[97] [98] [99] [100] | Fit for Purpose and Sufficiency | Unequal access and resource allocation<br><br>Unequal resource allocation resulting in well-funded hospitals being digitally mature and producing disproportionate data available for developing AI systems. |
| **Card 3**<br><br>An AI system used to diagnose autism spectrum disorder (ASD) based on clinical notes is developed using patient medical records across the UK. Research used to establish standards for diagnosing ASD have disproportionately studied boys, contributing to clinical biases towards diagnosing them.[101] [102] [103] The resulting system is prone to false negatives for girls, predicting that they are not autistic when they are. | Source Integrity and Measurement Accuracy | Discriminatory healthcare processes and biased clinical decision-making |

| Case Study Card | Relevant Element of Data Fairness | Contributing Real-World Pattern in Healthcare |
|---|---|---|
| **Card 4**<br><br>An AI system used to screen patients for risk of breast cancer is developed using patients' medical records. The system is trained using standard categories used in healthcare, such as patients' age and sex, to measure risk. The resulting model yields false negatives, identifying patients as risk-free when indeed they are at risk of breast cancer. This results in underdiagnosis, for instance, of cis men with malignant tumours in their breast tissue or transgender patients who are marked as male in their records and have breasts.[104] [105] | Relevance, Appropriateness and Domain Knowledge | Unequal access and resource allocation |
| **Card 5**<br><br>An AI system that has been designed to assist healthcare workers to judge the risk profile patients face upon discharge from an inpatient hospital stay. The data used to train the model includes information about patients' proximity to outpatient services like local health clinics and pharmacies. However, due to a lack of resources at the local level, the data collected from several areas with high levels of socioeconomic deprivation was out-of-date and failed to reflect various closures that have led to challenges for patients to access needed resources in those areas.[106] This has led to the model erroneously predicting safe discharge for patients who will face barriers to necessary care. | Timelines and Recency | Unequal access and resource allocation |

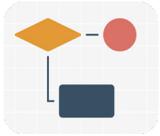

⏲ 60 mins | Participant Instructions

# Design Bias Reports

**Objective**
Practise understanding how different forms of bias may play out within different steps of the AI lifecycle.

**Overview**
In this activity, participants will be split into four groups, each assigned a step in the AI/ML project lifecycle and a case study illustrating an example of an AI system that produced discriminatory outcomes due to bias within their assigned step.

Using the guiding questions provided, each group is to analyse their case study and draft a short report on how bias likely had a discriminatory influence on system outcomes. Groups will present their explanations to the greater team.

**Team Instructions**

1. Once split into groups, take a moment to individually review your assigned case study and information about the related step in the AI/ML project lifecycle.

2. As a group, use the questions within the **Case Study Reports** section to draft your report. These questions will help you structure your presentation.

3. When answering the question titled 'Why', your group is to consider the possible biases within your assigned step of the lifecycle and determine which biases may have been present in your case study.

4. Reconvene as a team, having a volunteer from each group present their report.

5. When all teams have presented, have a group discussion about the different forms of bias present in the case studies.

**Case Study Reports**

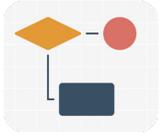

⏱ 60 mins | Facilitator Instructions

# Design Bias Reports

1. Give participants a moment to read over the activity instructions, answering any questions.

2. Let the team know that they will be split into four groups and that will have a certain amount of time to create their reports, giving them enough time to individually read over the materials provided.

3. Split the team into groups. Take a moment to individually review your assigned case study, and information related to the step in the lifecycle within which bias was present.

   - **Facilitators** and **co-facilitators** are to touch base with each of the groups throughout this activity, answering any questions and providing support using the considerations section of this activity.

4. Let the groups know when they have a few minutes to finish their reports.

5. When time is up, ask the team to reconvene, giving volunteer note-takers a few minutes to share their case study reports.

6. When all teams have presented, lead a group discussion about the different forms of bias present in the case studies.

7. Using the considerations section of this activity, revise all forms of bias discussed per case studies, ensuring the team is aware of the correct biases present at their assigned stage

8. Next, lead the discussion using the questions on the participant instructions.

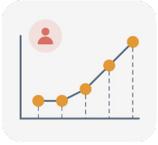

`⏱ 30 mins` | **Participant Instructions**

# Defining Metric-Based Fairness

**Objective**
Practise selecting definitions of fairness that fit specific use cases for which outcomes are being considered.

**Team Instructions**

1. In this activity, your team is to discuss about the advantages and disadvantages of different group fairness metrics that could govern the allocation of error rates of the model presented in the case study.

2. Take a moment to read over the case study.

3. Next, your team will be split into groups. In your groups, look over the **Definitions of Group Fairness** section.

4. Have a volunteer write notes about your group discussion in your group's section.

5. Have a discussion about the different definitions of group fairness, and the outcomes they may yield for this model, considering the questions in the section.

6. Having had a discussion, reconvene as a team.

7. Next, identify the most suitable fairness metric(s) to use for this system and fill out the **Fairness Position Statement** section.

**Definitions of Group Fairness**

**Fairness Position Statement**

## Case Study — Metric-Based Fairness

Most skin conditions do not lead to severe cases which GPs cannot diagnose or treat. However, a lack of rigorous diagnostic training across GP practices leads to a significant number of misdiagnoses, which delay patients' treatment. They also lead to unnecessary dermatologist referrals. Considering the recent rise in high-risk conditions such as skin cancer and the national shortage of dermatologists, unnecessary referrals strain the treatment of high-risk conditions.

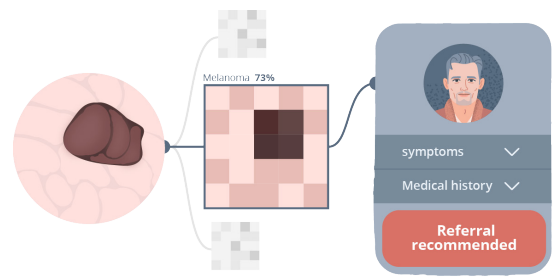

A decision-support machine learning application has been developed for use within GP practices. Its purpose is to assist the diagnoses of 20 of the most common skin conditions. The model processes images of patients' skin alongside case information (i.e. symptoms, self-identified medical history such as age, sex, existing conditions) using a deep learning algorithm. It flags if a patient has a high-risk condition and should be referred to a specialist. The model was developed using data from 30,000 cases diagnosed by 25 dermatologists across the UK. It has been trained and tested to match the average performance rates of senior dermatologists, and deliver equal performance metrics across ethnic groups. Deploying it in non-specialist settings will support GPs' diagnoses of skin conditions, promising to reduce misdiagnoses, streamline the delivery of treatment, and ensure that non-severe skin conditions remain within the care of GPs.

The system has been piloted within a variety of hospitals across the UK for external validation. Clinicians have reported that the model frequently misdiagnoses patients with dark skin tones. This could lead doctors to inaccurately dismissing high-risk cases and to fail to refer cases that aren't detected by the model. Upon review, the application's developers recognise that although the model performs equally across ethnic categories, it has less sensitivity for patients with darker skin. That is, it performs better for black patients with lighter skin than for black patients with darker skin. The developers decide to augment the training dataset to include more datapoints from patients with darker skin types. They also decide to update the fairness metrics that will govern the allocation of error rates. The fairness metrics will now be based on skin phenotypes types (linked to tone) rather than ethnicity.

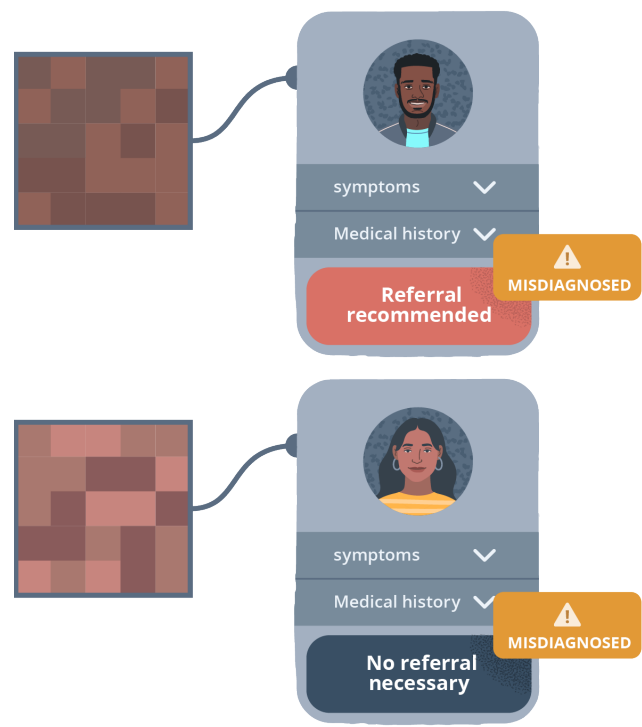

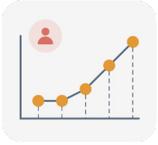

⏱ 30 mins  |  Facilitator Instructions

# Defining Metric-Based Fairness

1. Give the team a moment to individually read over the activity instructions, answering any questions.

2. Give them a few minutes to individually read over the case study.

3. When time is up, split the team into groups.

4. Give the groups some minutes to discuss the different definitions of group fairness, their technical limitations, and the outcomes they may yield for this model, considering the questions in the section. If needed, refer to Further Technical and Sociotechnical Considerations (page 42).

5. When time is up, reconvene as a group. Ask the team to identify the most suitable fairness metric or combination of fairness metrics to use for this system based on their moral positioning within their technical capacities. Ask them to fill out the **Fairness Position Statement** section.

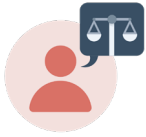

⏱ 25 mins | **Participant Instructions**

# Redressing System Implementation Bias

**Objective**

Practise identifying and redressing different forms of System Implementation Bias.

**Team Instructions**

1. In this activity, the team will be split into groups, each of which will be presented with a case study illustrating an instance of Automation Bias (overreliance) and Automation Distrust Bias.

2. In your groups, take a few minutes to come up with an initial solution to the bias in your assigned case study. Write your solution in your group's section.

   - Refer to the *Putting System Implementation Fairness Into Practice* section of this workbook for help in this activity.

3. When time is up, reconvene as a team, having volunteers share each group's solution.

4. Next, have a group discussion, deliberating how, if at all, the solution can be improved.

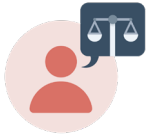

⏱ 25 mins | **Facilitator Instructions**

# Redressing System Implementation Bias

1. Give the group a few minutes to read over the instructions.

2. Next, split the team into groups.

3. Facilitators and co-facilitators are to touch base with each of the groups, using the considerations section of this activity to support group discussions.

4. When time is up, ask the group to reconvene, having a volunteer from each group share their problem and solution.

5. After each volunteer has shared, give the team a few minutes to deliberate ways to improve each solution.

    - **Co-facilitator:** write the improvements on the **Solutions** section of the board.

# Appendix A: Algorithmic Fairness Techniques Throughout the AI/ML Lifecycle

| Method Name | Method Description | Examples and Further Reading |
|---|---|---|
| **Type: Preprocessing**[107] [108] [109] | | |
| Resampling Reweighting Relabelling | Advantageous in highly regulated sectors like finance (e.g. credit scoring) and healthcare (e.g. predict disease risk given lifestyle and existing health conditions) because it's simpler to calculate and have oversight over. | Kamiran & Calders, 2009; Luong et al., 2011; Calders et al., 2009; Kamiran & Calders, 2012; Alessandro et al., 2017 |
| **Type: Preprocessing**[110] | | |
| Modifying feature representations | Makes it difficult for the model to differentiate between protected and unprotected groups by making both distributions similar or by using methods such as fair dimensionality reduction. | almon et al., 2017; Feldman et al., 2015; Samadi et al., 2018; Zemel et al., 2013 |
| **Type: Preprocessing + In-Processing**[111] | | |
| Variational fair autoencoder | Creates a model in which enough information is retained to solve the task at hand but information learned throughout the model training remains independent from sensitive attributes in the data. | Louizos et al., 2016 |
| **Type: Preprocessing + In-Processing** | | |
| Dynamic upsampling through a debiasing variational encoder | Rebalances the training data through learned representations so that the model "knows" which inputs are tied to underrepresented groups and in turn are sampled more frequently during the model training process. | Amini et al., 2019; Xu, 2019 |

| Method Name | Method Description | Examples and Further Reading |
|---|---|---|
| **Type: In-Processing**[112] [113] | | |
| Regularisation of objective function | A regularisation term (i.e. a tuning parameter to help the model make better predictions) penalises the information shared between the sensitive attributes and the predictions made by the classifier. | Kamishima et al., 2012; Berk et al., 2017 |
| **Type: In-Processing**[114] | | |
| Incorporating penalty terms into objective function | Enforces matching proxies of false positive rates and false negative rates. | Bechavod & Ligett, 2017a; Bechavod & Ligett, 2017b |
| **Type: In-Processing + Post-Processing**[115] [116] | | |
| Adjusting decision tree split criterion | Maximises information gain between label and split attributes while minimising the same with respect to the sensitive attribute in question. | Kamiran et al., 2010 |
| **Type: In-Processing + Post-pProcessing**[117] | | |
| Decoupled classifiers | Applies a different classifier to each group. The output of each single classifier is chosen to minimise a joint loss function which allows for relative treatments of different groups. This allows the modeler to make explicit the trade-offs between accuracy and fairness. | Dwork et al., 2018 |
| **Type: In-Processing**[118] | | |
| Privileged learning | Improves performance of the model where sensitive data is only available at model training not testing. | Quadrianto & Sharmanska, 2017; Vapnik & Izmailov, 2015 |

| Method Name | Method Description | Examples and Further Reading |
|---|---|---|
| **Type: In-Processing**[119] [120] | | |
| Loss functions with respect to fairness metrics | Minimises arbitrary loss function subject to individual fairness-constraint, disparate mistreatment, or additional fairness constraints. | Dwork et al., 2012; Zafar et al., 2017a; Zafar et al., 2017b; Woodworth, et al., 2017 |
| **Type: In-Processing** | | |
| Adversarial de-biasing | Prevents the model from representing a given input through reliance on a sensitive attribute while simultaneously predicting the correct output. An adversarial model (a model created to trick ML models by creating inputs that can be used to deceive classifiers) is used to test this. If the adversary cannot predict the sensitive attribute from the representation of the input, then the model has successfully learned a representation of the input data that does not depend on the sensitive attribute. | Zhang et al., 2018; Correa et al., 2021; Xu, 2019 |
| **Type: In-Processing**[121] | | |
| AdaFair | Ensemble method that accounts for class imbalance by analysing a balanced error as opposed to an overall error. | Iosifidis & Ntoutsi, 2019 |
| Fair-PCA approach | Requires reconstruction errors to be balanced across protected and unprotected groups. | Samadi et al., 2018 |
| Fair-k-means | Forces clustering to have equal representation for protected groups. | Chierichetti et al., 2017 |
| Regularizer | Reduces the mutual information between the class labels and sensitive attributes. | Kamishima et al., 2012 |

| Method Name | Method Description | Examples and Further Reading |
|---|---|---|
| **Type: Post-Processing**[122] [123] | | |
| Output score adjustments | Alters output scores of classifier to enhance equalized odds/opportunities or uses different thresholds to maximise accuracy while minimising demographic parity. | Hardt et al., 2016; Corbett-Davies et al., 2017; Menon & Williamson, 2018; Kamiran et al., 2018 |
| **Type: Post-Processing**[124] | | |
| White-box methods | Adjusts the internal workings of the model to create fairer results. Examples include changing confidence levels of classification rules or adjusting probabilities of specific model types such as Naïve Bayes. | Pedreschi, et al., 2009; Calders & Verwer, 2010 |

# Appendix B: Mapping Biases Across the AI Project Lifecycle

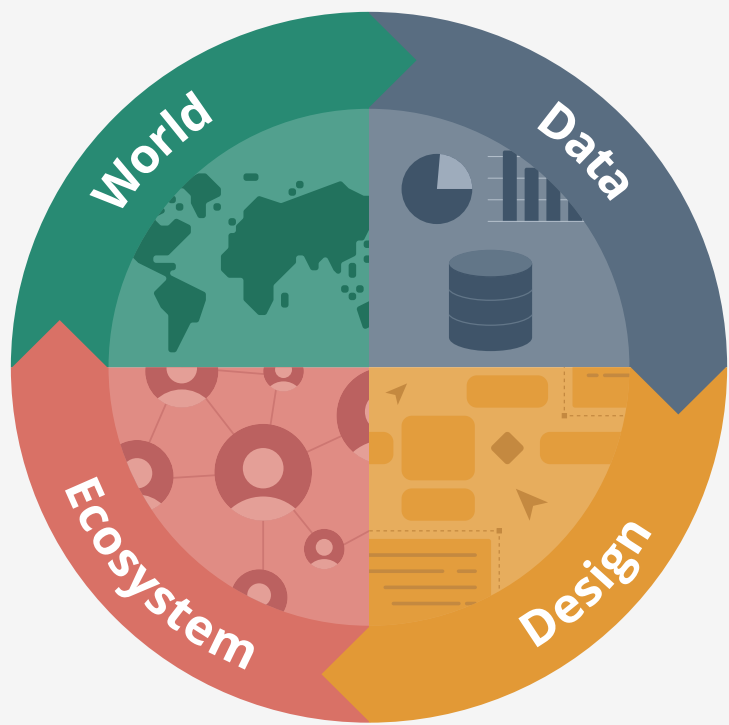

The World-Data-Design-Ecosystem model was developed to illustrate the cascading effects of inequity and unfair bias that move across these four dimensions of AI research and innovation practice. It is intended to provide you with a map with which to navigate the myriad entry points of inequity and unfair bias throughout the AI/ML project lifecycle. Starting from the top-left quadrant and moving clockwise:

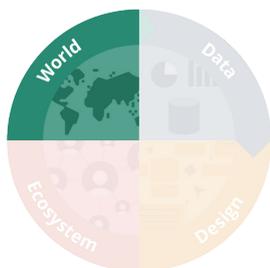

The **World** refers to the social and historical reality that precedes the design, development, and deployment of AI applications. It is the domain of lived experiences. It is where preexisting societal patterns of discrimination and social injustice— and the prejudices and discriminatory attitudes that correspond to such patterns— arise. These patterns, prejudices, and attitudes can be encoded in datasets. They are subsequently drawn into every stage of the AI/ML project lifecycle and perpetuated, reinforced, or exacerbated through inequitable innovation ecosystem dynamics and the pursuit of biased application choices and research agendas.

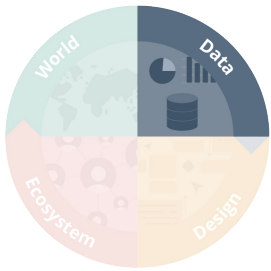

**Data**, in the widest sense, refer to 'any information collected together for reference or analysis'.[125] Data in general is viewed as humanly collected measurements of social, demographic, and biophysical information. Data is also viewed as historically generated and socially constructed products of human action and interaction. Acknowledging the social construction of data (and of the categories and classes through which humans name and organise them) is vital for the analysis of inequities and potential negative impacts that can arise through the AI/ML project lifecycle. This is because there are deep-seated tendencies in the design and development of data-driven technologies, to reify data.

To be sure, in the construction and categorisation of data, researchers, scientists, and developers can mistakenly treat socially contested and negotiated categories of identity as fixed, natural, or biological classes. When this happens, such categories can become erroneously biologised or reified. The producers of AI technologies who use these data run the risk of introducing discriminatory proxies into model inferences and erasing or rendering invisible categories or classes that they may have been excluded, aggregated, or missed altogether.

When referring to Data, considerable attentiveness is therefore needed to understand the complex and ever-changing social norms and practices, power relations, political, legal and economic structures, and human intentions that condition the production, construction, and interpretation of data. For example, in the medical field, sampling biases and lack of representativeness in the datasets can arise if people who have limited access to healthcare, mistrust clinical or research environments because of discrimination, or lack access to digital platforms and devices, are not adequately represented in electronic health records and other data sources. Likewise, the source integrity of data can be prejudiced both by implicitly biased clinical judgments reflected in clinical notes, screenings, tests and medications ordered, treatment decisions, and by inequitable tools and hardware that have been designed exclusively for dominant groups that resultantly mismeasure minority groups.

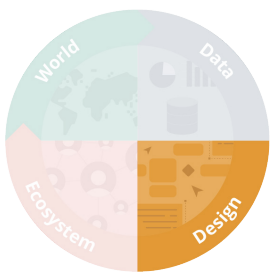

We use **Design** as an abbreviation for the sociotechnical design, development, and deployment lifecycle of AI systems. It is critical for project teams to identify and analyse how unfair bias and patterns of inequity and discrimination arise across the entirety of the lifecycle. There are several ways of carving up the project lifecycle for AI and data-driven systems. These range from the Cross Industry Standard Process for Data Mining (CRISP-DM), the Knowledge Discovery in Databases (KDD) and Sample, Explore, Modify, Model, and Assess (SEMMA) to other MLOps frameworks. Despite this diversity of approaches, it is important to work from a shared heuristic of the general workflow that typifies the AI project lifecycle so that the identification and mitigation of inequities and unfair biases across the AI/ML lifecycle can take into account the interwoven nature of its technical, social, and ethical aspects. In this workbooks series, we therefore break down the Design into stages of Project Design, Model Development, and System Deployment, each of which have a subset of activities. To review details of this model of the sociotechnical AI project lifecycle, please refer to the AI Ethics and Governance in Practice: An Introduction workbook.

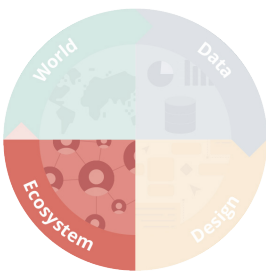

By **Ecosystem**, we mean the wider social system of economic, legal, cultural, and political structures or institutions-and the policies, norms, and procedures through which these structures and institutions influence human action. Inequities and biases at the ecosystem level can steer or shape AI research and innovation agendas in ways that can generate inequitable outcomes for minoritised, marginalised, vulnerable, historically discriminated against, or disadvantaged social groups. Such ecosystem level inequities and biases may originate in and further entrench asymmetrical power structures, unfair market dynamics, and skewed research funding schemes. These inequitable ecosystem dynamics can favour or bring disproportionate benefit to those in the majority, or those who wield disproportionate power in society, at the cost of those who are disparately impacted by the discriminatory outcomes of the design, development, and use of AI technologies.

# Taxonomy of Biases Across AI Lifecycle

This taxonomy of unfair biases across the AI innovation lifecycle is primarily organised into the four quadrants of the world-data-design-ecosystem model. This model was developed to signal the point of entry of each respective bias. A fifth dimension of cognitive bias has been added because some cognitive biases cut across the other four dimensions and are more appropriately understood in a space of their own.

The figure on the following page provides an 'at a glance' view of how of all the biases in the taxonomy hang together.

**KEY CONCEPT**

### Cognitive Bias

Systematic deviation from a norm of rationality that can occur in processes of thinking or judgement and that can lead to mental errors, misinterpretations of information, or flawed patterns of response to decision problems.

The biases outlined in this model require ongoing reflection and deliberation to minimise the possible negative impact upon downstream activities or the risk of discriminatory outcomes.

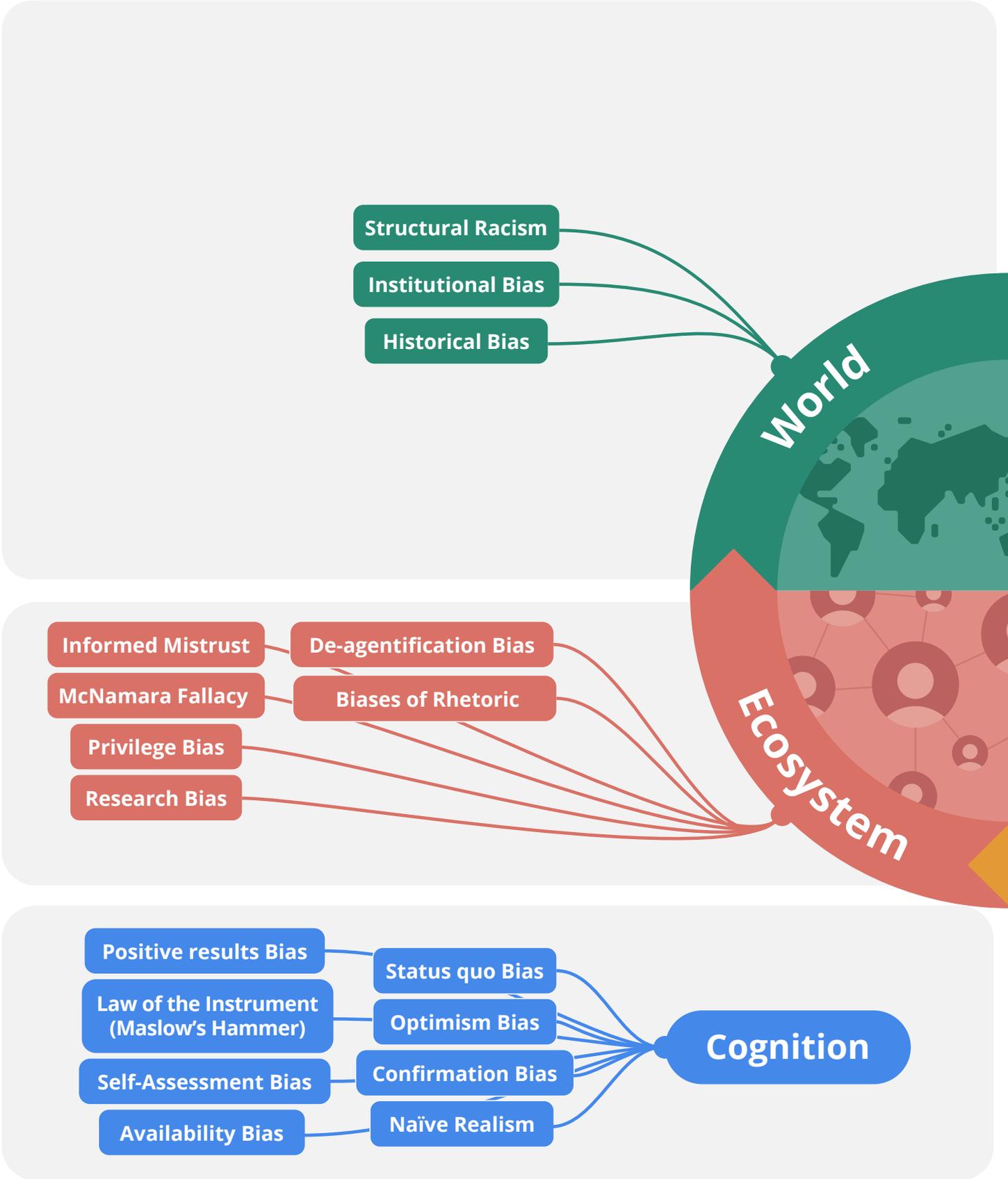

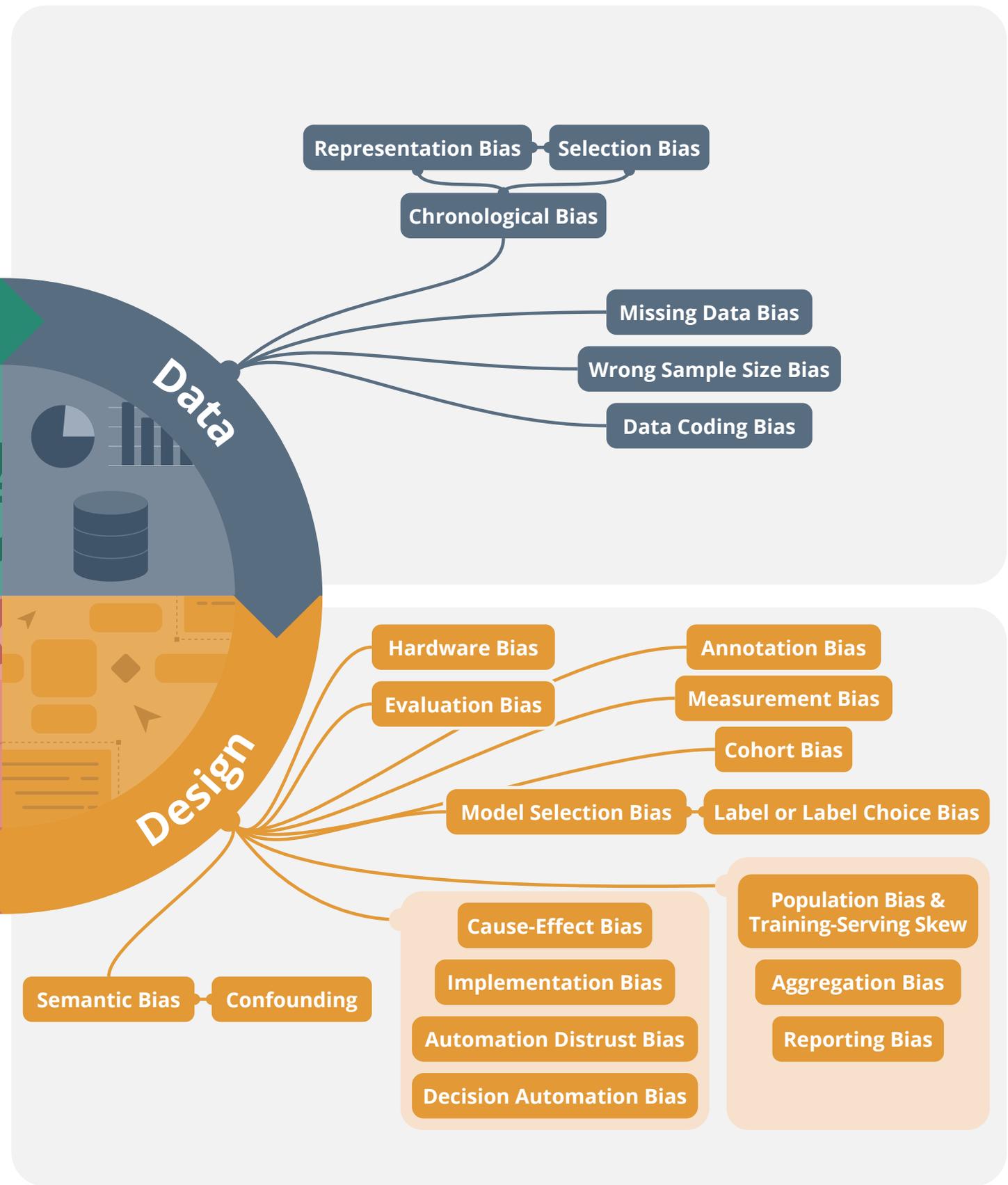

# World Biases

## Historical Bias

Historical Bias concerns pre-existing societal patterns of discrimination and social injustice—and the prejudices and discriminatory attitudes that correspond to such patterns. These patterns, prejudices, and attitudes can be drawn into every stage of the AI innovation lifecycle and be perpetuated, reinforced, or exacerbated through inequitable innovation ecosystem dynamics and the pursuit of biased application choices and research agendas. They can also arise in AI innovation contexts when historical patterns of inequity or discrimination are inadvertently or unintentionally reproduced, or even augmented, in the development and use of an AI system—even when the system is functioning to a high standard of accuracy and reliability.[126] [127] For instance, even with scientifically sound sampling and feature selection, a project will exhibit Historical Bias where it perpetuates (or exacerbates) socioeconomic inequalities through the outcomes it generates.

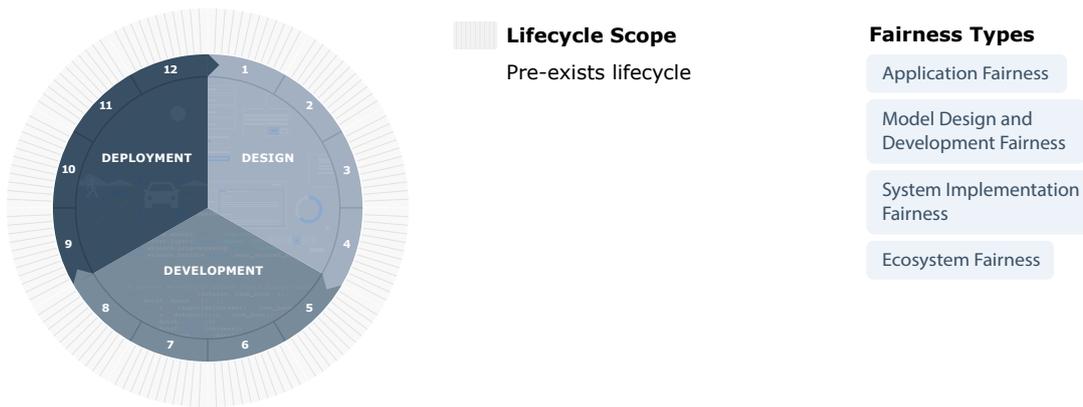

**Lifecycle Scope**

Pre-exists lifecycle

**Fairness Types**

Application Fairness

Model Design and Development Fairness

System Implementation Fairness

Ecosystem Fairness

## Structural Racism

Structural Racism (also sometimes called systemic racism) is a form of racial discrimination that is 'not simply the result of private prejudices held by individuals, but is also produced and reproduced by laws, rules, and practices, sanctioned and even implemented by various levels of government, and embedded in the economic system as well as in cultural and societal norms'.[128] Other forms of discrimination such as sexism, classism, ableism, ageism, antisemitism, and transphobia can also similarly have structural or systemic aspects.

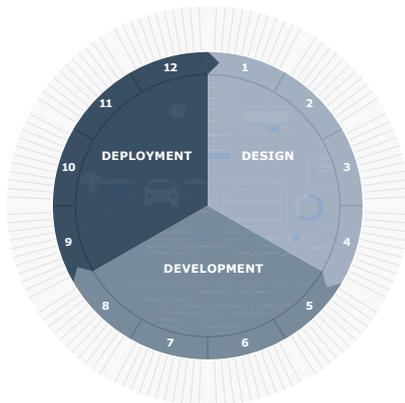

**Lifecycle Scope**
Pre-exists lifecycle

**Fairness Types**
Data Fairness
Application Fairness
Model Design and Development Fairness
System Implementation Fairness
Ecosystem Fairness
Metric-Based Fairness

## Institutional Bias

Institutional Bias is 'a tendency for the procedures and practices of particular institutions to operate in ways which result in certain social groups being advantaged or favoured and others being disadvantaged or devalued. This need not be the result of any conscious prejudice or discrimination but rather of the majority simply following existing rules or norms'.[129]

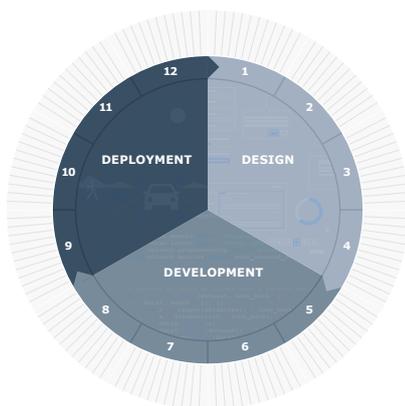

**Lifecycle Scope**
Pre-exists lifecycle

**Fairness Types**
Application Fairness
System Implementation Fairness
Ecosystem Fairness
Metric-Based Fairness

# Data Biases

## Representation Bias

When a population is either inappropriately represented (e.g. not allowing sufficient self-representation in demographic variables) or a subgroup is underrepresented in the dataset, the model may subsequently fail to generalise and underperform for a subgroup.[130] [131] [132] For example, representation biases could arise in a symptom checking application that has been trained on a data collected exclusively through smartphone use or online interaction as this dataset would likely underrepresent groups within the general population like elderly people who may lack access to smartphones or connectivity.

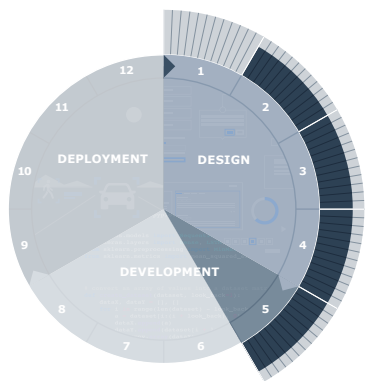

**Lifecycle Scope**
Project Planning – Preprocessing & Feature Engineering

**Significant Stages**
2. Problem Formulation
3. Data Extraction or Procurement
4. Data Analysis
5. Preprocessing & Feature Engineering

**Fairness Types**
Data Fairness
Model Design and Development Fairness
System Implementation Fairness

## Data Coding Bias

Data Coding Bias occurs when the misrepresentation or erasure of demographic characteristics such as gender or ethnicity by biased coding systems obscures needs and interests of impacted people, adversely impacts their access to needed good and services, and, subsequently, prejudices the datasets in which biased coding is embedded. For instance when AI system, which are shaped by data coding bias, are deployed in the context of health and healthcare, they can obscure patient needs, adversely impacts patient's access to appropriate screenings, diagnosis, and treatments, and, subsequently, prejudices the datasets in which biased coding is embedded.

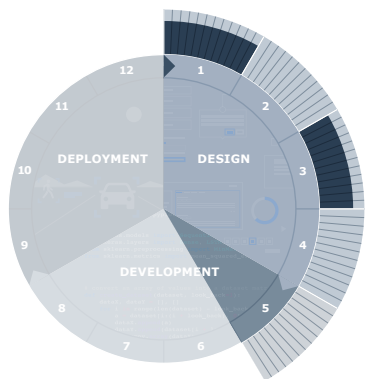

**Lifecycle Scope**
Project Planning – Preprocessing & Feature Engineering

**Significant Phases**
1. Project Planning
3. Data Extraction or Procurement

**Fairness Types**
Data Fairness
Model Design and Development Fairness

## Selection Bias

Selection Bias is a term used for a range of biases that affect the selection or inclusion of data points within a dataset. In general, this bias arises when an association is present between the variables being studied and additional factors that make it more likely that some data will be present in a dataset when compared to other possible data points in the space.[133] For instance, where individuals differ in their geographic or socioeconomic access to an activity or service that is the site of data collection, this variation may result in their exclusion from the corresponding dataset.

Likewise, where certain socioeconomically deprived or marginalised social groups are disproportionately dependent on a social service to fulfil basic needs, it may be oversampled if data is collected from the provision of that service.

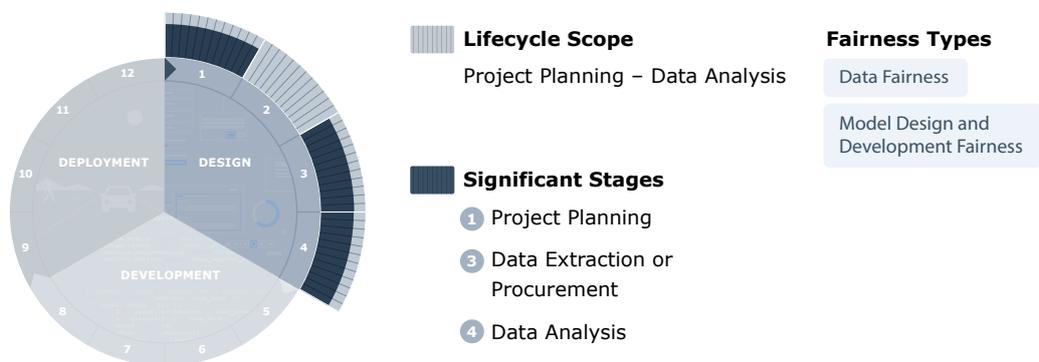

**Lifecycle Scope**
Project Planning – Data Analysis

**Significant Stages**
1. Project Planning
3. Data Extraction or Procurement
4. Data Analysis

**Fairness Types**
Data Fairness

Model Design and Development Fairness

## Chronological Bias

Chronological Bias arises when individuals in the dataset are added at different times, and where this chronological difference results in individuals being subjected to different methods or criteria of data extraction based on the time their data were recorded.[134] For instance, if the dataset used to build a predictive risk model in childrens' social care spans over several years, large-scale care reforms, policy changes, adjustments in relevant statutes (such as changes to legal thresholds or definitions), and changes in data recording methods may create major inconsistencies in the data points extracted from person to person.

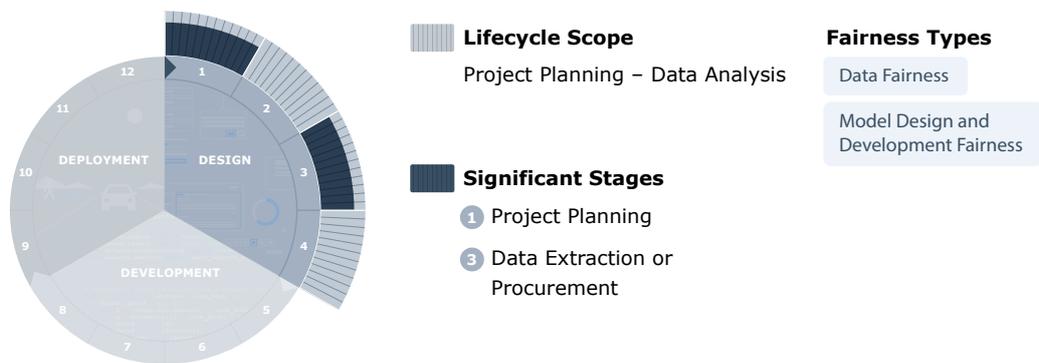

**Lifecycle Scope**
Project Planning – Data Analysis

**Significant Stages**
1. Project Planning
3. Data Extraction or Procurement

**Fairness Types**
Data Fairness

Model Design and Development Fairness

## Missing Data Bias

Missing data can cause a wide variety of issues within an AI project. These data may be missing for a variety of reasons related to broader social factors. Missingness can lead to inaccurate inferences and affect the validity of the model where it is the result of non-random but statistically informative events.[135] [136] For instance, Missing Data Bias may arise in predictive risk models used in social care to detect potentially harmful behaviour in adolescents. This is the case when interview responses and data collected from the same respondents over extended periods of time are used as part of the dataset. This can be seen in cases where interview questions about socially stigmatised behaviours or traits like drug use or sexual orientation trigger fears of punishment, humiliation, or reproach and thus prompt nonresponses, and in cases where data collection over time leads to the inconsistent involvement and drop-out of study participants.

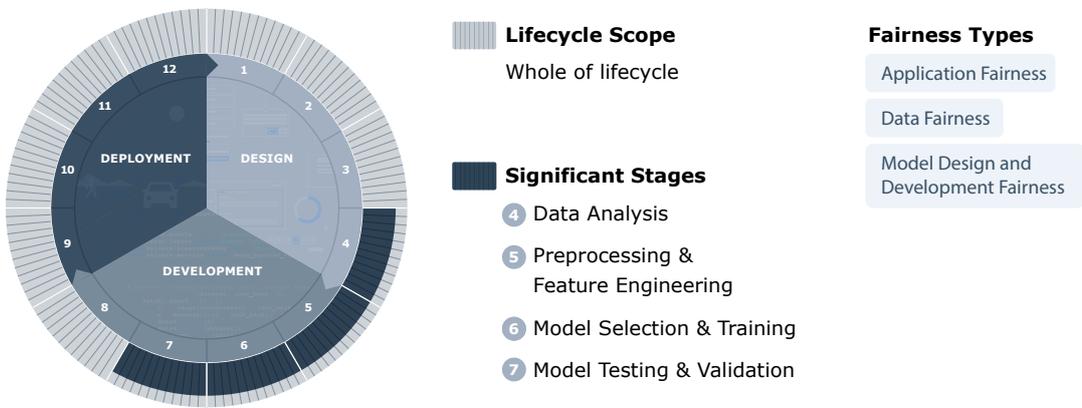

**Lifecycle Scope**
Whole of lifecycle

**Significant Stages**
4 Data Analysis
5 Preprocessing & Feature Engineering
6 Model Selection & Training
7 Model Testing & Validation

**Fairness Types**
Application Fairness
Data Fairness
Model Design and Development Fairness

## Wrong Sample Size Bias

Using the wrong sample size for the study can lead to chance findings that fail to adequately represent the variability of the underlying data distribution, in the case of small samples.[137] [138] Or it can lead to chance findings that are statistically significant but not relevant or actionable, in the case of larger samples. Wrong Sample Size Bias may occur in cases where model designers have included too many features in a machine learning algorithm. This is often referred to as the "curse of dimensionality". It is a mathematical phenomenon wherein increases in the number of features or "data dimensions" included in an algorithm means that exponentially more data points need to be sampled to enable good predictive or classificatory performance.[139] [140]

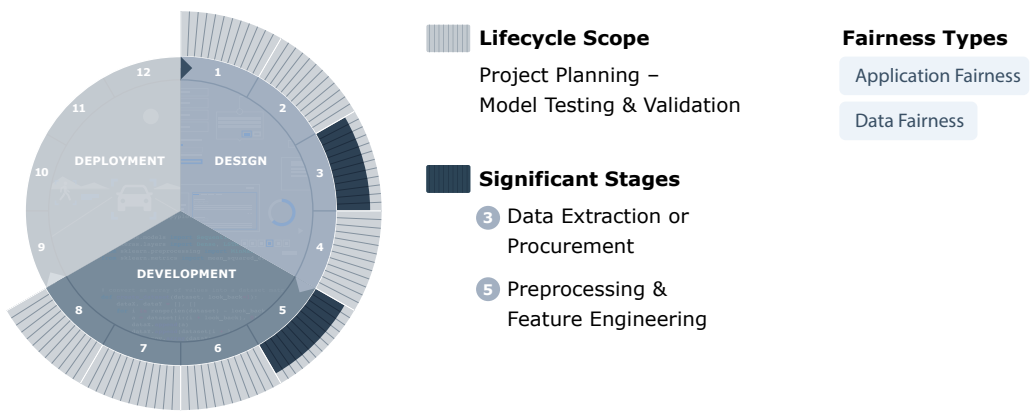

**Lifecycle Scope**
Project Planning – Model Testing & Validation

**Significant Stages**
3 Data Extraction or Procurement
5 Preprocessing & Feature Engineering

**Fairness Types**
Application Fairness
Data Fairness

# Design Biases

## Label or Label Choice Bias

A label (or target variable) used within an algorithmic model may not have the same meaning for all data subjects. There may be a discrepancy between what sense the designers are seeking to capture in a label, or what they are trying to measure in it, and the way that affected individuals understand its meaning.[141] [142] [143] [144] [145] Where there is this kind of variation in meaning for different groups within a population, adverse consequences and discriminatory impact could follow. For example, designers of a predictive model in public health may choose 'patient wellbeing' as their label, defining it in terms of disease prevalence and hospitalisation. However, subpopulations who suffer from health disparities and socioeconomic deprivation may understand wellbeing more in terms of basic functionings, the food security needed for health promotion, and the absence of the social environmental stressors that contribute to the development of chronic medical conditions. Were this predictive model is to be used to develop public health policy, members of this latter group could suffer from a further entrenchment of poor health outcomes.

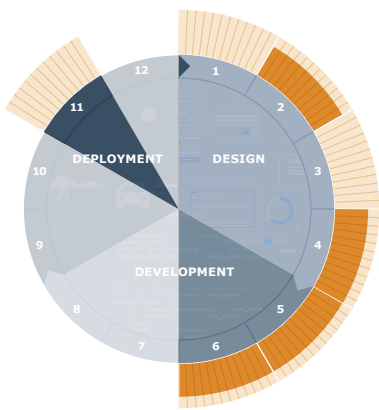

**Lifecycle Scope**

Project Planning – Model Selection & Training; System Use & Monitoring

**Significant Stages**

- ② Problem Formulation
- ④ Data Analysis
- ⑤ Preprocessing & Feature Engineering
- ⑥ Model Selection & Training

**Fairness Types**

- Application Fairness
- Model Design and Development Fairness
- System Implementation Fairness
- Ecosystem Fairness

## Measurement Bias

Measurement Bias addresses the choice of how to measure the labels or features being used. It arises when the measurement scale or criteria being applied fails to capture data pertaining to the concepts or constructs that are being measured in a fair and equitable manner.[146] [147] [148] [149] [150] For example, a recidivism risk model that uses prior arrests or arrested relatives as proxies to measure criminality may surface Measurement Bias insofar as patterns of arrest can reflect discriminatory tendencies to overpolice certain protected social groups or biased assessments on the part of arresting officers rather than true criminality.[151]

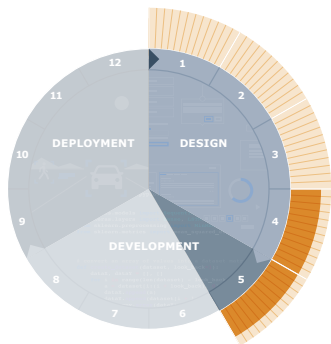

**Lifecycle Scope**
Project Planning – Preprocessing & Feature Engineering

**Significant Stages**
4. Data Analysis
5. Preprocessing & Feature Engineering

**Fairness Types**
Data Fairness
Model Design and Development Fairness

## Cohort Bias

Cohort Bias is a subtype of Measurement Bias where model designers use categories that default 'to traditional or easily measured groups without considering other potentially protected groups or levels of granularity (e.g. whether sex is recorded as male, female, or other or more granular categories)'.[152]

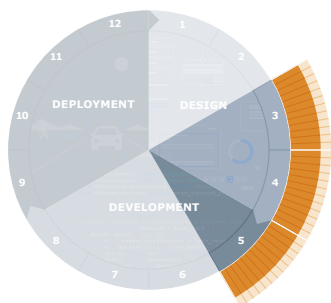

**Lifecycle Scope**
Data Extraction or Procurement - Preprocessing & Feature Engineering

**Significant Stages**
3. Data Extraction or Procurement
4. Data Analysis
5. Preprocessing & Feature Engineering

**Fairness Types**
Data Fairness
Model Design and Development Fairness
Ecosystem Fairness

## Annotation Bias

Annotation Bias occurs when annotators incorporate their subjective perceptions into their annotations.[153] [154] Data is often annotated by trained expert annotators or crowdsourced annotators. In its simplest form, annotators can choose an inaccurate label due to fatigue or lack of focus[155] but Annotation Bias can also result from positionality limitations that derive from demographic features, such as age, education, or first language,[156] and other systemic cultural or societal biases that influence annotators.[157] For instance, annotators may label differently facial expressions of different ethnic, age, or gender groups,[158] have different levels of familiarity with communication norms,[159] or different understandings of what should be annotated as harmful content.[160] When the role of annotator subjectivity is unacknowledged or annotators are not specifically trained to mitigate biases, there are greater chances that the model will incorporate Annotation Biases and be unfair.[161] [162]

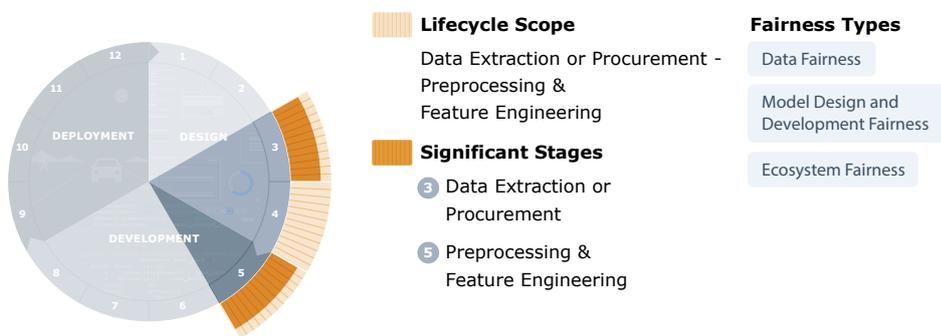

**Lifecycle Scope**
Data Extraction or Procurement - Preprocessing & Feature Engineering

**Significant Stages**
3. Data Extraction or Procurement
5. Preprocessing & Feature Engineering

**Fairness Types**
Data Fairness
Model Design and Development Fairness
Ecosystem Fairness

## Hardware Bias

Hardware Bias arises where physically instantiated algorithmic systems or measurement devices are not designed to consider the diverse physiological needs of minoritised, marginalised, disadvantaged, or other non-majority stakeholders. When deployed, such systems will therefore perform less effectively for members of these groups due to their design. For instance, mechanical design of implants for hip replacements that disregard the bone structure of people with wider hips, could lead to alignment problems. Similarly, poor outcomes, including higher rate of stroke risk, could result from ventricular assistive devices that may be too large for many people with smaller body frames.

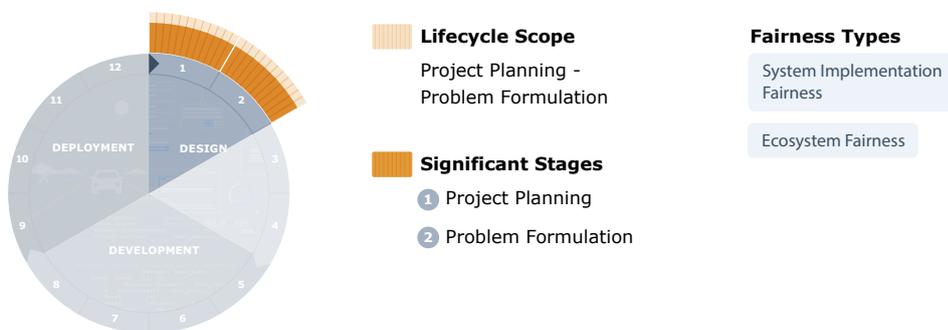

**Lifecycle Scope**
Project Planning - Problem Formulation

**Significant Stages**
1. Project Planning
2. Problem Formulation

**Fairness Types**
System Implementation Fairness
Ecosystem Fairness

## Model Selection Bias

Model Selection Bias occurs when AI designers and developers choose a model that does not sufficiently respond to the needs and requirements of the research question or problem and the domain context or use case. This may result in a lack of appropriate transparency and explainability, that is, where a complex model is chosen and the context demands interpretable results. Model Selection Bias may also result in outcomes based upon model inferences that do not reflect the level of nuance needed to confront the question or problem itself.

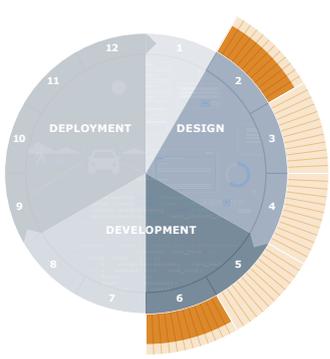

**Lifecycle Scope**
Problem Formulation - Model Selection & Training

**Significant Stages**
2 Problem Formulation
6 Model Selection & Training

**Fairness Types**
Metric-Based Fairness
Model Design and Development Fairness
Ecosystem Fairness

## Evaluation Bias

Evaluation Bias occurs during model iteration and evaluation as a result of the application of performance metrics that are insufficient given the intended use of the model and the composition of the dataset on which it is trained.[163] For example, an Evaluation Bias may occur where performance metrics that measure only overall accuracy are applied to a trained computer vision system that performs differentially for subgroups that have different skin tones.[164] Likewise, Evaluation Biases arise where the external benchmark datasets that are used to evaluate the performance of trained models are insufficiently representative of the populations to which they will be applied.[165] In the case of computer vision, this may occur where established benchmarks overly represent a segment of the populations (such as adult light-skinned males) and thus reinforce the biased criteria for optimal performance.

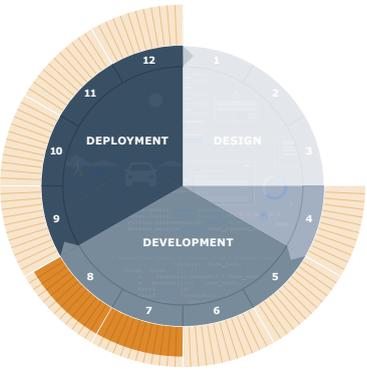

**Lifecycle Scope**
Data Analysis – Model Updating or Deprovisioning

**Significant Stages**
7 Model Testing & Validation
8 Model Reporting

**Fairness Types**
Metric-Based Fairness
Model Design and Development Fairness
Ecosystem Fairness

## Semantic Bias

Semantic bias occurs when discriminatory inferences are allowed to arise in the architecture of a trained AI model and to remain an element of the productionalised system. When Historical Biases are baked into datasets in the form of discriminatory proxies or embedded prejudices (e.g. word embeddings that pick up on racial or gender biases), these biases can be semantically encoded in the model's covariates and parameters.[166] Semantic biases occur when model design and evaluation processes fail to detect and mitigate such discriminatory aspects.

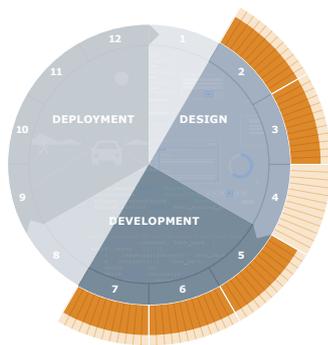

**Lifecycle Scope**
Problem Formulation - model Testing & Validation

**Significant Stages**
2. Problem Formulation
3. Data Extraction or Procurement
5. Preprocessing & Feature Engineering
6. Model Selection & Training
7. Model Testing & Validation

**Fairness Types**
Data Fairness
Model Design and Development Fairness
Ecosystem Fairness

## Confounding

In statistics, Confounding refers to a distortion that occurs when a certain variable independently influences both the dependent and independent variables. It can suggest there is a connection between the dependent and independent variables when there really isn't one. This can also skew the output.[167] Clear examples of confounding can be found in the use of electronic health records (EHRs) that arise in clinical environments and healthcare processes.[168] [169] Not only do EHRs reflect the health status of patients, but also patients' interactions with the healthcare system. This can introduce confounders such as the frequency of inpatient medical testing reflecting the busyness or labour shortages of medical staff rather than the progression of a disease during hospitalisation, differences between the onset of a disease and the date of diagnosis, and health conditions that are missing from the EHRs of a patient due to a non-random lack of testing. Contextual awareness and domain knowledge are crucial elements for identifying and redressing confounders.

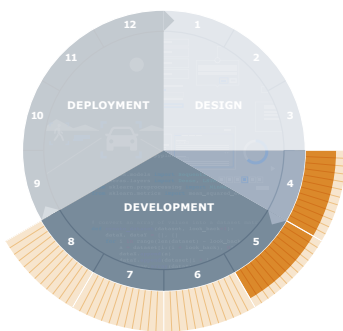

**Lifecycle Scope**
Data Analysis – Model Reporting

**Significant Stages**
4. Data Analysis
5. Preprocessing & Feature Engineering

**Fairness Types**
Data Fairness
Model Design and Development Fairness
Ecosystem Fairness

## Aggregation Bias

Aggregation Bias arises when a "one-size-fits-all" approach is taken to the outputs of a trained algorithmic model. Model results apply evenly to all members of the impacted population, even where variations in characteristics between subgroups mean that mapping functions from inputs to outputs are not consistent across these subgroups.[170] [171] In other words, in a model where Aggregation Bias is present, even when combinations of features affect members of different subgroups differently, the output of the system disregards the relevant variations in condition distributions for the subgroups. This results in the loss of information and lowered performance. In cases where data from one subgroup is more prevalent than those of others, the model will be more effective for that subgroup. For example, in clinical decision-support systems, clinically significant variations between sexes and ethnicities—in terms of disease aetiology, expression, complications, and treatment—mean that systems which aggregate results by treating all data points similarly will not perform optimally for any subgroup.[172]

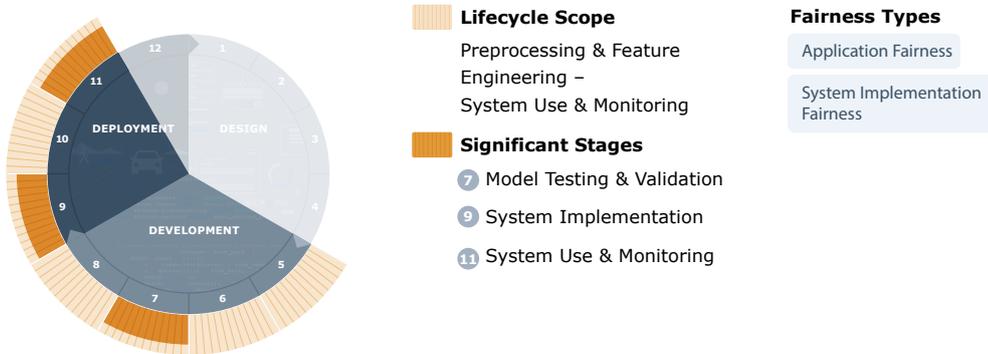

**Lifecycle Scope**
Preprocessing & Feature Engineering –
System Use & Monitoring

**Significant Stages**
7 Model Testing & Validation
9 System Implementation
11 System Use & Monitoring

**Fairness Types**
Application Fairness
System Implementation Fairness

## Reporting Bias

Reporting Bias arises when systems are produced without transparently reported evidence of effectiveness across demographic categories or testing for differential performance across sensitive, protected, or intersectional subgroups.[173] This can lead system developers and users to disregard or to attempt to bypass public accountability for equity, fairness, and bias mitigation. This means that transparency and accountability measures that should be in place to ensure that there are no discriminatory harms resulting from the use of the system are obscured or deprioritised.

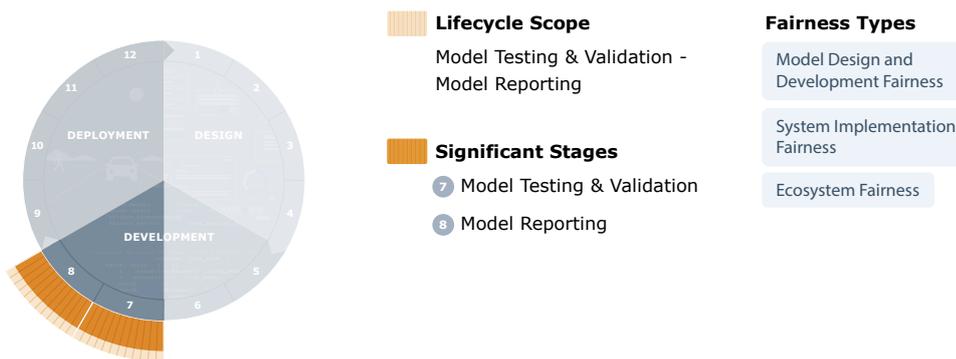

**Lifecycle Scope**
Model Testing & Validation - Model Reporting

**Significant Stages**
7. Model Testing & Validation
8. Model Reporting

**Fairness Types**
Model Design and Development Fairness
System Implementation Fairness
Ecosystem Fairness

## Population Bias and Training-Serving Skew

Population Bias occurs when the demographics or characteristics of the cohort that comprises the training dataset differ from those of the original target population to whom the model is applied.[174] [175] For instance, a polygenic risk model that was trained primarily on data from a cohort of people of European ancestry in a country that has significant ethnic diversity is applied to people of non-European descent in that country. Consequently, the model performs worse for them. Another example arises in the social media research context. The demographic composition of internet platform users is skewed towards certain population subgroups and hence can differ from the studied target population.[176]

Training-Serving Skew occurs when a model is deployed for individuals whose data is dissimilar to the dataset used to train, test, and validate the model.[177] This can occur, for instance, where a trained model is applied to a population in a geographical area different from where the original data was collected. Or where the model is applied to the same population but at a time much later than when the training data was collected. In both cases, the trained model may fail to generalise. The new, out-of-sample inputs are being drawn from populations with different underlying distributions.

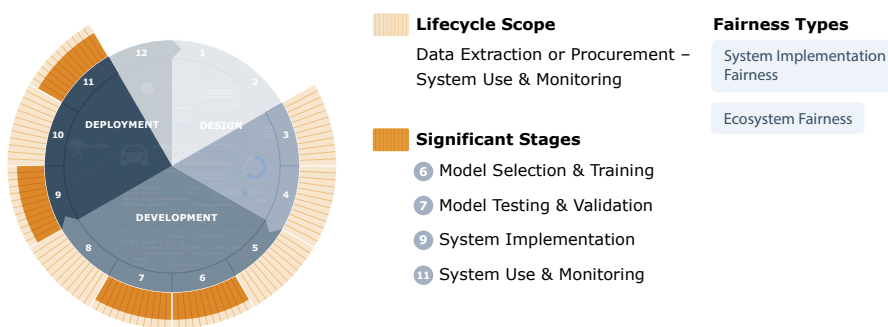

**Lifecycle Scope**
Data Extraction or Procurement – System Use & Monitoring

**Significant Stages**
6. Model Selection & Training
7. Model Testing & Validation
9. System Implementation
11. System Use & Monitoring

**Fairness Types**
System Implementation Fairness
Ecosystem Fairness

## Cause-Effect Bias

Decision-support systems generate inferences based upon statistical correlations. Cause-Effect Bias arises when users or implementers of these systems assume that correlation implies causation without examining the validity of such an attribution.[178] [179] For instance, a model for predicting pneumonia risk and hospital re-admission shows that patients with asthma have a lower risk of death than non-asthmatics. This leads users to erroneously conclude that having asthma is a protective factor in the risk scenario. However, a closer look at the model's inferences demonstrates a different situation. The correlation of asthma to lower risk was attributable to the fact that patients in the cohort known to have asthma were, as a common practice, automatically triaged to the Intensive Care Unit as a precautionary measure. This meant that they had a 50% reduction in mortality risk.[180]

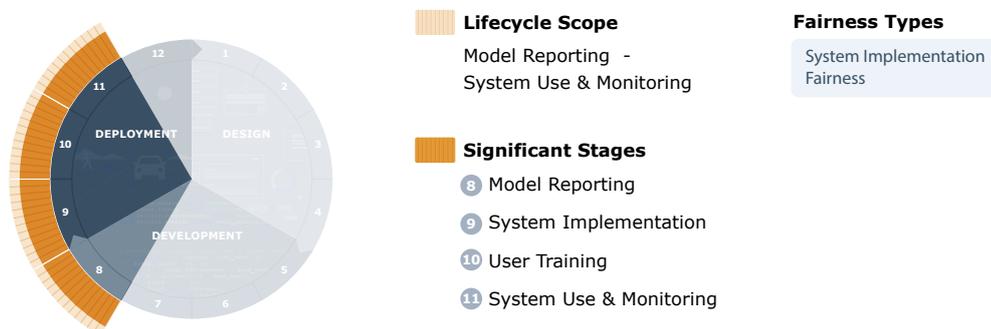

**Lifecycle Scope**

Model Reporting -
System Use & Monitoring

**Significant Stages**

8 Model Reporting
9 System Implementation
10 User Training
11 System Use & Monitoring

**Fairness Types**

System Implementation Fairness

## Implementation Bias

Implementation Bias refers to any bias that arises when a system is used or repurposed in ways that were not intended by its original designers or developers. This kind of unintended use is made more hazardous when the affordances that the system provides enables out-of-scope applications that pose higher risks of negative impact than originally envisioned. For example, a biometric identification system that is used by a public authority to assist in the detection of potential terrorist activity could be repurposed to target and monitor activists or political opponents.[181] [182]

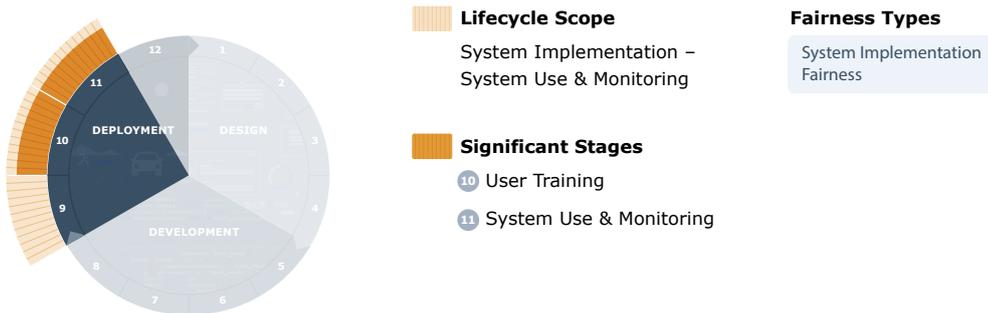

**Lifecycle Scope**
System Implementation –
System Use & Monitoring

**Significant Stages**
10 User Training
11 System Use & Monitoring

**Fairness Types**
System Implementation Fairness

## Decision-Automation Bias

Decision-Automation Bias arises when deployers of AI systems may tend to become hampered in their critical judgement, rational agency, and situational awareness.[183] [184] This happens because they believe that the AI system is objective, neutral, certain, or superior to humans. When this occurs, implementers may rely too much on the system or miss important faults, errors, or problems that arise over the course of the use of an automated system. Implementers become complacent and overly deferent to its directions and cues.[185] Decision-Automation Bias may also lead to overcompliance or errors of commission. This occurs when implementers defer to the perceived infallibility of the system and thereby become unable to detect problems emerging from its use for reason of a failure to hold the results against available information.

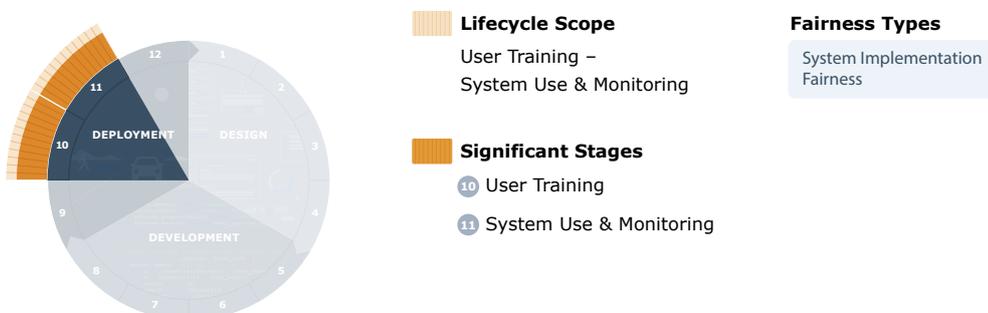

**Lifecycle Scope**
User Training –
System Use & Monitoring

**Significant Stages**
10 User Training
11 System Use & Monitoring

**Fairness Types**
System Implementation Fairness

## Automation Distrust Bias

Automation Distrust Bias arises when users of an automated decision-support system disregard its salient contributions to evidence-based reasoning either as a result of their distrust or scepticism about AI technologies in general or as a result of their over-prioritisation of the importance of prudence, common sense, and human expertise.[186] An aversion to the non-human and amoral character of automated systems may also influence decision subjects' hesitation to consult these technologies in high impact contexts such as healthcare, transportation, and law.[187]

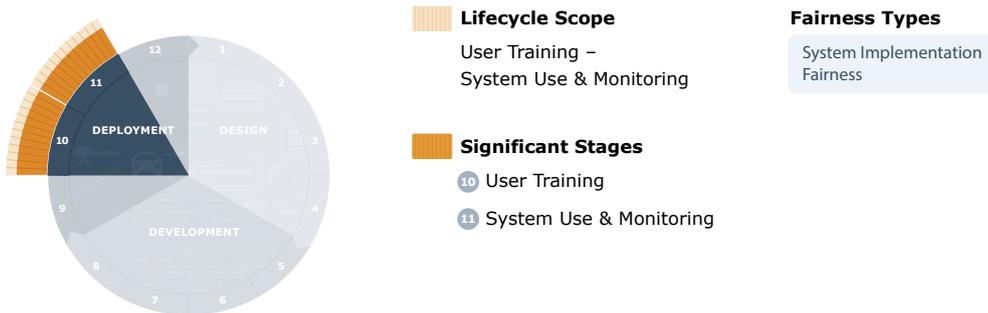

**Lifecycle Scope**
User Training –
System Use & Monitoring

**Significant Stages**
10 User Training
11 System Use & Monitoring

**Fairness Types**
System Implementation Fairness

# Ecosystem Biases

## Privilege Bias

Privilege Bias occurs when policies, institutions, and infrastructures skew the benefits of public service technologies disproportionately towards privileged social groups. For example, in the medical context, 'models may be unavailable in settings where protected groups receive care or require technology/sensors disproportionately available to the nonprotected class'.[188]

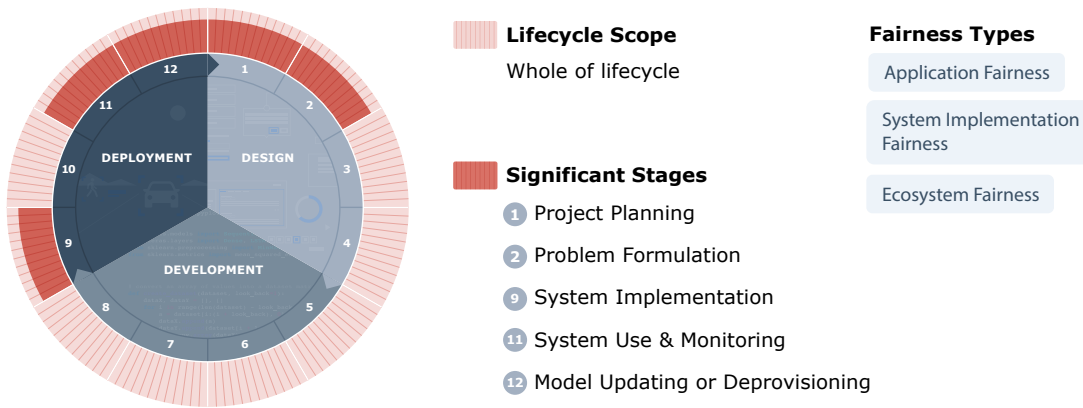

**Lifecycle Scope**
Whole of lifecycle

**Significant Stages**
1 Project Planning
2 Problem Formulation
9 System Implementation
11 System Use & Monitoring
12 Model Updating or Deprovisioning

**Fairness Types**
Application Fairness
System Implementation Fairness
Ecosystem Fairness

## Research Bias

Research Bias occurs where there is a deficit in social equity standards to guide how AI research and innovation is funded, conducted, reviewed, published, and disseminated. This can cause gaps and deficits in public interest oriented research. Research Bias can also manifest in a lack of inclusion and diversity on research teams, and limited studies incorporating representative real-world data for data-driven insights.[189] Research Bias additionally includes inequitable manifestations of funding structures or in incentives set by investors or funding institutions.

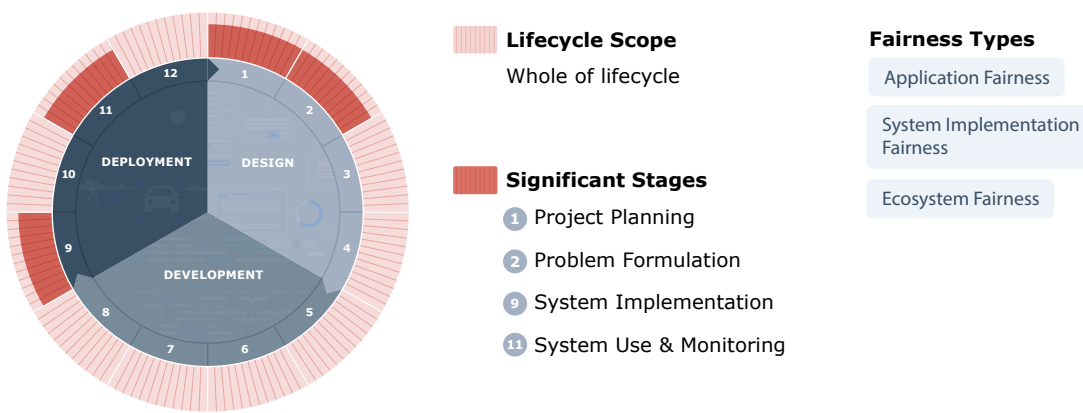

**Lifecycle Scope**
Whole of lifecycle

**Significant Stages**
1 Project Planning
2 Problem Formulation
9 System Implementation
11 System Use & Monitoring

**Fairness Types**
Application Fairness
System Implementation Fairness
Ecosystem Fairness

## McNamara Fallacy

McNamara Fallacy describes the belief that quantitative information is more valuable than other information.[190] This can lead to scientistic reductivism,[191] technochauvinism,[192] or technological solutionism.[193] The McNamara Fallacy plays a significant role in biasing AI innovation processes when AI researchers, designers, and developers view algorithmic techniques and statistical insights as the only inputs capable of solving societal problems. This leads to them actively disregarding interdisciplinary understandings of the subtle historical and sociocultural contexts of inequity and discrimination.

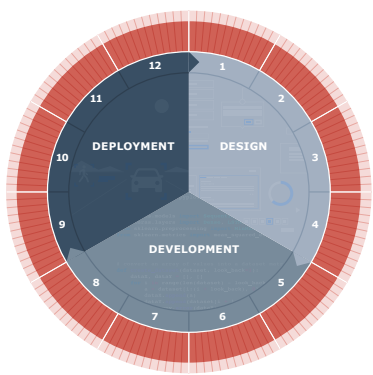

**Lifecycle Scope**
Whole of lifecycle

**Significant Stages**
All stages

**Fairness Types**
Data Fairness
Application Fairness
Model Design and Development Fairness
System Implementation Fairness
Ecosystem Fairness

## Biases of Rhetoric

Biases of Rhetoric occur during the communication of research or innovation results (e.g. model performance). It refers to the use of unjustified or illegitimate forms of persuasive language that lacks meaningful content or substantive evidence.[194] These biases relate to overemphasis of the performance or efficacy of a technique or intervention. For example, when showing comparative preference for the favoured technique to the detriment of feasible alternatives.

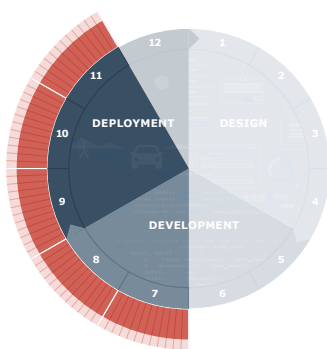

**Lifecycle Scope**
System Implementation – System Use & Monitoring

**Significant Stages**
7 Model Testing & Validation
8 Model Reporting
9 System Implementation
10 User Training
11 System Use & Monitoring

**Fairness Types**
Application Fairness
Model Design and Development Fairness
System Implementation Fairness
Ecosystem Fairness

### Informed Mistrust

Informed Mistrust occurs where protected groups believe that a model is biased against them because they have been mistreated in the past. Because of this, they may choose not to seek the needed social service or care (in the medical context) from the organisation which uses the model. Similarly, they may choose to not give all the necessary information to the organisation. The protected group may be harmed because they do not get the social service or care they need.[195]

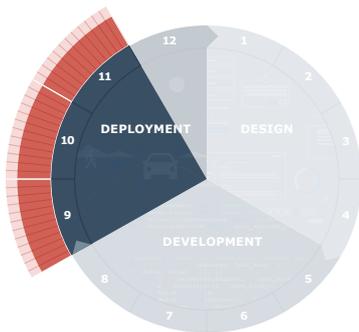

**Lifecycle Scope**
System Implementation - System Use & Monitoring

**Significant Stages**
9 System Implementation
10 User Training
11 System Use & Monitoring

**Fairness Types**
Application Fairness
Model Design and Development Fairness
Ecosystem Fairness

### De-Agentification Bias

De-Agentification Bias occurs when social structures and innovation practices systemically exclude minoritised, marginalised, vulnerable, historically discriminated against, or disadvantaged social groups from participating or providing input in AI innovation ecosystems. Protected groups may not have input into the different stages of the model. Likewise, they may not have 'the resources, education, or political influence to detect biases, protest, and force correction'.[196]

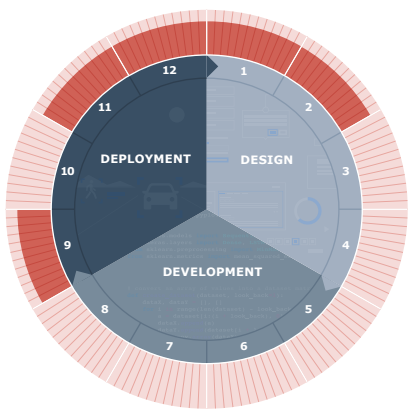

**Lifecycle Scope**
Whole of lifecycle

**Significant Stages**
1 Project Planning
2 Problem Formulation
9 System Implementation
11 System Use & Monitoring
12 Model Updating or Deprovisioning

**Fairness Types**
Application Fairness
Model Design and Development Fairness
Ecosystem Fairness

# Cognition Biases

## Availability Bias

The tendency to make judgements or decisions based on the information that is most readily available (e.g. more easily recalled). When this information is recalled on multiple occasions, the bias can be reinforced through repetition—known as a 'cascade'. This bias can cause issues for project teams throughout the project lifecycle where decisions are influenced by available or oft-repeated information (e.g. hypothesis testing during data analysis).

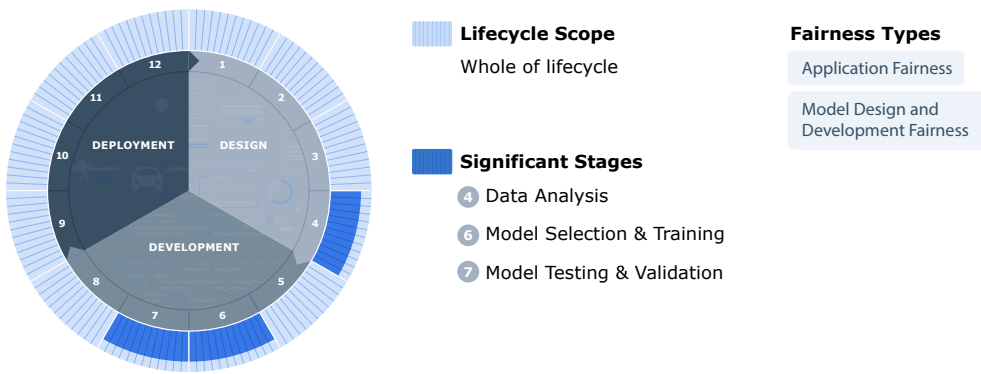

**Lifecycle Scope**
Whole of lifecycle

**Significant Stages**
- 4 Data Analysis
- 6 Model Selection & Training
- 7 Model Testing & Validation

**Fairness Types**
- Application Fairness
- Model Design and Development Fairness

## Self-Assessment Bias

A tendency to evaluate one's abilities in more favourable terms than others, or to be more critical of others than oneself. In the context of a project team, this could include the overly-positive assessment the group's abilities (e.g. through reinforcing groupthink). For instance, during Project Planning, a project team may believe that their resources and capabilities are sufficient for the objective of the project, but in fact be forced to either cut corners or deliver a product below its standard.

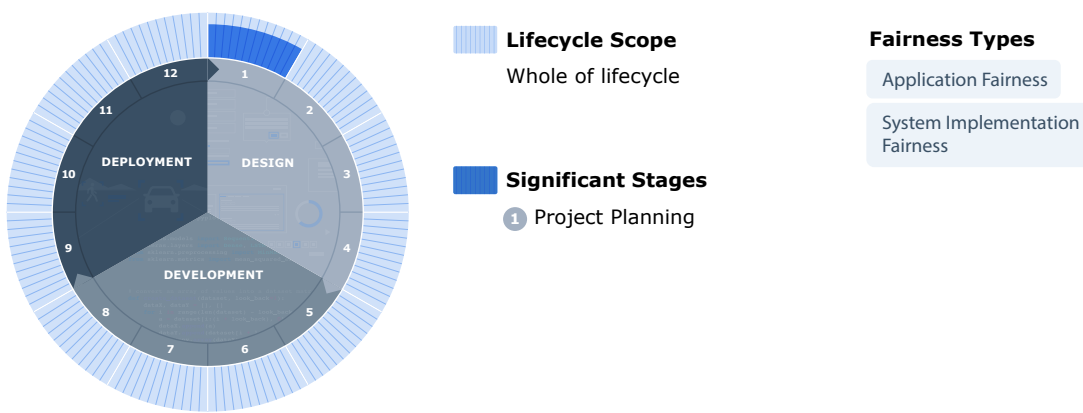

**Lifecycle Scope**
Whole of lifecycle

**Significant Stages**
- 1 Project Planning

**Fairness Types**
- Application Fairness
- System Implementation Fairness

## Confirmation Bias

Confirmation Biases arise from tendencies to search for, gather, or use information that confirms preexisting ideas and beliefs, and to dismiss or downplay the significance of information that disconfirms one's favoured hypothesis. This can be the result of motivated reasoning or sub-conscious attitudes. They in turn may lead to prejudicial judgements that are not based on reasoned evidence. For example, Confirmation Biases could surface in the judgment of the user of an AI decision-support application, who believes in following common sense intuitions acquired through professional experience rather than the outputs of an algorithmic model. For this reason, they dismiss its recommendations regardless of their rational persuasiveness or veracity.

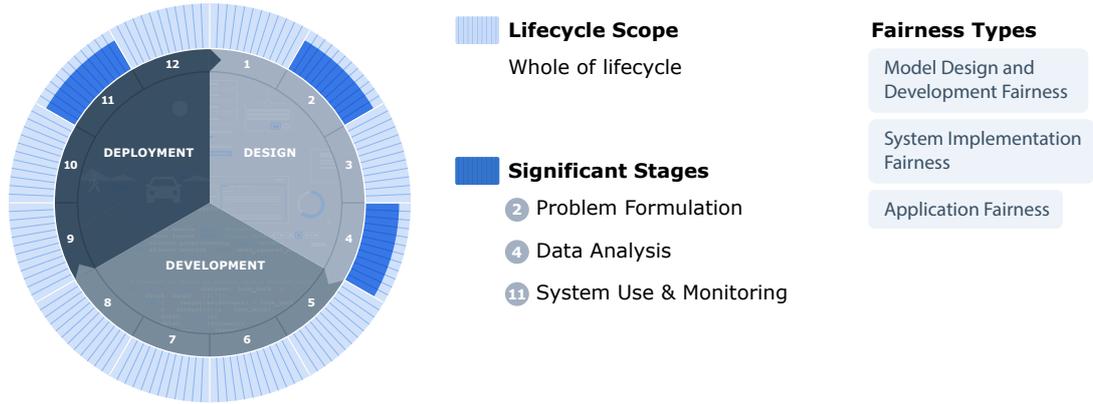

**Lifecycle Scope**
Whole of lifecycle

**Significant Stages**
2 Problem Formulation
4 Data Analysis
11 System Use & Monitoring

**Fairness Types**
Model Design and Development Fairness
System Implementation Fairness
Application Fairness

## Naïve Realism

A disposition to perceive the world in objective terms that can inhibit recognition of socially constructed categories. For instance, treating 'employability' something that is objectively measurable and, therefore, able to be predicted by a machine learning algorithm on the basis of objective factors (e.g. exam grades, educational attainment).

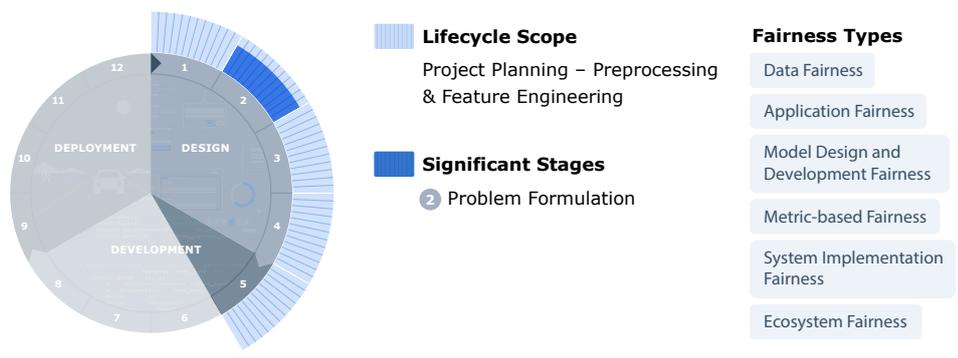

**Lifecycle Scope**
Project Planning – Preprocessing & Feature Engineering

**Significant Stages**
2 Problem Formulation

**Fairness Types**
Data Fairness
Application Fairness
Model Design and Development Fairness
Metric-based Fairness
System Implementation Fairness
Ecosystem Fairness

## Law of the Instrument (Maslow's Hammer)

This bias is best captured by the popular phrase 'If all you have is a hammer, everything looks like a nail'.[197] The phrase cautions against the cognitive bias of overreliance on a particular tool or method, perhaps one that is familiar to members of the project team. For example, a project team that is composed of experts in a specific ML technique, may overuse the technique and misapply it in a context where a different technique would be better suited. Or, in some cases, where it would be better not to use ML/AI technology at all.

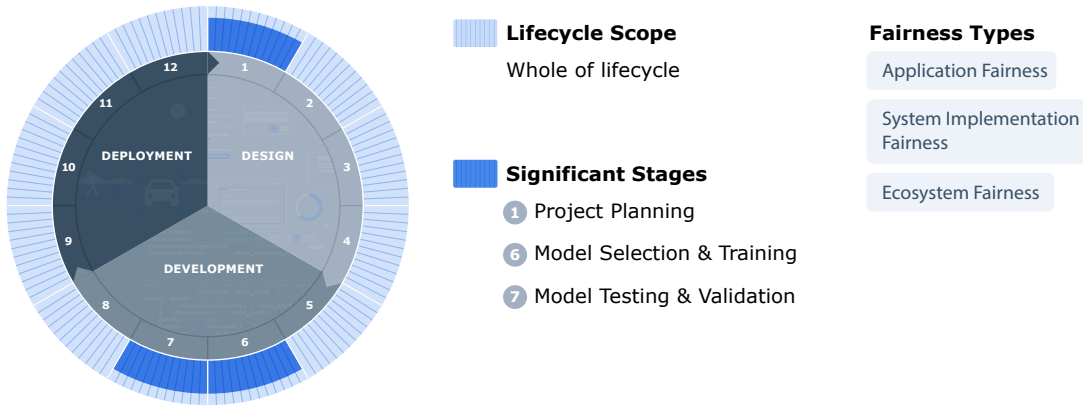

**Lifecycle Scope**
Whole of lifecycle

**Significant Stages**
1. Project Planning
6. Model Selection & Training
7. Model Testing & Validation

**Fairness Types**
Application Fairness
System Implementation Fairness
Ecosystem Fairness

## Optimism Bias

Optimism Bias can lead project teams to underestimate the amount of time required to adequately implement a new system or plan.[198] This is also known as the planning fallacy. In the context of the project lifecycle, Optimism Bias may arise during Project Planning. However, it can create downstream issues when implementing a model during the System Implementation stage, due to a failure to recognise possible system engineering barriers.

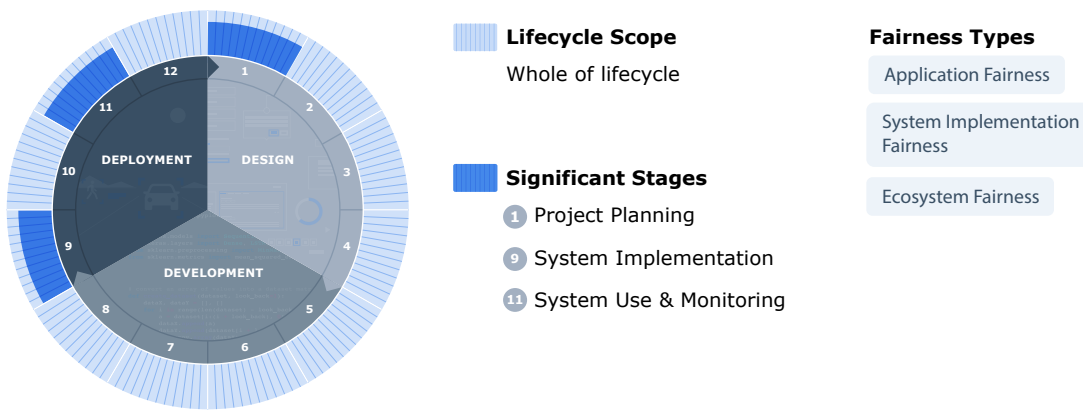

**Lifecycle Scope**
Whole of lifecycle

**Significant Stages**
1. Project Planning
9. System Implementation
11. System Use & Monitoring

**Fairness Types**
Application Fairness
System Implementation Fairness
Ecosystem Fairness

## Status Quo Bias

An affectively motivated preference for 'the way things are currently'. This can prevent more innovative or effective processes or services being implemented. Status Quo Bias is most acutely felt during the transition between projects. For instance, in the choice to deprovision a system and begin a new project, in spite of deteriorating performance from the existing solution.[199] Although this bias is often treated as a cognitive bias, we highlight it here as a social bias to draw attention to the broader social or institutional factors that in part determine the status quo.

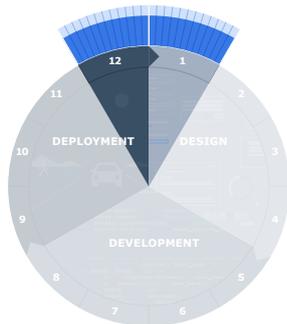

**Lifecycle Scope**
Model Updating or Deprovisioning – Project Planning

**Significant Stages**
- ⑫ Model Updating or Deprovisioning
- ① Project Planning

**Fairness Types**
- Application Fairness
- Model Design and Development Fairness
- Ecosystem Fairness

## Positive Results Bias

Positive Results Bias is also known as publication bias. It refers to the phenomenon of observing a skewed level of positive results published in journals because negative or null results tend to go unpublished.[200] This can cause people to think that certain techniques or methods or technique works better than they actually do. It can also lead to researchers duplicating studies that have already been done but not published. An example of this was observed in the well-known 'reproducibility crisis' that affected the social psychology literature.

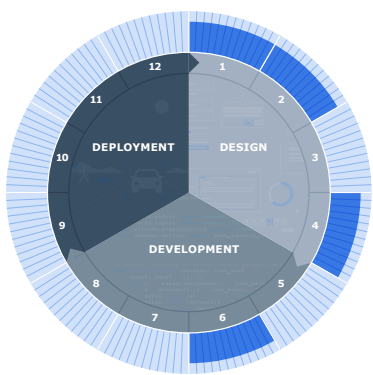

**Lifecycle Scope**
Whole of lifecycle

**Significant Stages**
- ① Project Planning
- ② Problem Formulation
- ④ Data Analysis
- ⑥ Model Selection & Training

**Fairness Types**
- Application Fairness
- System Implementation Fairness
- Ecosystem Fairness

# Endnotes

156 Kuwatly, H., Wich, M., & Groh, G. (2020). Identifying and measuring annotator bias based on annotators' demographic characteristics. *Proceedings of the Fourth Workshop on Online Abuse and Harms,* 184–190. https://doi.org/10.18653/v1/2020.alw-1.21

157 Chen, Y., & Joo, J. (2021). *Understanding and mitigating Annotation Bias in facial expression recognition (arXiv:2108.08504).* arXiv. http://arxiv.org/abs/2108.08504

158 Chen, Y., & Joo, J. (2021). *Understanding and mitigating Annotation Bias in facial expression recognition (arXiv:2108.08504).* arXiv. http://arxiv.org/abs/2108.08504

159 Hovy, D., & Prabhumoye, S. (2021). Five sources of bias in natural language processing. *Language and Linguistics Compass, 15*(8), e12432. https://doi.org/10.1111/lnc3.12432

160 Röttger, P., Vidgen, B., Hovy, D., & Pierrehumbert, J. B. (2022). Two contrasting data annotation paradigms for subjective NLP tasks. *arXiv.* 1–10. http://arxiv.org/abs/2112.07475

161 Chen, Y., & Joo, J. (2021). *Understanding and mitigating Annotation Bias in facial expression recognition (arXiv:2108.08504).* arXiv. http://arxiv.org/abs/2108.08504

162 Röttger, P., Vidgen, B., Hovy, D., & Pierrehumbert, J. B. (2022). Two contrasting data annotation paradigms for subjective NLP tasks. *arXiv.* 1–10. http://arxiv.org/abs/2112.07475

163 Suresh, H., & Guttag, J. (2021). A Framework for Understanding Sources of Harm throughout the Machine Learning Life Cycle. In Equity and Access in Algorithms, Mechanisms, and Optimization (EAAMO '21). *Association for Computing Machinery, USA,* 17, 1–9. https://doi.org/10.1145/3465416.3483305

164 Seyyed-Kalantari, L., Zhang, H., McDermott, M. B., Chen, I. Y., & Ghassemi, M. (2021). Underdiagnosis bias of artificial intelligence algorithms applied to chest radiographs in under-served patient populations. *Nature medicine, 27*(12), 2176-2182. https://doi.org/10.1038/s41591-021-01595-0

165 Malgady, R. G., Rogler, L. H., & Costantino, G. (1987). Ethnocultural and linguistic bias in mental health evaluation of Hispanics. *American Psychologist, 42*(3), 228. https://doi.org/10.1037//0003-066x.42.3.228

166 Hovy, D., & Prabhumoye, S. (2021). Five sources of bias in natural language processing. *Language and Linguistics Compass, 15*(8), e12432. https://doi.org/10.1111/lnc3.12432

167 McNamee R. (2003). Confounding and confounders. *Occupational and environmental medicine, 60*(3), 227–234. https://doi.org/10.1136/oem.60.3.227

168 Abramson, J. H., & Abramson, Z. H. (2001). *Making sense of data: a self-instruction manual on the interpretation of epidemiological data.* Oxford university press.

169 Farmer, R., Mathur, R., Bhaskaran, K., Eastwood, S. V., Chaturvedi, N., & Smeeth, L. (2018). *Promises and pitfalls of electronic health record analysis. Diabetologia, 61*(6), 1241–1248. https://doi.org/10.1007/s00125-017-4518-6

# Bibliography and Further Readings

## Algorithmic Fairness Techniques

To find out more about the AI Ethics and
Governance in Practice Programme please visit:

turing.ac.uk/ai-ethics-governance

The Alan Turing Institute